\documentclass[aps,reprint,notitlepage,nofootinbib,superscriptaddress]{revtex4-1}
\usepackage{pdfsync}
\usepackage{amsmath}    
\usepackage{amsfonts} 
\usepackage{amssymb}
\usepackage{MnSymbol}
\usepackage{mathrsfs} 
\usepackage{textcomp} 
\usepackage{bbm}
\usepackage{float}    
\usepackage{accents} 
\usepackage{natbib}
\usepackage{graphicx}   
\usepackage[caption=false]{subfig}
\usepackage{mathtools}
\usepackage{graphics}
\usepackage{epsfig}
\usepackage{multirow}
\usepackage{bibentry}
\usepackage{braket}
\usepackage{algorithm}
\usepackage{algorithmic}
\usepackage{listings}
\usepackage{setspace}
\usepackage{fancyhdr}
\usepackage{slashed}
\usepackage{xfrac}
\usepackage{xcolor}
\usepackage{cases}
\usepackage[retainorgcmds]{IEEEtrantools}
\usepackage[pdftex, citecolor=blue, urlcolor=blue, linkcolor=blue, colorlinks=true, bookmarksopen=true]{hyperref}
\allowdisplaybreaks  

\newcommand{\Cc}{\mathbb{C}}
\newcommand{\Ss}{\mathbb{S}}
\newcommand{\pT}{\hat{+}}
\newcommand{\mT}{\hat{-}}
\newcommand{\muT}{\hat{\mu}}
\newcommand{\nuT}{\hat{\nu}}

\newcommand{\uniT}[1]{\mathring{#1}}
\newcommand{\itP}[1]{\hat{#1}}

\newcommand{\Pp}{\mathbb{P}}
\newcommand{\Qq}{\mathbb{Q}}

\begin{document}
\title{The Electromagnetic Gauge Field Interpolation between the Instant Form and the Front Form of the Hamiltonian Dynamics}
\author{Chueng-Ryong Ji}
\affiliation{Department of Physics, North Carolina State University, Raleigh, North Carolina 27695-8202}
\author{Ziyue Li}
\affiliation{Department of Physics, North Carolina State University, Raleigh, North Carolina 27695-8202}
\author{Alfredo Takashi Suzuki}
\affiliation{Instituto de F\'{\i}sica Te\'orica-UNESP Universidade Estadual Paulista, Rua Dr. Bento Teobaldo Ferraz, 271 - Bloco II - 01140-070, S\~ao Paulo, SP, Brazil.}

\begin{abstract}
We present the electromagnetic gauge field interpolation between the instant form and the front form of the relativistic Hamiltonian dynamics and extend our interpolation of the scattering amplitude presented in the simple scalar field theory to the case of the electromagnetic gauge field theory with the scalar fermion fields known as the sQED theory.
We find that the Coulomb gauge in the instant form dynamics (IFD) and the light-front gauge in the front form dynamics,
or the light-front dynamics (LFD), are naturally linked by the unified general physical gauge that interpolates between
these two forms of dynamics and derive the spin-1 polarization vector for the photon that can be generally applicable for any
interpolation angle.
Corresponding photon propagator for an arbitrary interpolation angle is found and examined in terms of the gauge field polarization and the interpolating time ordering.
Using these results, we calculate the lowest-order scattering processes for an arbitrary interpolation angle in sQED.
We provide an example of breaking the reflection symmetry under the longitudinal boost, $P^z \leftrightarrow -P^z$,
for the time-ordered scattering amplitude in any interpolating dynamics except the LFD and clarify the
confusion in the prevailing notion of the equivalence between the infinite momentum frame (IMF) and the LFD.
The particular correlation found in our previous analysis of the scattering amplitude in the simple scalar field theory, coined as the J-shaped correlation, between the total momentum of the system and the interpolation angle persists in the present analysis of the sQED scattering amplitude. We discuss the singular behavior of this correlation in conjunction with the zero-mode issue in the LFD.
\end{abstract}

\maketitle

\section{Introduction}
\label{sec:introduction}

In 1949, Dirac \cite{Dirac1949} proposed three forms of relativistic dynamics: the instant form ($x^{0}=0$), the front form ($x^{+}=(x^{0}+x^{3})/\sqrt{2}$=0) and the point form ($x^{\mu}x_{\mu}=a^{2}>0, x^{0}>0$).
While the quantization at the equal time $t=x^{0}$ produces the instant form dynamics (IFD) of quantum field theory, the quantization at equal light-front time $\tau \equiv (t+z/c)/\sqrt{2}=x^{+}$ ($c$ is taken to be one unit in this work) yields the front form dynamics, known as the light-front dynamics (LFD).
Although the point form dynamics has also been explored \cite{Glozman1998a, *Wagenbrunn2001, *Melde2007}, the IFD and the LFD are still the most popular choices.

One of the reasons why the LFD is useful may be attributed to the energy-momentum dispersion relation.
For a particle of mass $m$ that has four-momentum $k=(k^{0},k^{1},k^{2},k^{3})$, its energy-momentum dispersion relation at equal-$t$ (instant form) is given by
\begin{align}
  k^{0}=\sqrt{\mathbf{k}^{2}+m^{2}}, \label{eqn:E-P_relation_IF}
\end{align}
where the energy $k^{0}$ is conjugate to $t$ and the three momentum vector $\mathbf{k}$ is given by $\mathbf{k}=(k^{1},k^{2},k^{3})$.
On the other hand, the corresponding energy-momentum relation at equal-$\tau$ (light-front form) is given by
\begin{align}
  k^{-}=\dfrac{\mathbf{k}_{\perp}^{2}+m^{2}}{k^{+}}, \label{eqn:E-P_relation_LF}
\end{align}
where the light-front energy $k^{-}=(k^{0}-k^{3})/\sqrt{2}$ is conjugate to $\tau$, and the light-front momenta $k^{+}=(k^{0}+k^{3})/\sqrt{2}$ and $\mathbf{k}_{\perp}=(k^{1},k^{2})$ are orthogonal to $k^{-}$.
In contrast to the irrational dispersion relation Eq.~(\ref{eqn:E-P_relation_IF}) in the IFD, this rational energy-momentum relation Eq.~(\ref{eqn:E-P_relation_LF}) in the LFD not only makes the relation simpler, but also correlates the sign of $k^{-}$ and $k^{+}$.
When the system is evolving to the future direction (i.e. positive $\tau$), in order for $k^{-}$ to be positive, $k^{+}$ also has to be positive.
This feature prevents certain processes from happening in the LFD, for example, the spontaneous pair production from vacuum is forbidden unless $k^{+}=0$ for both particles due to the momentum conservation.
Some dynamic processes are therefore eliminated in the LFD  and correspondingly the required computation may be simplified,
although we shall discuss the zero-mode issue involving the case of $k^{+}=0$ later.
In the IFD, however, this type of sign correlation does not exist, and the vacuum structure appears much more complicated than in the case of LFD due to the quantum fluctuations. This difference in the energy-momentum dispersion relation makes the LFD quite distinct from other forms of the relativistic Hamiltonian dynamics.

Furthermore, the Poincar\'e algebra is drastically changed in the LFD compared to the IFD.
In LFD, we have the maximum number (seven) of kinematic (i.e. interaction independent) operators out of the ten Poincar\'e generators and they leave the state at $\tau=0$ unchanged.
In particular, the longitudinal boost operator joins the stability group of kinematic operators in LFD.
This built-in boost invariance together with the simpler vacuum property makes the LFD quite appealing and may save substantial computational efforts to get the QCD solutions that reflect the full Poincar\'e symmetries.

The light-front quantization \cite{Dirac1949, Steinhardt1980} has been applied successfully in the context of current algebra \cite{Weinberg1966,*Jersak1968} and the parton model \cite{Bjorken1969, *DRELL1969} in the past.
With further advances in the Hamiltonian renormalization program \cite{Pauli1985, *Brodsky, Perry1990, *Perry1991, *Mustaki1991}, the LFD appears to be even more promising for the relativistic treatment of hadrons.
In the work of Brodsky et al. \cite{Brodsky1998a, *Hiller1998}, it is demonstrated how to solve the problem of renormalizing light-front Hamiltonian theories while maintaining Lorentz symmetry and other symmetries.
The genesis of the work presented in Ref.~\cite{Brodsky1998a, *Hiller1998} may be found in Ref.~\cite{Robertson1992,*McCartor1992} and additional examples including the use of LFD methods to solve the bound-state problems in field theory can be found in the review of QCD and other field theories on the light front \cite{Brodsky1998}.
A possible realization of chiral symmetry breaking in the light-front vacuum has also been discussed in the literature \cite{Susskind1994a,*Wilson}.

However, the transverse rotation whose direction is perpendicular to the direction of the quantization axis $z$ at equal $\tau$ becomes a dynamical problem in the LFD because the quantization surface $\tau$ is not invariant under the transverse rotation and the transverse angular momentum operator involves the interaction that changes the particle number \cite{JiSurya1992}.

As an effort to understand the conversion of the dynamical problem from boost to rotation as well as the link between the IFD and the LFD, we interpolate the two forms of dynamics by introducing an interpolation angle that changes the ordinary time $t$ to the light front time $\tau$ or vice versa.
The same method of interpolating hypersurfaces has been used by Hornbostel \cite{Hornbostel1992} to analyze various aspects of field theories including the issue of nontrivial vacuum. The same vein of application to study the axial anomaly in the Schwinger model has also been presented \cite{Ji1996}, and other related works \cite{Chen1971, Elizalde1976, Frishman1977, Sawicki1991} can also be found in the literature.

Our interpolation between the IFD and the LFD provides the whole picture of landscape between the two and clarifies the issue, if any, in linking them to each other. We started out by studying the Poincar\'e algebra for any arbitrary interpolation angle \cite{Ji2001}, and provided the physical meaning of the kinematic vs. dynamic operators by introducing the interpolating time-ordered scattering amplitudes \cite{Ji2012}. Although we want ultimately to obtain a general formulation for the QED and the QCD using the interpolation between the IFD and the LFD, we start from the simpler theory to discuss first the bare-bone structure that will persist even in the more complicated theories. Since we have studied the simple scalar field theory \cite{Ji2012} involving just the fundamental degrees of freedom such as the momenta of particles in scattering processes, we now consider involving the electromagnetic gauge degree of freedom interpolated between the IFD and the LFD in the present work. We develop the electromagnetic gauge field propagator interpolated between the IFD and the LFD and extend our interpolation of the scattering amplitude presented in the simple scalar field theory to the case of the electromagnetic gauge field theory but still with the scalar fermion fields known as the sQED theory.

In LFD, the light-front gauge ($A^{+}=(A^{0}+A^{3})/\sqrt{2}=0$) is commonly used, since the transverse polarizations of the gauge field can be immediately identified as the dynamical degrees of freedom, and ghost fields can be ignored in the quantum
action of non-Abelian gauge theory~\cite{Lepage1980, Bassetto1987, *Bassetto1998, Leibbrandt1987, *Leibbrandt1988,*Leibbrandt2000}. This makes it especially attractive in various QCD applications.
We find that the light-front gauge in the LFD is naturally linked to the Coulomb gauge in the IFD through the interpolation angle.
The corresponding gauge propagator that interpolates between the IFD and the LFD also sheds light on the debate about whether the gauge propagator should be the two-term form \cite{Mustaki1991} or the three-term form \cite{Leibbrandt1984, Srivastava2001, Suzuki2003, *Suzuki2004a, *Suzuki2004b, *Misra2005}. For example, by analyzing the lowest-order sQED Feynman amplitude, one may typically get the corresponding three time-ordered amplitudes, one of which corresponds to the contribution from the instantaneous interaction of the gauge field. This contribution from the instantaneous interaction is, however, precisely canceled by one of the terms in the three-term gauge propagator and thus the two-term gauge propagator may also be used effectively for the calculation of the same Feynman amplitude without involving the instantaneous interaction of the gauge field. Otherwise, to maintain the equivalence to the covariant formulation, the three-term propagator should be used including the instantaneous interaction of the gauge field.

Our work also clarifies the singular nature of the correlation between the total momentum of system and the interpolation angle and provides a deeper understanding of the treacherous zero-mode issue in the LFD. We find that the particular correlation between the total momentum of the system and the interpolation angle, coined as the J-shaped correlation in our previous analysis, persists even in the sQED scattering amplitude involving the gauge field. We discuss the universal nature of the J-shaped correlation which appears completely independent from the nature of particles (i.e. mass, spin, etc.) involved in the scattering process.

Although the interpolation method has been introduced before \cite{Hornbostel1992,Ji1996,Ji2001,Ji2012}, it has not yet been widely explored and a brief description of the method is still necessary for the presentation of our work. In Sec.~\ref{sec:interpolation_between_instant_form_and_light_front_form}, we thus provide a brief review of the interpolation angle method essential for the rest of this article.
In Sec.~\ref{sec:Link_between_Coulomb_gauge_and_Light-Front_gauge}, we derive the photon polarization vector for any interpolation angle.
Using this derivation, we present the general gauge that links the light-front gauge to the Coulomb gauge, and construct the corresponding photon propagator for an arbitrary interpolation angle.  In Sec.~\ref{sec:Time_Ordered_Photon_Exchange},
we decompose this interpolating gauge propagator according to the time ordering and apply it to the lowest scattering process such as an analogue of the well-known QED process $e \mu \rightarrow e \mu$ in sQED without involving the fermion spins.
We also take a close look at the limiting cases of $\Cc \rightarrow 0$ and compare the results with the exact $\Cc = 0$ (LFD) results in this section.
In Sec.~\ref{sec:the_frame_dependence_and_interpolation_angle_dependence_of_time_ordered_amplitudes}, we plot the time-ordered amplitudes in terms of the total momentum of the system and the interpolation angle to reveal both the frame dependence and interpolation angle dependence of these amplitudes.
We then give a detailed discussion of the universal J-shaped correlation curve that emerges from these time-ordered diagrams and how it gives rise to the zero-mode contributions at $P^{z}=-\infty$.
A summary and conclusion follows in Sec.\ref{sec:conclusions}.

In Appendix~\ref{sec:derivation_of_photon_polarization_vectors}, we list the explicit matrix representation of the boost $\mathbf{K}$ and rotation $\mathbf{J}$ generators and provide the explicit description of the steps involved in deriving the photon polarization vectors for an arbitrary interpolation angle.
In Appendix~\ref{sec:photon_propagator_derived_directly_from_polarization_vectors_in_the_interpolating_form},
we derive the numerator of the photon propagator from the photon polarization vectors for an arbitrary interpolation angle.
In Appendix~\ref{sec:photon_propagator_decomposition_on_the_light_front}, we decompose the photon propagator on the light-front in terms of the transverse and longitudinal components.
For the completeness and the comparison with our analysis of sQED scattering process (analogous to $e \mu \rightarrow e \mu$ in QED) presented in Secs.~\ref{sec:Time_Ordered_Photon_Exchange} and
\ref{sec:the_frame_dependence_and_interpolation_angle_dependence_of_time_ordered_amplitudes},
we present in Appendices~\ref{Append:time_ordered_scattering_amplitudes_for_pure_scalar_theory} and
\ref{Append:time_ordered_annihilation_amplitudes_for_photon_exchange_process}
our calculations for the same process ``$e \mu \rightarrow e \mu$" in the scalar field  theory and
the process related by the crossing symmetry ``$e^+ e^- \rightarrow \mu^+ \mu^-$" in sQED.
Together with our previous work for the scalar field theory discussed in \cite{Ji2012},
this work completes our study of the lowest order scattering processes related by the crossing symmetry
in the scalar field theory as well as in sQED.

\section{Method of Interpolation Angle}
\label{sec:interpolation_between_instant_form_and_light_front_form}

In this section, we briefly review the interpolation angle method presenting just the necessary formulae for the present work.
For more detailed introduction and review of this method, the readers may consult our previous works presented in \cite{Ji2001} and \cite{Ji2012}.

The interpolating space-time coordinates may be defined as a transformation from the ordinary space-time coordinates, $x^{\muT}=\mathcal{R}^{\muT}_{\phantom{\mu}{\nu}}x^{\nu}$, i.e.
\begin{align}\label{eqn:interpolation_angle_definition}
  \begin{pmatrix}
    x^{\pT}\\
    x^{\itP{1}}\\
    x^{\itP{2}}\\
    x^{\mT}
  \end{pmatrix}=
  \begin{pmatrix}
    \cos\delta & 0  & 0  & \sin\delta \\
    0          & 1  & 0  & 0 \\
    0          & 0  & 1  & 0 \\
    \sin\delta & 0  & 0  & -\cos\delta
  \end{pmatrix}
  \begin{pmatrix}
    x^{0}\\
    x^{1}\\
    x^{2}\\
    x^{3}
  \end{pmatrix},
\end{align}
where $0\leq\delta\leq\pi/4$ is the interpolation angle.
Following \cite{Ji2012}, we use ``\textasciicircum'' on the indices to denote the interpolating variables with the parameter $\delta$.
In the limits $\delta\rightarrow0$ and $\delta\rightarrow\pi/4$, we recover the corresponding variables in the instant form and the front form, respectively.  For example, the interpolating coordinates $x^{\itP{\pm}}$ in the limit $\delta\rightarrow\pi/4$ become the light-front coordinates $x^{\pm}=(x^{0}\pm x^{3})/\sqrt{2}$ without  ``\textasciicircum''.

In this interpolating basis, the metric becomes
\begin{align}\label{eqn:g_munu_interpolation}
  g^{\muT\nuT}
  = g_{\muT\nuT}
  =
  \begin{pmatrix}
    \Cc & 0  & 0  & \Ss \\
    0          & -1 & 0  & 0 \\
    0          & 0  & -1 & 0 \\
    \Ss & 0  & 0  & -\Cc
  \end{pmatrix},
\end{align}
where $\Ss=\sin2\delta$ and $\Cc=\cos2\delta$.
The covariant interpolating space-time coordinates are then easily obtained as
\begin{align}\label{eqn:covariant_x_interpolation_definition}
  x_{\muT}=g_{\muT\nuT}x^{\nuT}=
  \begin{pmatrix}
    x_{\pT}\\
    x_{\itP{1}}\\
    x_{\itP{2}}\\
    x_{\mT}
  \end{pmatrix}
  =
  \begin{pmatrix}
    \cos\delta & 0  & 0  & -\sin\delta \\
    0          & -1 & 0  & 0 \\
    0          & 0  & -1 & 0 \\
    \sin\delta & 0  & 0  & \cos\delta
  \end{pmatrix}
  \begin{pmatrix}
    x^{0}\\
    x^{1}\\
    x^{2}\\
    x^{3}
  \end{pmatrix}.
\end{align}
The same transformations also apply to the momentum:
\begin{subequations}
  \label{eqn:P_interpolation}
  \begin{align}
    P^{\pT}&=P^{0}\cos\delta + P^{3}\sin\delta,\label{eqn:P_interpolation_1}\\
    P^{\mT}&=P^{0}\sin\delta - P^{3}\cos\delta,\label{eqn:P_interpolation_2}\\
    P_{\pT}&=P^{0}\cos\delta - P^{3}\sin\delta,\label{eqn:P_interpolation_3}\\
    P_{\mT}&=P^{0}\sin\delta + P^{3}\cos\delta.\label{eqn:P_interpolation_4}
  \end{align}
\end{subequations}

Since the perpendicular components remain the same ($a^{\itP{j}}=a^{j},a_{\itP{j}}=a_{j}, j=1,2$), we will omit the ``\textasciicircum''  notation unless necessary from now on for the perpendicular indices $j=1,2$ in a four-vector.

Using $g^{\muT\nuT}$ and $g_{\muT\nuT}$, we see that the covariant and contravariant components are related by
\begin{alignat}{2}
  a_{\pT}=\Cc a^{\pT}+\Ss a^{\mT}; &\quad a^{\pT}=\Cc a_{\pT}+\Ss a_{\mT} \label{eqn:relation_between_covariant_and_contravariant_components_with_any_interpolation}\\
  a_{\mT}=\Ss a^{\pT}-\Cc a^{\mT}; &\quad a^{\mT}=\Ss a_{\pT}-\Cc a_{\mT} \nonumber\\
  a_{j}=-a^{j},&\quad (j=1,2) \nonumber.
\end{alignat}

The inner product of two four-vectors must be interpolation angle independent as
one can verify
\begin{align}\label{eqn:inner_product_of_four_vectors_interpolation_angle}
  a^{\muT}b_{\muT}&=(a_{\pT}b_{\pT}-a_{\mT}b_{\mT})\Cc+(a_{\pT}b_{\mT}+a_{\mT}b_{\pT})\Ss-a_{1}b_{1}-a_{2}b_{2}\nonumber\\
  &=a^{\mu}b_{\mu}.
\end{align}
In particular, we have the energy-momentum dispersion relation given by
\begin{align}\label{eqn:on_mass_shell_4_momentum_inner_product}
  P^{\muT}P_{\muT}=P_{\pT}^{2}\Cc-P_{\mT}^{2}\Cc+2P_{\pT}P_{\mT}\Ss-\mathbf{P}_{\perp}^{2}.
\end{align}
Another useful relation
\begin{align}\label{eqn:on_shell_4_momentum_useful_relation}
  P^{\muT}P_{\muT}\Cc=P^{\pT2}-P_{\mT}^{2}-\mathbf{P}_{\perp}^{2}\Cc
\end{align}
can also be easily verified.
For the particles of mass $M$, $P^{\mu}P_{\mu}$ on the mass shell equals $M^{2}$ of course.

\begin{table*}[t]
  \caption{\label{tab:Kinematic_and_dynamic_generators_for_different_interoplation_angles}Kinematic and dynamic generators for different interpolation angles}
    \begin{ruledtabular}
      \begin{tabular}{lcc}
	& Kinematic & Dynamic \\
	\hline
	\rule{0pt}{3ex} $\delta=0$ & $\mathcal{K}^{\hat{1}}=-J^{2}, \mathcal{K}^{\hat{2}}=J^{1}, J^{3}, P^{1}, P^{2}, P^{3}$ & $\mathcal{D}^{\hat{1}}=-K^{1}, \mathcal{D}^{\hat{2}}=-K^{2}, K^{3}, P^{0}$\\
	$0\leq\delta<\pi/4$ & $\mathcal{K}^{\hat{1}}, \mathcal{K}^{\hat{2}}, J^{3}, P^{1}, P^{2}, P_{\mT}$ & $\mathcal{D}^{\hat{1}}, \mathcal{D}^{\hat{2}}, K^{3}, P_{\pT}$\\
	$\delta=\pi/4$ & $\mathcal{K}^{\hat{1}}=-E^{1}, \mathcal{K}^{\hat{2}}=-E^{2}, J^{3}, K^{3}, P^{1}, P^{2}, P^{+}$ & $\mathcal{D}^{\hat{1}}=-F^{1}, \mathcal{D}^{\hat{2}}=-F^{2}, P^{-}$\\
      \end{tabular}
    \end{ruledtabular}
\end{table*}

Accordingly, the Poincar\'e matrix
\begin{align}\label{eqn:J_mu_nu_IF}
  M^{\mu\nu}&=
  \begin{pmatrix}
    0 & K^{1} & K^{2} & K^{3}\\
    -K^{1} & 0 & J^{3} & -J^{2}\\
    -K^{2} & -J^{3} & 0 & J^{1}\\
    -K^{3} & J^{2} & -J^{1} & 0
  \end{pmatrix}
\end{align}
transforms as well, so that
\begin{align}\label{eqn:Poincare_Matrix_Interpolation_superscripts}
  M^{\muT\nuT}
  &=
  \mathcal{R}^{\muT}_{\alpha}M^{\alpha\beta}\mathcal{R}^{\nuT}_{\beta}
  =
  \begin{pmatrix}
    0 & {E}^{\itP{1}} & {E}^{\itP{2}} & -{K}^{3}\\
    -{E}^{\itP{1}} & 0 & {J}^{3} & -{F}^{\itP{1}}\\
    -{E}^{\itP{2}} & -{J}^{3} & 0 & -{F}^{\itP{2}}\\
    {K}^{3} & {F}^{\itP{1}} & {F}^{\itP{2}} & 0
  \end{pmatrix}
\end{align}
and
\begin{align}\label{eqn:Poincare_Matrix_Interpolation_subscripts}
  M_{\muT\nuT}
  =
  g_{\muT\itP{\alpha}}M^{\itP{\alpha}\itP{\beta}}g_{\itP{\beta}\nuT}
  =
  \begin{pmatrix}
    0 & {\mathcal{D}}^{\itP{1}} & {\mathcal{D}}^{\itP{2}} & {K}^{3}\\
    -{\mathcal{D}}^{\itP{1}} & 0 & {J}^{3} & -{\mathcal{K}}^{\itP{1}}\\
    -{\mathcal{D}}^{\itP{2}} & -{J}^{3} & 0 & -{\mathcal{K}}^{\itP{2}}\\
    -{K}^{3} & {\mathcal{K}}^{\itP{1}} & {\mathcal{K}}^{\itP{2}} & 0
  \end{pmatrix},
\end{align}
where
\begin{align}\label{eqn:E_F_D_K_Definition_Interpolation_Angle}
  &E^{\itP{1}}=J^{2}\sin\delta+K^{1}\cos\delta,
  &&\mathcal{K}^{\itP{1}}=-K^{1}\sin\delta-J^{2}\cos\delta, \nonumber\\
  &E^{\itP{2}}=K^{2}\cos\delta-J^{1}\sin\delta,
  &&\mathcal{K}^{\itP{2}}=J^{1}\cos\delta-K^{2}\sin\delta, \nonumber\\
  &F^{\itP{1}}=K^{1}\sin\delta-J^{2}\cos\delta,
  &&\mathcal{D}^{\itP{1}}=-K^{1}\cos\delta+J^{2}\sin\delta, \nonumber\\
  &F^{\itP{2}}=K^{2}\sin\delta+J^{1}\cos\delta,
  &&\mathcal{D}^{\itP{2}}=-J^{1}\sin\delta-K^{2}\cos\delta.
\end{align}
The interpolating $E^{\itP{j}}$ and $F^{\itP{j}}$ will coincide with the usual $E^{j}$ and $F^{j}$ of LFD in the limit $\delta=\pi/4$.
Note here that the ``\textasciicircum'' notation is reinstated for $1, 2$ to emphasize the angle $\delta$ dependence and that the position of the indices on $K, J, E, F, \mathcal{D}, \mathcal{K}$ won't matter as they are not the four-vectors: i.e. $E_{\itP{1}}=E^{\itP{1}}$, etc. Of course, $M^{\muT\nuT}$ and $M_{\muT\nuT}$ should be distinguished in any case.

The generalized Poincar\'e Algebra for any interpolation angle can be found in \cite{Ji2001}.
Among the ten Poincar\'e generators, the six generators ($\mathcal{K}^{\itP{1}}, \mathcal{K}^{\itP{2}}, J^{3}, P_{1}, P_{2}, P_{\mT}$) are always kinematic in the sense that the $x^{\pT}=0$ plane is intact under the transformations generated by them.
As discussed in \cite{Ji2001, Ji2012}, the operator $K^{3}=M_{\pT\mT}$ is dynamical in the region where $0\leq\delta<\pi/4$ but becomes kinematic in the light-front limit ($\delta=\pi/4$).
The set of kinematic and dynamic generators depending on the interpolation angle are summarized in Table.~\ref{tab:Kinematic_and_dynamic_generators_for_different_interoplation_angles}.
Since the kinematic transformations don't alter $x^{\pT}$, the individual time-ordered amplitude must be invariant under the kinematic transformations. This can be seen explicitly in the example of scattering process discussed
in Secs.~\ref{sec:Time_Ordered_Photon_Exchange} and
\ref{sec:the_frame_dependence_and_interpolation_angle_dependence_of_time_ordered_amplitudes}.

Using the kinematic transformations defined above and following the procedure presented by Jacob and Wick \cite{Jacob1959}
to define the helicity in the IFD, we may define the helicity applicable to any arbitrary interpolation angle $\delta$.
For this purpose, we introduced the transformation $T$ \cite{Ji2001, Ji2012} given by
\begin{align}\label{eqn:T_transformation_for_any_interpolation_angle}
  T=T_{12}T_{3}=e^{i\beta_{1}\mathcal{K}^{\itP{1}}+i\beta_{2}\mathcal{K}^{\itP{2}}}e^{-i\beta_{3}K^{3}},
\end{align}
where we consider the operation on the state such as $T_{12}T_{3}| \psi \rangle$ in this work rather than
the operation on the operator as discussed in our previous work \cite{Ji2001, Ji2012}.
As shown in the textbook example of body-fixed frame vs. space-fixed frame in the Euler angle rotation
in the rigid body problem, the order of operation on the operator can be reversed in constructing the Euler
angle rotation depending on the choice of frame in the operation \cite{Sakurai}. The similar type of reverse in the
order of operation can occur in the case that the operation of $T$ is applied to the operator rather than to the state.
Thus, one should be careful in applying the operation with respect to the space that it applies to
when one considers the operation of $T$ to the operator rather than to the state as we have discussed in our
previous work \cite{Ji2012}.
In this work, we simplify the discussion by considering the operation of $T$
to the state but not to the operator. We also adopt the sign convention of the exponents $i\beta_{1}\mathcal{K}^{\hat{1}}$
and $i\beta_{2}\mathcal{K}^{\hat{2}}$ of $T_{12}$ to make the resulted momentum be positive
for the infinitesimal positive values of $\beta_{1}$ and $\beta_{2}$. Our sign convention in this work
turns out to be consistent with Soper's notation \cite{Soper1971} in the LFD.

Using the transformation given by Eq.(\ref{eqn:T_transformation_for_any_interpolation_angle}), we obtain the following
four-momentum components ($P'^{\pT}, P'^{1}, P'^{2}$ and $P'_{\mT}$) from the initial four-momentum components
($P^{\pT}, P^{1}, P^{2}$ and $P_{\mT}$)  \cite{Ji2001, Ji2012}:
\begin{subequations}
\begin{align}
  P'^{\pT}= & P^{\pT}\cosh\beta_{3}+P_{\mT}\sinh\beta_{3},\label{eqn:P_sup_ph_under_T_any_interpolation_angle}
\\
  P'^{1}= & P^{1}+\beta_{1}\frac{\sin\alpha}{\alpha}\left(P_{\mT}\cosh\beta_{3}+P^{\pT}\sinh\beta_{3}\right) \nonumber \\
  &+\frac{\cos\alpha-1}{\alpha^{2}}\Cc\beta_{1}\left(\beta_{1}P^{1}+\beta_{2}P^{2}\right), \label{eqn:P_1_T_any_interpolation_angle}\\
  P'^{2}= & P^{2}+\beta_{2}\frac{\sin\alpha}{\alpha}\left(P_{\mT}\cosh\beta_{3}+P^{\pT}\sinh\beta_{3}\right) \nonumber \\
  &+\frac{\cos\alpha-1}{\alpha^{2}}\Cc\beta_{2}\left(\beta_{1}P^{1}+\beta_{2}P^{2}\right),\label{eqn:P_2_T_any_interpolation_angle}\\
  P'_{\mT}=&\left(P_{\mT}\cosh\beta_{3}+P^{\pT}\sinh\beta_{3}\right)\cos\alpha \nonumber \\
  &+\frac{\sin\alpha}{\alpha}\Cc\left(\beta_{1}P^{1}+\beta_{2}P^{2}\right),\label{eqn:P_mh_T_any_interpolation_angle}
\end{align}
\end{subequations}
where $\alpha=\sqrt{\Cc(\beta_{1}^{2}+\beta_{2}^{2}})$. Although these four-momentum components ($P'^{\pT}, P'^{1}, P'^{2}$ and $P'_{\mT}$) given by Eqs.(\ref{eqn:P_sup_ph_under_T_any_interpolation_angle})-(\ref{eqn:P_mh_T_any_interpolation_angle})
turn out to be most convenient in carrying out our calculation, other choice of components such as $P'_{\pT}$ and $P'^{\mT}$
can be easily obtained using Eq.~(\ref{eqn:relation_between_covariant_and_contravariant_components_with_any_interpolation}).

As an example of using Eqs.(\ref{eqn:P_sup_ph_under_T_any_interpolation_angle})-(\ref{eqn:P_mh_T_any_interpolation_angle}),
we may find that the particle of mass $M$ at rest gains the following four-momentum components under the transformation $T$:
\begin{subequations}
  \label{eqn:4_momentum_transformation_from_rest_frame_any_interpolation_angle}
  \begin{align}
    P^{\pT}&=\left(\cos\delta\cosh\beta_{3}+\sin\delta\sinh\beta_{3}\right)M, \label{eqn:P_plus_hat_transformation_from_rest_frame}\\
    P^{1}&=\beta_{1}\frac{\sin\alpha}{\alpha}\left(\sin\delta\cosh\beta_{3}+\cos\delta\sinh\beta_{3}\right)M, \label{eqn:4_momentum_transformation_from_rest_b}\\
    P^{2}&=\beta_{2}\frac{\sin\alpha}{\alpha}\left(\sin\delta\cosh\beta_{3}+\cos\delta\sinh\beta_{3}\right)M, \label{eqn:4_momentum_transformation_from_rest_c}\\
    P_{\mT}&=\cos\alpha\left(\sin\delta\cosh\beta_{3}+\cos\delta\sinh\beta_{3}\right)M, \label{eqn:4_momentum_transformation_from_rest_d}
  \end{align}
\end{subequations}
where the factor $\left(\sin\delta\cosh\beta_{3}+\cos\delta\sinh\beta_{3}\right)$ in the three momentum ($P^{1}, P^{2}$, $P_{\mT}$)
is due to the first boost $T_{3}=e^{-i\beta_{3}K^{3}}$. Our convention taking this factor to be positive, i.e.
$\left(\sin\delta\cosh\beta_{3}+\cos\delta\sinh\beta_{3}\right)>0$, is consistent with the convention taken by
Jacob and Wick \cite{Jacob1959} in their procedure to define the helicity in the IFD.
We also note that $P_{\mT}=M\sin\delta$, $P^{\pT}=M\cos\delta$, and $P^{1}=P^{2}=0$ in the particle rest frame.

Solving Eqs.(\ref{eqn:P_plus_hat_transformation_from_rest_frame})-(\ref{eqn:4_momentum_transformation_from_rest_d}) for
$\beta_1, \beta_2, \beta_3$, we further note that
\begin{subequations}
  \label{eqn:beta123_relation_with_P}
 \begin{align}
  \sin\delta\cosh\beta_{3}+\cos\delta\sinh\beta_{3}&=\dfrac{ \Pp}{M},\\
  \cos\delta\cosh\beta_{3}+\sin\delta\sinh\beta_{3}&=\dfrac{ P^{\pT} }{M},
 \end{align}
\end{subequations}
and
\begin{subequations}
 \label{eqn:useful_beta_P_relation}
 \begin{align}
  \cos\alpha&=\dfrac{P_{\mT}}{\Pp},\\
  \sin\alpha&=\dfrac{\sqrt{\mathbf{P}_{\perp}^{2}\Cc}}{\Pp},\\
  e^{\beta_{3}}&=\dfrac{P^{\pT}+ {\Pp}}{M\left(\sin\delta+\cos\delta\right)},\\
  e^{-\beta_{3}}&=\dfrac{P^{\pT}- {\Pp}}{M\left(\cos\delta-\sin\delta\right)},\\
  \dfrac{\beta_{j}}{\alpha}&=\dfrac{P^{j}}{\sqrt{\mathbf{P}_{\perp}^{2}\Cc}},(j=1,2),
 \end{align}
\end{subequations}
where $\Pp\equiv\sqrt{P_{\mT}^{2}+\mathbf{P}_{\perp}^{2}\Cc}$ corresponds to the magnitude of the particle's three-momentum
in the IFD while it becomes identical to $P^+$ in the limit $\delta \rightarrow \pi/4$ so that $\alpha = 0$, i.e. $\cos\alpha=1$ and $\sin\alpha=0$, in the LFD as one can see from Eqs.(\ref{eqn:useful_beta_P_relation}a) and (\ref{eqn:useful_beta_P_relation}b), respectively. Multiplying Eqs.(\ref{eqn:useful_beta_P_relation}c) and (\ref{eqn:useful_beta_P_relation}d), we get the on-mass-shell condition consistent with Eq. (\ref{eqn:on_shell_4_momentum_useful_relation}) for the particle of rest mass $M$:
\begin{align}\label{eqn:P+_P-_Pperp_relation}
  P_{\mT}^{2}+\mathbf{P}_{\perp}^{2}\Cc=(P^{\pT})^2-M^{2}\Cc.
\end{align}
Consequently, the quantity denoted by $\Pp=\sqrt{P_{\mT}^{2}+\mathbf{P}_{\perp}^{2}\Cc}$ can also be written as $\Pp=\sqrt{(P^{\pT})^2-M^{2}\Cc}$.
We note the correspondence of $\Pp$ to $\sqrt{(P^3)^{2}+\mathbf{P}_{\perp}^{2}}=|\mathbf{P}|$ in the limit $\delta \rightarrow 0$ (or $\Cc \rightarrow 1$) and $\Pp$ to $P^+$ in the limit $\delta \rightarrow \pi/4$ (or $\Cc \rightarrow 0$).


\section{The Link between the Coulomb Gauge and the Light-Front Gauge}
\label{sec:Link_between_Coulomb_gauge_and_Light-Front_gauge}

We now discuss the gauge field in an arbitrary interpolation angle. Rather than fixing the gauge first,
we start from the explicit physical polarization four-vectors of the spin-1 particle and then identify the corresponding
gauge that these explicit representations of the gauge field polarization satisfy.
This procedure is possible because we can follow the Jacob-Wick procedure discussed
in Sec. \ref{sec:interpolation_between_instant_form_and_light_front_form} and apply the corresponding
$T$ transformation [Eq.~(\ref{eqn:T_transformation_for_any_interpolation_angle})] to the rest frame
spin-1 particle polarization vectors which may be naturally given by the
spherical harmonics. Although this procedure applies to the physical spin-1 particle with a non-zero mass $M$ such as the $\rho$-meson, we may take advantage of the Lorentz invariance of the four-momentum squared $P^{\muT}P_{\muT} = M^2$ and
replace $M^2$ by $P^{\muT}P_{\muT}$ to extend the obtained polarization four-vectors to the virtual gauge particle. For the real photon, of course $M=0$ and the longitudinal polarization vector should be discarded.
From this procedure, we find that the identified gauge interpolates between the Coulomb gauge in the instant form and
the light-front gauge in the front form.

\subsection{Spin-1 Polarization Vector for Any Interpolation Angle}
\label{sec:polarization_vector_for_any_interpolation_angle}
We use the four-vector representation of Lorentz group for the spin-1 particle.
The polarization vector in a specific frame is obtained by boosting the four-vectors to that frame.
In the rest frame, where the four-momentum is $(M,0,0,0)$, the polarization vectors are taken to be
\begin{align}\label{eqn:polarization_vectors_in_rest_frame}
  \epsilon(\pm)=\mp\frac{1}{\sqrt{2}}(0,1,\pm i,0), &\quad \epsilon(0)=(0,0,0,1),
\end{align}
since the spherical harmonics $Y_1^{\pm 1}$ and $Y_1^0$ naturally correspond to the
transverse and longitudinal polarization vectors, respectively.

To find the polarization vectors of the spin-1 particle with an arbitrary momentum $P^{\muT}$, we use the Jacob-Wick procedure discussed
in Sec. \ref{sec:interpolation_between_instant_form_and_light_front_form} and apply the $T$ transformation [Eq.~(\ref{eqn:T_transformation_for_any_interpolation_angle})] with the explicit four-vector representation of the operators $\mathbf{K}$ and $\mathbf{J}$.
The details of the calculation are summarized in Appendix~\ref{sec:derivation_of_photon_polarization_vectors}.
The result of the polarization vectors written in the form of $(\epsilon_{\pT},\epsilon_{1},\epsilon_{2},\epsilon_{\mT})$ is given by
\begin{widetext}
  \begin{subequations}
    \label{eqn:polarization_vector_in_P_any_interpolation}
    \begin{align}
    \epsilon_{\muT}(P,+)&=
   - \dfrac{1}{\sqrt{2}\Pp}
    \left( \Ss |{\bf P}_{\perp}|,  \dfrac{P_{1}P_{\mT}-iP_{2}\Pp}{|{\bf P}_{\perp}|}, \dfrac{P_{2}P_{\mT}+i P_{1}\Pp}{|{\bf P}_{\perp}|},
    -\Cc |{\bf P}_{\perp}| \right), \label{eqn:polarization_vector_+_in_P_any_interpolation}\\
    \epsilon_{\muT}(P,-)&=
    \dfrac{1}{\sqrt{2}\Pp}
    \left(
    \Ss|{\bf P}_{\perp}|,
    \dfrac{P_{1}P_{\mT}+i P_{2}\Pp}{|{\bf P}_{\perp}|},
    \dfrac{P_{2}P_{\mT}-i P_{1}\Pp}{|{\bf P}_{\perp}|},
    -\Cc|{\bf P}_{\perp}|
    \right), \label{eqn:polarization_vector_-_in_P_any_interpolation}\\
    \epsilon_{\muT}(P,0)&=
    \dfrac{P^{\pT}}{M \Pp}
    \left(
    P_{\pT}-\dfrac{M^{2}}{P^{\pT}},
    P_{1},
    P_{2},
    P_{\mT}
    \right). \label{eqn:polarization_vector_0_in_P_any_interpolation}
  \end{align}
  \end{subequations}
\end{widetext}
They satisfy the transversality and orthorgonality constraints
\begin{align}\label{eqn:epsilon_constraints}
  \epsilon_{\muT}(P,\lambda)P^{\muT}=0, \quad \epsilon^{*}(P,\lambda)\cdot\epsilon(P,\lambda')=-\delta_{\lambda\lambda'}.
\end{align}
It is also obvious that $\epsilon(P,0)$ is ``parallel'' to the three momentum $\mathbf{P}$, since $(\epsilon_{1},\epsilon_{2},\epsilon_{\mT})\sim(P_{1},P_{2},P_{\mT})$.
Noticing that $\Ss\rightarrow0$, $\Cc\rightarrow1$, $\Pp\rightarrow|\mathbf{P}|$ when $\delta\rightarrow0$, and $\Ss\rightarrow1$, $\Cc\rightarrow0$, $\Pp\rightarrow P^{+}$ when $\delta\rightarrow\pi/4$, one can easily check that these polarization vectors have correct limits of the instant form and the light-front form at $\delta=0$ and $\delta=\pi/4$, respectively.

\subsection{Interpolating Transverse Gauge}
\label{sec:gauge_condition_for_any_interpolation_angle_and_photon_propagator}

Having obtained the explicit polarization four-vectors of the spin-1 particle with mass $M$, we notice that the transverse polarizations
given by Eqs.~(\ref{eqn:polarization_vector_+_in_P_any_interpolation}) and (\ref{eqn:polarization_vector_-_in_P_any_interpolation})
are independent of the particle mass $M$. Thus, they can be used also as the transverse polarization four-vectors of the gauge field such as the photon.

We observe that the transverse polarization vectors ($\lambda = \pm$) in Eqs.~(\ref{eqn:polarization_vector_+_in_P_any_interpolation}) and (\ref{eqn:polarization_vector_-_in_P_any_interpolation}) satisfy the following conditions:
\begin{align}
  \epsilon^{\pT}(\lambda)=\Cc\epsilon_{\pT}(\lambda)+\Ss\epsilon_{\mT}(\lambda)=0,\\
	\epsilon_{\mT}(\lambda)P_{\mT}+\boldsymbol\epsilon_{\perp}(\lambda)\mathbf{P}_{\perp}\Cc=0, \label{eqn:epsilon_gauge_condition}
\end{align}
where we used $\epsilon_{\muT}(\lambda)=\epsilon_{\hat{\mu}}(P,\lambda)$ for convenience.
So, we can write the gauge condition for transverse photons as
\begin{align}
  A^{\pT}=0,\label{eqn:gauge_condition_for_any_interpolation_angle_a}\\
  A_{\mT}P_{\mT}+\mathbf{A}_{\perp}\mathbf{P}_{\perp}\Cc=0, \label{eqn:gauge_condition_for_any_interpolation_angle_b}
\end{align}
where the second condition can also be written as
\begin{align}\label{eqn:general_gauge_condition_position_space}
  \partial_{\mT}A_{\mT}+\boldsymbol\partial_{\perp}\mathbf{A}_{\perp}\Cc=0.
\end{align}

We now demonstrate that these two conditions are closely related to each other.
First of all, the Lorentz condition $\partial_{\muT}A^{\muT}=0$ is always satisfied, as already verified in Eq.~(\ref{eqn:epsilon_constraints}).
But the Lorentz condition can be rewritten in the following way
\begin{eqnarray}
  0&=&\partial_{\muT}A^{\muT}\nonumber\\
	&=&\partial_{\pT}A_{\pT}\Cc-\partial_{\mT}A_{\mT}\Cc+\partial_{\pT}A_{\mT}\Ss+\partial_{\mT}A_{\pT}\Ss-\boldsymbol\partial_{\perp}\mathbf{A}_{\perp}\nonumber\\
	&=&\partial_{\pT}A^{\pT}-\partial_{\mT}A_{\mT}\Cc+\partial_{\mT}A_{\pT}\Ss-\boldsymbol\partial_{\perp}\mathbf{A}_{\perp}\nonumber\\
	&=&\partial_{\pT}A^{\pT}-\left(\frac{1-\Ss^{2}}{\Cc}\right)\partial_{\mT}A_{\mT}+\partial_{\mT}A_{\pT}\Ss-\boldsymbol\partial_{\perp}\mathbf{A}_{\perp}\nonumber\\
	&=&\partial_{\pT}A^{\pT}+\frac{\Ss}{\Cc}\left(\Ss\partial_{\mT}A_{\mT}+\Cc\partial_{\mT}A_{\pT}\right)-\left(\frac{\partial_{\mT}A_{\mT}}{\Cc}+\boldsymbol\partial_{\perp}\mathbf{A}_{\perp}\right)\nonumber\\
  &=&\frac{\Cc}{\Cc}\partial_{\pT}A^{\pT}+\frac{\Ss}{\Cc}\left(\partial_{\mT}A^{\pT}\right)-\left(\frac{\partial_{\mT}A_{\mT}}{\Cc}+\boldsymbol\partial_{\perp}\mathbf{A}_{\perp}\right)\nonumber\\
  &=&\frac{1}{\Cc}\partial^{\pT}A^{\pT}-\left(\frac{\partial_{\mT}A_{\mT}}{\Cc}+\boldsymbol\partial_{\perp}\mathbf{A}_{\perp}\right)\nonumber\\
	&=&\frac{1}{\Cc}\left[ \partial^{\pT}A^{\pT}-\left(\partial_{\mT}A_{\mT}+\boldsymbol\partial_{\perp}\mathbf{A}_{\perp}\Cc\right) \right],
	\label{eqn:Par_A_decomposition}
\end{eqnarray}
where Eq.~(\ref{eqn:inner_product_of_four_vectors_interpolation_angle}) is used in the second line and the relations in Eq.~(\ref{eqn:relation_between_covariant_and_contravariant_components_with_any_interpolation}) are used to go between the superscript components and the subscript components.
Eq.~(\ref{eqn:Par_A_decomposition}) indicates that if Eq.~(\ref{eqn:gauge_condition_for_any_interpolation_angle_a}) holds, then we have Eq.~(\ref{eqn:general_gauge_condition_position_space}).
On the other hand, if Eq.~(\ref{eqn:general_gauge_condition_position_space}) holds, then we have
\begin{align}
  \partial^{\pT}A^{\pT}=0. \label{eqn:Gauge_condition_A+_differential_form}
\end{align}
However, this is a differential equation, and we have the freedom to specify the boundary condition.
And we can choose our boundary condition to make $A^{\pT}=0$.
This same trick was used in the instant form \cite{Ryder} where the Coulomb gauge $\boldsymbol\nabla \cdot \mathbf{A}=0$ gives $\partial_{0}A^{0}=0$, but we can choose our initial condition, or gauge, so that $A^{0}=0$.
Therefore, these two gauge conditions are effectively equivalent.

Similarly, Eq.~(\ref{eqn:gauge_condition_for_any_interpolation_angle_b}) by itself does not eliminate one degree of freedom.
We also need to specify a boundary condition for this differential equation.
Since we are not focusing on writing out $A_{\muT}$ explicitly, however, we will not dwell on this subject here.

The above discussion should make it clear that Eq.~(\ref{eqn:gauge_condition_for_any_interpolation_angle_a}) and Eq.~(\ref{eqn:general_gauge_condition_position_space}) are really two sides of the same coin.
However, in order to be consistent with conventions used in the instant form, where the radiation gauge condition is specified as ``$A^{0}=0$ and $\boldsymbol\nabla \cdot \mathbf{A}=0$'', here we say that the radiation gauge condition for any interpolating angle is Eq.~(\ref{eqn:gauge_condition_for_any_interpolation_angle_a}) and Eq.~(\ref{eqn:general_gauge_condition_position_space}).
In the instant form limit ($\delta=0$), $A_{\mT}\rightarrow A^{3}, \Cc\rightarrow1$, and Eq.~(\ref{eqn:general_gauge_condition_position_space}) becomes $\boldsymbol\nabla \cdot \mathbf{A}=0$ while Eq.~(\ref{eqn:gauge_condition_for_any_interpolation_angle_a}) becomes $A^{0}=0$, which is the familiar Coulomb gauge.
In the light-front limit ($\delta=\pi/4$), $A_{\mT}\rightarrow A^{+}, \Cc\rightarrow0$, and the gauge conditions reduce to ``$A^{+}=0$ and $\partial^{+}A^{+}=0$'', which is just $A^{+}=0$, i.e. the light-front gauge.

\subsection{Propagator for Transverse Photons}
\label{sec:transverse_photon_propagator}

Choosing the transverse gauge fields as the dynamical degrees of freedom, we get the photon propagator in the interpolating transverse gauge given by
\begin{equation}\label{eqn:transverse_photon_propagator}
  \langle 0 |T(A_{\muT}(y)A_{\nuT}(x))| 0 \rangle = i \int \dfrac{d^{4}q}{(2\pi)^{4}}e^{-iq(y-x)}\dfrac{\mathscr{T}_{\muT\nuT}}{q^{2}+i\epsilon},
\end{equation}
where $\mathscr{T}_{\muT\nuT}\equiv \sum_{\lambda=\pm}{\epsilon_{\muT}^{*}(\lambda)\epsilon_{\nuT}(\lambda)}$
and $\epsilon_{\muT}(\lambda)=\epsilon_{\hat{\mu}}(q,\lambda)$ taking the photon momentum $P=q$.
Here, we use the obvious familiar notation $q^2= q_{\muT}q^{\muT}$.
Although we can compute $\mathscr{T}_{\muT\nuT}$ directly using Eqs.~(\ref{eqn:polarization_vector_+_in_P_any_interpolation}) and (\ref{eqn:polarization_vector_-_in_P_any_interpolation}) as shown in Appendix~\ref{sec:photon_propagator_derived_directly_from_polarization_vectors_in_the_interpolating_form},
we demonstrate here the method of vierbein as a cross-check to our result.

To construct a vierbein, we just need a temporal basis four-vector and a longitudinal basis four-vector which we denote as $\uniT{n}_{\muT}$ and $\uniT{q}_{\muT}$, respectively, since the transverse basis four-vectors are already given by $\epsilon_{\muT}{(\pm)}$.
The temporal basis four-vector can be taken as a unit timelike four-vector given by $\uniT{n}_{\muT}=\frac{1}{\sqrt{\Cc}}n_{\muT}=\frac{1}{\sqrt{\Cc}}(1,0,0,0)$ whose dual vector is $\uniT{n}^{\muT}=(\sqrt{\Cc},0,0, \Ss/\sqrt{\Cc})$. The $\frac{1}{\sqrt{\Cc}}$ in $n_{\muT}$ is a normalization factor to get $\uniT{n}_{\muT}\uniT{n}^{\muT}=1$.
The longitudinal basis four-vector $\uniT{q}_{\muT}$ can also be rather easily found from the gauge condition given by Eq.~(\ref{eqn:epsilon_gauge_condition}) which can be written as $\epsilon_{\muT}(\lambda)\uniT{q}^{\muT}=0$ with
$\uniT{q}^{\muT}=N(0, -q_{1}\Cc,-q_{2}\Cc,-q_{\mT})$. The normalization factor $N$ is determined by $\uniT{q}_{\muT}\uniT{q}^{\muT}=\uniT{q}^{\muT}g_{\muT\nuT}\uniT{q}^{\nuT}=-1$ with $g_{\muT\nuT}$ given by Eq.~(\ref{eqn:g_munu_interpolation}), so that we have
\begin{align}\label{eqn:hat_q_logitudinal_unit_four_vector}
  \uniT{q}^{\muT}=\dfrac{1}{\sqrt{\Cc(\mathbf{q}_{\perp}^{2}\Cc+q_{\mT}^{2}})}(0,-q_{1}\Cc,-q_{2}\Cc,-q_{\mT})
\end{align}
and
\begin{align}\label{eqn:hat_q_covector_logitudinal_unit_four_vector}
  \uniT{q}_{\muT}=\dfrac{1}{\sqrt{\Cc(\mathbf{q}_{\perp}^{2}\Cc+q_{\mT}^{2}})}\left( -\Ss q_{\mT}, q_{1}\Cc, q_{2}\Cc, q_{\mT}\Cc \right).
\end{align}
These four basis four-vectors are of course mutually orthogonal: {\it i.e.} $\uniT{q}_{\muT}\uniT{n}^{\muT}=0$, $\epsilon_{\muT}(\lambda)\uniT{n}^{\muT}=0$ and $\epsilon_{\muT}(\lambda)\uniT{q}^{\muT}=0$, where $\lambda=\pm$.

Since $\epsilon_{\muT}(\pm),\uniT{n}_{\muT},\uniT{q}_{\muT}$ form a vierbein, we may start from
\begin{align}\label{eqn:tetrad_construction_of_gmunu}
  g_{\muT\nuT}=\uniT{n}_{\muT}\uniT{n}_{\nuT}-\sum_{\lambda=\pm}\epsilon_{\muT}^{*}(\lambda)\epsilon_{\nuT}(\lambda)-\uniT{q}_{\muT}\uniT{q}_{\nuT},
\end{align}
and obtain
\begin{align}\label{eqn:transverse_photon_propagator_numerator_in_radiation_gauge}
    \mathscr{T}_{\muT\nuT}\equiv &\sum_{\lambda=\pm}{\epsilon_{\muT}^{*}(\lambda)\epsilon_{\nuT}(\lambda)}\nonumber\\
    =&-g_{\muT\nuT}+\uniT{n}_{\muT}\uniT{n}_{\nuT}-\uniT{q}_{\muT}\uniT{q}_{\nuT}\nonumber\\
    =&-g_{\muT\nuT}+\dfrac{(q\cdot n)(q_{\muT}n_{\nuT}+q_{\nuT}n_{\muT})}{\mathbf{q}_{\perp}^{2}\Cc+q_{\mT}^{2}} \nonumber\\
    &- \dfrac{\Cc q_{\muT}q_{\nuT}}{\mathbf{q}_{\perp}^{2}\Cc+q_{\mT}^{2}} - \dfrac{q^{2}n_{\muT}n_{\nuT}}{\mathbf{q}_{\perp}^{2}\Cc+q_{\mT}^{2}},
\end{align}
where we used $\uniT{n}_{\muT}=\frac{1}{\sqrt{\Cc}}n_{\muT}$ and rewrote $\uniT{q}_{\muT}$ in terms of $q_{\muT}$ and $n_{\muT}$ as
\begin{align}\label{eqn:hat_q_in_terms_of_n_and_q}
  \uniT{q}_{\muT}=\dfrac{\Cc q_{\muT}-(q\cdot n)n_{\muT}}{\sqrt{\Cc(\mathbf{q}_{\perp}^{2}\Cc+q_{\mT}^{2}})}
\end{align}
in the last step to remove any artifact of divergence in $\uniT{q}_{\muT}$ and $\uniT{n}_{\muT}$ in the light-front limit.
The photon propagator constructed out of $n_{\muT}$ and $q_{\muT}$ interpolates smoothly and correctly to the one corresponding to the light-front gauge.

In the instant form limit, $\Cc\rightarrow1$, $\mathbf{q}_{\perp}^{2}\Cc+q_{\mT}^{2}\rightarrow \mathbf{q}^{2}=(q\cdot n)^{2}-q^{2}$, and this reduces to the well-known photon propagator in Coulomb gauge ($\boldsymbol\nabla\cdot\mathbf{A}=0$) \cite{Ryder}:
\begin{align}\label{eqn:transverse_photon_propagator_numerator_in_coulomb_gauge}
  \mathscr{T}_{\mu\nu}=
  -\eta_{\mu\nu}+\dfrac{(q\cdot n)(q_{\mu}n_{\nu}
  + q_{\nu}n_{\mu})}{(q\cdot n)^{2}-q^{2}}
  \nonumber\\
  - \dfrac{q_{\mu}q_{\nu}}{(q\cdot n)^{2}-q^{2}}
  - \dfrac{q^{2}n_{\mu}n_{\nu}}{(q\cdot n)^{2}-q^{2}},
\end{align}
where $\eta_{\mu\nu}={\rm diag}(1,-1,-1,-1)$, and $\mu$ and $\nu$ run for $0,1,2,3$.

In the light-front limit, $\Cc\rightarrow0$, $q_{\mT}^{2}\rightarrow q^{+2}=(q\cdot n)^{2}$, and this becomes precisely the photon propagator under light-front gauge ($A^{+}=0$), and we have
\begin{align}\label{eqn:transverse_photon_propagator_numerator_in_LF_gauge}
  \mathscr{T}_{\mu\nu}=-g_{\mu\nu}+\dfrac{(q\cdot n)(q_{\mu}n_{\nu}+q_{\nu}n_{\mu})}{(q\cdot n)^{2}} - \dfrac{q^{2}n_{\mu}n_{\nu}}{(q\cdot n)^{2}},
\end{align}
where $g_{\mu\nu}$ is given by Eq.~(\ref{eqn:g_munu_interpolation}) with $\delta=\pi/4$, and $\mu$ and $\nu$ run for $+,1,2,-$.
From this derivation, we see that the appropriate photon propagator for the interpolating gauge given by Eqs.~(\ref{eqn:gauge_condition_for_any_interpolation_angle_a}) and (\ref{eqn:gauge_condition_for_any_interpolation_angle_b}) (or
(\ref{eqn:general_gauge_condition_position_space}))
has three terms, consistent with \cite{Leibbrandt1984, Srivastava2001, Suzuki2003, *Suzuki2004a, *Suzuki2004b, *Misra2005}.
As we will see in the next section, the last term in Eq.~(\ref{eqn:transverse_photon_propagator_numerator_in_LF_gauge}) is canceled by the instantaneous interaction.
Therefore, the two term gauge propagator \cite{Mustaki1991} can be used effectively without involving
the instantaneous interaction.

\subsection{Longitudinal Photons}
\label{sec:longitudinal_photons}

Having used the transverse gauge fields as the dynamical degrees of the freedom, we are left with the longitudinal degree of freedom which also deserves a physical interpretation. This leftover longitudinal degree of freedom is necessary to describe the virtual photon. First, to find the longitudinal polarization ($\lambda=0$) of the virtual photon, we need to generalize
Eq.~(\ref{eqn:polarization_vector_0_in_P_any_interpolation}) replacing $M^2$ by $P^{\muT}P_{\muT}=q^2$
which can be either positive (timelike) or negative (spacelike).  From the direct computation shown in Appendix~\ref{sec:photon_propagator_derived_directly_from_polarization_vectors_in_the_interpolating_form} with this replacement, we find
\begin{align}
  \mathscr{L}_{\muT\nuT}
  \equiv&\epsilon_{\muT}^{*}(0)\epsilon_{\nuT}(0)\nonumber\\
  =&-\dfrac{(q\cdot n)(q_{\muT}n_{\nuT}+q_{\nuT}n_{\muT})}{\mathbf{q}_{\perp}^{2}\Cc+q_{\mT}^{2}} \nonumber\\
   &+ \dfrac{(q^{\pT})^{2} q_{\muT}q_{\nuT}}{(q)^{2}(\mathbf{q}_{\perp}^{2}\Cc+q_{\mT}^{2})} + \dfrac{q^{2} n_{\muT}n_{\nuT}}{\mathbf{q}_{\perp}^{2}\Cc+q_{\mT}^{2}}. \label{eqn:Longitudinal_photon_propagator_numerator}
\end{align}
Together with $\mathscr{T}_{\muT\nuT}$ given by Eq.~(\ref{eqn:transverse_photon_propagator_numerator_in_radiation_gauge}),
it verifies the well-known completeness relation \cite{Ryder}
\begin{align}
  \mathscr{T}_{\muT\nuT}+\mathscr{L}_{\muT\nuT}=-g_{\muT\nuT}+\dfrac{q_{\muT}q_{\nuT}}{q^{2}}. \label{eqn:Transverse_+_Longitudinal_photon_propagator_numerator}
\end{align}
We note that the last term in Eq.~(\ref{eqn:Longitudinal_photon_propagator_numerator}) without the dependence on $q_{\muT}$ or $q_{\nuT}$ provides the instantaneous contribution.
As we show in the next section, the instantaneous interaction appearing in the transverse gauge corresponds to the contribution from
the longitudinal polarization.

\section{Time-ordered Photon Exchange}
\label{sec:Time_Ordered_Photon_Exchange}

As Kogut and Soper~\cite{Kogut1970c} regarded the theory of quantum electrodynamics as being defined by the usual perturbation expansion of the S-matrix in Feynman diagrams, we rewrite the sQED theory here by systematically decomposing each covariant Feynman diagram into a sum of interpolating $x^{\pT}$-ordered diagrams. Since we consider the Feynman expansion as a formal expansion as Kogut and Soper did, we also shall not be concerned in this paper with the convergence of the perturbation series, or convergence and regularization of the integrals.

For clarity, we split this section into three subsections.
In the first subsection, we decompose the covariant photon propagator in an arbitrary interpolation angle as a sum of $x^{\pT}$-ordered terms.
In the second part, we use the obtained propagator to derive the $x^{\pT}$-ordered diagrams and their amplitudes for the lowest order photon exchange process. In the final subsection, we verify the invariance of the corresponding total amplitude and discuss
about the instant form and light-front limits of the $x^{\pT}$-ordered photon propagators.

\subsection{Photon Propagator Decomposition}
\label{sub:Photon_Propagator_Decomposition}

The completeness relation given by Eq.~(\ref{eqn:Transverse_+_Longitudinal_photon_propagator_numerator}) corresponds to the
numerator of the covariant photon propagator in Landau gauge. We start from the covariant photon propagator in position space
\begin{align}\label{eqn:photon_propagator_feynman}
  D_{F}(x)_{\muT\nuT}=\int \dfrac{d^{4}q}{(2\pi)^{4}}\dfrac{-i g_{\muT\nuT} + i \dfrac{q_{\muT}q_{\nuT}}{q^{2}}}{q^{\muT}q_{\muT}+i\epsilon} e^{-i q_{\muT}x^{\muT}}.
\end{align}
Due to the current conservation, however, the term involving $q_{\muT}$ doesn't contribute to any physical process and our starting point is equivalent to the Feynman photon propagator that Kogut and Soper used for their starting point\cite{Kogut1970c}. For the same reason,
the first and second terms that involve $q_{\muT}$ can be dropped in Eq.~(\ref{eqn:Longitudinal_photon_propagator_numerator}) and
the covariant photon propagator can be written as
\begin{multline}
  D_{F}(x)_{\muT\nuT}=i\int \dfrac{d^{4}q}{(2\pi)^{4}}\dfrac{\mathscr{T}_{\muT\nuT}}{q^{\muT}q_{\muT}+i\epsilon} e^{-i q_{\muT}x^{\muT}}\\
  +\int \dfrac{d^{4}q}{(2\pi)^{4}} \dfrac{n_{\muT}n_{\nuT}}{\mathbf{q}_{\perp}^{2}\Cc+q_{\mT}^{2}} e^{-i q_{\muT}x^{\muT}}\\
  =\int \dfrac{d^{2}\mathbf{q}_{\perp} d q_{\mT} d q_{\pT}} {(2\pi)^{4}}\exp[-i(q_{\pT}x^{\pT}+q_{\mT}x^{\mT}+\mathbf{q}_{\perp}\mathbf{x}^{\perp})]\\
  \left[ \dfrac{i\mathscr{T}_{\muT\nuT}}{\Cc q_{\pT}^{2}+2 \Ss q_{\mT}q_{\pT}-\Cc q_{\mT}^{2}-\mathbf{q}_{\perp}^{2}+i\epsilon}
  + \dfrac{i n_{\muT}n_{\nuT}}{\mathbf{q}_{\perp}^{2}\Cc+q_{\mT}^{2}}\right],
  \label{eqn:photon_propagator_gauge_decomposition_2}
\end{multline}
where we used Eq.~(\ref{eqn:inner_product_of_four_vectors_interpolation_angle}) for $q^{\muT}q_{\muT}$ in the denominator.

To get the $x^{\pT}$-ordered contributions, we now evaluate the $q_{\pT}$ integral in Eq.~(\ref{eqn:photon_propagator_gauge_decomposition_2}). We note here that the case of $\Cc=0$ should be distinguished from the case of
$\Cc\neq0$ because the pole structures in the $\mathscr{T}_{\muT\nuT}$ term of Eq.~(\ref{eqn:photon_propagator_gauge_decomposition_2}) are different between the two cases, i.e. a single pole for $\Cc=0$ vs. two poles for $\Cc\neq0$.

For $\Cc\neq0$, the $\mathscr{T}_{\muT\nuT}$ term of Eq.~(\ref{eqn:photon_propagator_gauge_decomposition_2}) has two poles at
\begin{subequations}
		\begin{align}
				A-i\epsilon'=\left(-\Ss q_{\mT}+\sqrt{q_{\mT}^{2}+\Cc \mathbf{q}_{\perp}^{2}}\right)/\Cc-i\epsilon',\label{eqn:A_Pole}\\
				B+i\epsilon'=\left(-\Ss q_{\mT}-\sqrt{q_{\mT}^{2}+\Cc \mathbf{q}_{\perp}^{2}}\right)/\Cc+i\epsilon',\label{eqn:B_Pole}
		\end{align}
\end{subequations}
where $\epsilon'>0$. In calculating the contour integration, we close the contour in the lower (upper) half plane for $x^{\pT}>0$ ($x^{\pT}<0$).
This produces a term proportional to the step function $\Theta(x^{\pT})$ and the other term proportional to $\Theta(-x^{\pT})$.
We then make changes of the variables $\mathbf{q}_{\perp}\rightarrow - \mathbf{q}_{\perp}$ and $q_{\mT}\rightarrow -q_{\mT}$ which lead to $B\rightarrow-A$ for the $\Theta(-x^{\pT})$ term and simplify the result expressing $q_{\pT}$ in terms of $A$.
The last term in Eq.~(\ref{eqn:photon_propagator_gauge_decomposition_2}) immediately gives a delta function of $x^{\pT}$ after the  $q_{\pT}$ integration and provides the instantaneous contribution. We note that the instantaneous contribution stems from the longitudinal polarization of the virtual photon.  Putting all terms together, we obtain the following result for the case of $\Cc\neq0$:
\begin{multline}\label{eqn:photon_time_ordered_propagator}
				D_{F}(x)_{\muT\nuT}= \int \dfrac{d^{2}\mathbf{q}_{\perp} }{(2\pi)^{3}}\int_{-\infty}^{\infty}d q_{\mT} \dfrac{\mathscr{T}_{\muT\nuT}}{2 \sqrt{q_{\mT}^{2}+\Cc \mathbf{q}_{\perp}^{2}}} \\
				\left[ \Theta(x^{\pT})e^{-i q_{\muT}x^{\muT}}+\Theta(-x^{\pT})e^{i q_{\muT}x^{\muT}} \right]\\
				+i\delta(x^{\pT})\int \dfrac{d^{2}\mathbf{q}_{\perp} }{(2\pi)^{3}}\int_{-\infty}^{\infty}d q_{\mT} \dfrac{n_{\muT}n_{\nuT}}{\mathbf{q}_{\perp}^{2}\Cc+q_{\mT}^{2}} e^{-i(q_{\mT}x^{\mT}+\mathbf{q}_{\perp}\mathbf{x}^{\perp})},
\end{multline}
where $q_{\pT}$ in the exponent of the first two terms should be taken as $A$ given by Eq.~(\ref{eqn:A_Pole}).

Now let's look at the light-front case.
Because $\Cc=0$, there's only one pole at $\mathbf{q}_{\perp}^{2}/2q_{\mT}-i\epsilon/2q_{\mT}=\mathbf{q}_{\perp}^{2}/2q^{+}-i\epsilon/2q^{+}$.
It depends on the sign of $q^{+}$ whether this pole is in the upper half plane or the lower half plane.
As the integration over $q^{-}$ needs to be done in $q^{+}>0$ and $q^{+}<0$ regions separately,
each region will get a step function after closing the contour in the plane where the arc contribution is absent.
We again make changes of the variables $q^{+}\rightarrow -q^{+}$ and $\mathbf{q}_{\perp}\rightarrow - \mathbf{q}_{\perp}$ in the term proportional to $\Theta(-x^{+})$.
We then obtain the result at the light-front~\cite{Kogut1970c}:
\begin{multline}
  D_{F}(x)_{\mu\nu}\\
	= \int\frac{d^{2}\mathbf{q}_{\perp}}{(2\pi)^{3}}
	\int_{0}^{\infty}\frac{dq^{+}}{2q^{+}}\mathscr{T}_{\mu\nu}
	\left[\Theta(x^{+})e^{-iq_{\mu}x^{\mu}}+\Theta(-x^{+})e^{iq_{\mu}x^{\mu}}\right]\\
		+i\delta(x^{+})\int \dfrac{d^{2}\mathbf{q}_{\perp} }{(2\pi)^{3}}\int_{-\infty}^{\infty}d q^{+} \dfrac{n_{\mu}n_{\nu}}{(q^{+})^{2}} e^{-i(q^{+}x^{-}
		-\mathbf{q}^{\perp}\mathbf{x}^{\perp})},
  \label{eqn:LF_propogator_decomposition}
\end{multline}
where the indices $\mu$ and $\nu$ run for $+,1,2,-$ and $q^{-}$ in the exponent of the first two terms should be taken as $\mathbf{q}_{\perp}^{2}/2q^{+}$.
More details of the derivation for this equation can be found in Appendix~\ref{sec:photon_propagator_decomposition_on_the_light_front}.

We note that the result for $\Cc\neq0$ given by Eq.~(\ref{eqn:photon_time_ordered_propagator}) doesn't coincide with
the result for $\Cc=0$ given by Eq.~(\ref{eqn:LF_propogator_decomposition}) as we take the limit $\Cc \rightarrow 0$,
because the integration range ($-\infty$ $\infty$) in $q_{\mT}$ is different from the integration range $(0,\infty)$ in $q^+$
due to the difference in the pole structure between the two cases, $\Cc\neq0$ and $\Cc=0$, as we mentioned earlier.
Nevertheless, it can be written in a unified form by introducing an interpolating step function $\hat\Theta(q_{\mT})$ given by
\begin{eqnarray}
\hat\Theta(q_{\mT}) &=& \Theta(q_{\mT}) + (1-\delta_{\Cc 0})\Theta(-q_{\mT}) \nonumber \\
&=&
\begin{cases}
1 \hspace{1cm} &(\Cc\neq0)\\
\Theta(q^+) &(\Cc=0)
\end{cases}
\label{eqn:Interpolating_Theta_Function}
\end{eqnarray}
and realizing that $A$ given by Eq.~(\ref{eqn:A_Pole}) coincides with $q^{-}=\mathbf{q}_{\perp}^{2}/2q^{+}$ in the limit $\Cc \rightarrow 0$, i.e.
\begin{multline}\label{eqn:unified_photon_time_ordered_propagator}
				D_{F}(x)_{\muT\nuT}= \int \dfrac{d^{2}\mathbf{q}_{\perp} }{(2\pi)^{3}}\int_{-\infty}^{\infty}d q_{\mT} \hat\Theta(q_{\mT}) \dfrac{\mathscr{T}_{\muT\nuT}}{2 \sqrt{q_{\mT}^{2}+\Cc \mathbf{q}_{\perp}^{2}}} \\
				\left[ \Theta(x^{\pT})e^{-i q_{\muT}x^{\muT}}+\Theta(-x^{\pT})e^{i q_{\muT}x^{\muT}} \right]\\
				+i\delta(x^{\pT})\int \dfrac{d^{2}\mathbf{q}_{\perp} }{(2\pi)^{3}}\int_{-\infty}^{\infty}d q_{\mT} \dfrac{n_{\muT}n_{\nuT}}{\mathbf{q}_{\perp}^{2}\Cc+q_{\mT}^{2}} e^{-i(q_{\mT}x^{\mT}+\mathbf{q}_{\perp}\mathbf{x}^{\perp})},
\end{multline}
where again $q_{\pT}$ in the exponent of the first two terms should be taken as $A$ given by Eq.~(\ref{eqn:A_Pole}).
From now on, we will use Eq.~(\ref{eqn:unified_photon_time_ordered_propagator}) for all interpolation angles including $\Cc=0$.
\subsection{Time-ordered Diagrams}
\label{sub:diagrams}
The Lagrangian for sQED can be written as
\begin{align}
  \mathscr{L}=&D_{\mu}\phi D^{\mu}\phi^{*}-m^{2}\phi^{*}\phi-\dfrac{1}{4}F^{\mu\nu}F_{\mu\nu} \nonumber\\
  =&\partial_{\mu}\phi\partial^{\mu}\phi^{*}-m^{2}\phi^{*}\phi-\dfrac{1}{4}F^{\mu\nu}F_{\mu\nu}\nonumber\\
  &-eJ^{\mu}A_{\mu}+e^{2}A_{\mu}A^{\mu}\phi^{*}\phi,\label{eqn:sQED_Lagrangian}
\end{align}
where
\begin{align}
  D_{\mu}=&\partial_{\mu}+ieA_{\mu}\label{eqn:D_mu},\\
  J^{\mu}=&-i(\phi\partial^{\mu}\phi^{*}-\phi^{*}\partial^{\mu}\phi).\label{eqn:J_mu}
\end{align}
To examine the contribution of each term in Eq.~(\ref{eqn:unified_photon_time_ordered_propagator}),
we compute the lowest order tree level scattering amplitude starting from the usual Feynman amplitude in coordinate space.
For the lowest tree level scattering diagram, $e^{2}A_{\mu}A^{\mu}\phi^{*}\phi$ doesn't contribute, and the amplitude can be written as
\begin{align}
  i\mathcal{M} = (-i e)^{2}\int d^{4}x d^{4}y &[J^{\muT}(y)D_{F}(y-x)_{\muT\nuT}J^{\nuT}(x)].  \label{eqn:scattering_feynman_amplitude}
\end{align}
The scalar wave functions used here are the plane waves
\begin{align}\label{eqn:fermion_external_WF}
  \phi(x)=e^{-ip_{\muT}x^{\muT}}.
\end{align}

\begin{figure*}[t!]
  \subfloat{\includegraphics[width=0.5\columnwidth]{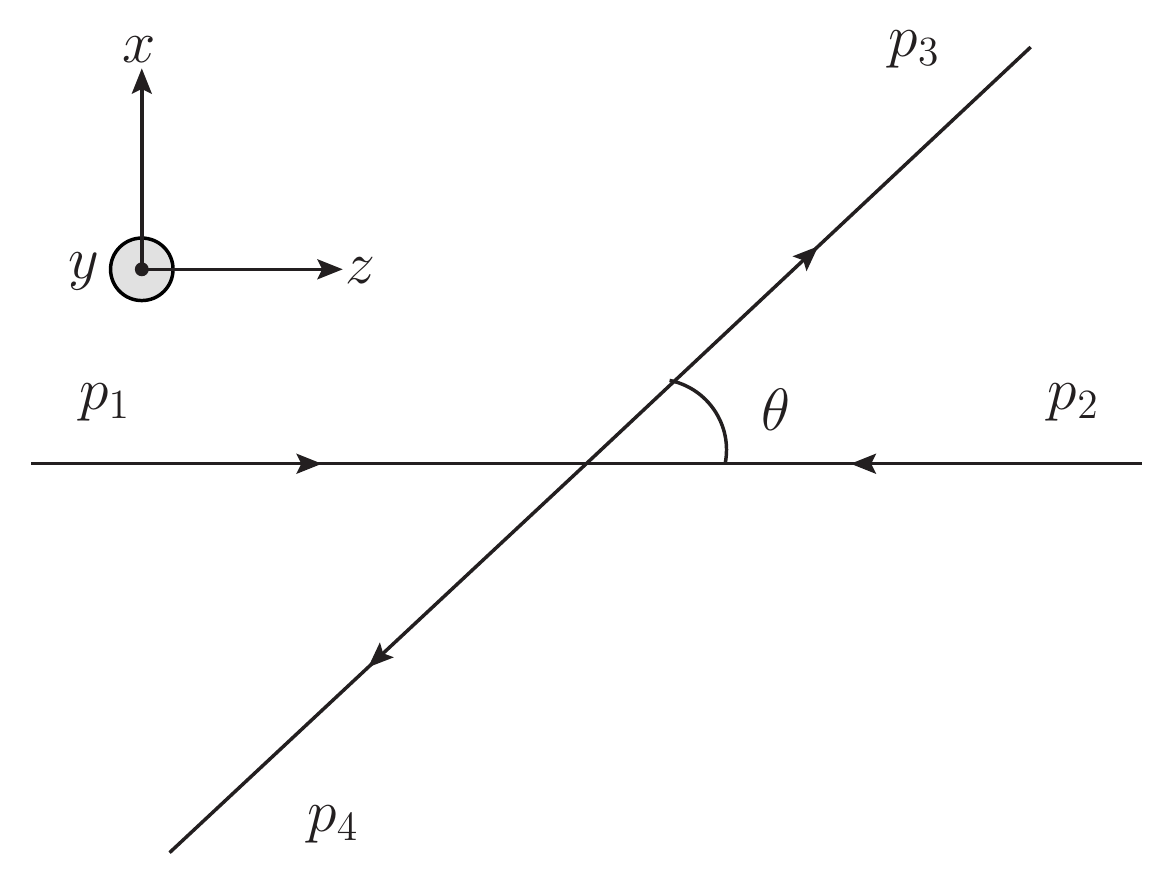}
  }
  \subfloat{\includegraphics[width=0.7\textwidth]{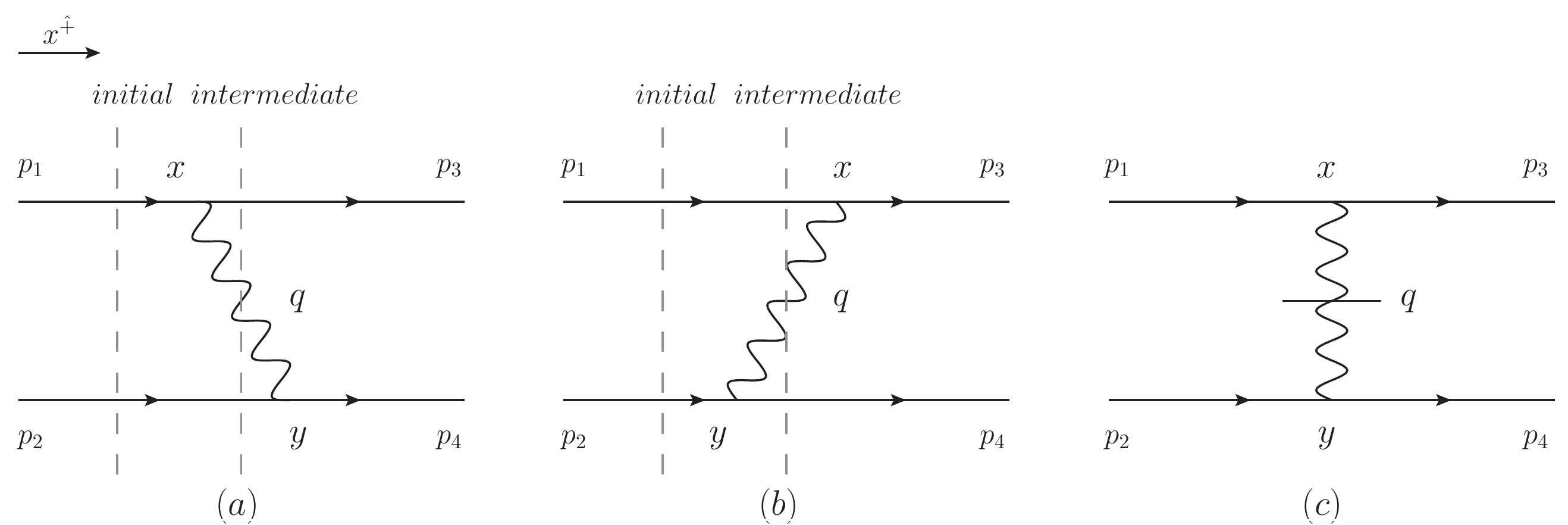}}
  \caption{The scattering of two particles in their center of mass frame (left most figure) and the three corresponding $x^{\pT}$-ordered diagrams.}
  \label{fig:time_ordered_photon_propagator}
\end{figure*}

For a specific scattering process shown in FIG.~\ref{fig:time_ordered_photon_propagator}, the currents from
$p_{1}$ to $p_{3}$ and from $p_{2}$ to $p_{4}$ are respectively given by
\begin{align}
  J^{\nuT}=&-i(\phi_{1}\partial^{\nuT}\phi_{3}^{*}-\phi_{3}^{*}\partial^{\nuT}\phi_{1})
  =(p_{1}^{\nuT}+p_{3}^{\nuT})e^{i(p_{3}-p_{1})x},\label{eqn:J_13}\\
  J^{\muT}=&-i(\phi_{2}\partial^{\muT}\phi_{4}^{*}-\phi_{4}^{*}\partial^{\muT}\phi_{2})
  =(p_{2}^{\muT}+p_{4}^{\muT})e^{i(p_{4}-p_{2})y}.\label{eqn:J_24}
\end{align}
With the change of variables
\begin{alignat}{2}\label{eqn:xy_to_xT}
  &x\rightarrow x, &\quad&y\rightarrow T=y-x,
\end{alignat}
Eq.~(\ref{eqn:scattering_feynman_amplitude}) becomes
\begin{align}
  i\mathcal{M}=(-ie)^{2}\int &d^{4}x d^{4}T e^{i(p_{4}-p_{2})T}e^{i(p_{4}+p_{3}-p_{2}-p_{1})x}\nonumber\\
  &(p_{2}^{\muT}+p_{4}^{\muT})D_{\muT\nuT}(T)(p_{1}^{\nuT}+p_{3}^{\nuT}).
  \label{eqn:scattering_amp_variable}
\end{align}
The $x$ integration resulting in $(2\pi)^{4}\delta^{4}(p_{4}+p_{3}-p_{2}-p_{1})$ provides the total energy and momentum conservation.
For the $T$ integration, we use $D_{\muT\nuT}$ given by Eq.~(\ref{eqn:unified_photon_time_ordered_propagator}) as well as the following relations:
\begin{align}\label{eqn:M_evaluation_with}
  \int_{-\infty}^{\infty}dT^{\pT}\Theta(T^{\pT})e^{iP_{\pT}T^{\pT}}&=\dfrac{i}{P_{\pT}},\\
  \int_{-\infty}^{\infty}dT^{\pT}\Theta(-T^{\pT})e^{iP_{\pT}T^{\pT}}&=-\dfrac{i}{P_{\pT}},\\
  \int_{-\infty}^{\infty}dT^{\pT}e^{iP_{\pT}T^{\pT}}&=2\pi\delta(P_{\pT}).
\end{align}
After the $x$ and $T$ integration, we get
\begin{align}\label{eqn:M_after_x_T_integration}
  i\mathcal{M}=&(-ie)^{2} (p_{4}^{\muT}+p_{2}^{\muT}) \Pi_{\muT\nuT} (p_{3}^{\nuT}+p_{1}^{\nuT})  \nonumber\\
     &(2\pi)^{4}\delta^{4}(p_{4}+p_{3}-p_{2}-p_{1}),
\end{align}
where
\begin{widetext}
\begin{align}\label{eqn:M_propagator_after_x_T_integration}
  \Pi_{\muT\nuT}=\int \dfrac{d^{2}\mathbf{q}_{\perp} d q_{\mT} {\hat\Theta(q_{\mT})}}{2 \sqrt{q_{\mT}^{2}+\Cc \mathbf{q}_{\perp}^{2}}}
&\left[\dfrac{i\mathscr{T}_{\muT\nuT}}{p_{4\pT}-p_{2\pT}-A}
  \delta(p_{4\mT}-p_{2\mT}-q_{\mT})\delta^{2}(\mathbf{p}_{4\perp}-\mathbf{p}_{2\perp}-\mathbf{q}_{\perp})\right.\nonumber\\
&\left. -\dfrac{i\mathscr{T}_{\muT\nuT}}{p_{4\pT}-p_{2\pT}+A}
  \delta(p_{4\mT}-p_{2\mT}+q_{\mT})\delta^{2}(\mathbf{p}_{4\perp}-\mathbf{p}_{2\perp}+\mathbf{q}_{\perp})\right]\nonumber\\
    &+\int d^{2}\mathbf{q}_{\perp} d q_{\mT} \dfrac{i n_{\muT}n_{\nuT}}{\mathbf{q}_{\perp}^{2}\Cc+q_{\mT}^{2}}
  \delta(p_{4\mT}-p_{2\mT}-q_{\mT})\delta^{2}(\mathbf{p}_{4\perp}-\mathbf{p}_{2\perp}-\mathbf{q}_{\perp}).
\end{align}
\end{widetext}

The three terms in $\Pi_{\muT\nuT}$ corresponds to three different ``time'' orderings $y^{\pT}>x^{\pT}$, $y^{\pT}<x^{\pT}$ and $y^{\pT}=x^{\pT}$ respectively~\footnote{In this paper, the ``time'' means the generalized interpolation time $x^{\pT}$ unless specified otherwise.}.
The associated delta functions provide the momentum conservation at each vertex as well as the conservation of total energy and momentum between the initial and final particles. In Fig.~\ref{fig:time_ordered_photon_propagator}, the three ``time" ordered diagrams are depicted with
the momentum conservation at each vertex.

The corresponding photon propagators for $T^{\pT}>0$ (or $y^{\pT}>x^{\pT}$) and $T^{\pT}<0$ (or $y^{\pT}<x^{\pT}$) are respectively given by
\begin{align}
  \Pi_{\muT\nuT}^{(a)}
    =&\dfrac{1}{2 Q^{\pT(a)}}
\dfrac{i\mathscr{T}_{\muT\nuT}{\hat\Theta(p_{4\mT}-p_{2\mT})}}{p_{4\pT}-p_{2\pT}-Q_{\pT}^{(a)}}\nonumber\\
    =&\dfrac{1}{2 Q^{\pT(a)}}
\dfrac{i\mathscr{T}_{\muT\nuT}{\hat\Theta(p_{1\mT}-p_{3\mT})}}{p_{1\pT}-p_{3\pT}-Q_{\pT}^{(a)}}, \label{eqn:Sigma_propagator_a}
\end{align}
and
\begin{align}
  \Pi_{\muT\nuT}^{(b)}
    =-&\dfrac{1}{2 Q^{\pT(b)}}
\dfrac{i\mathscr{T}_{\muT\nuT}{\hat\Theta(p_{2\mT}-p_{4\mT})}}{p_{4\pT}-p_{2\pT}+Q_{\pT}^{(b)}}\nonumber\\
    =&\dfrac{1}{2 Q^{\pT(b)}}
\dfrac{i\mathscr{T}_{\muT\nuT}{\hat\Theta(p_{2\mT}-p_{4\mT})}}{p_{2\pT}-p_{4\pT}-Q_{\pT}^{(b)}}\nonumber\\
=&\dfrac{1}{2 Q^{\pT(b)}}
\dfrac{i\mathscr{T}_{\muT\nuT}{\hat\Theta(p_{3\mT}-p_{1\mT})}}{p_{3\pT}-p_{1\pT}-Q_{\pT}^{(b)}}, \label{eqn:Sigma_propagator_b}
\end{align}
where
\begin{align}
  Q^{\pT(i)}&=\sqrt{[q^{(i)}_{\mT}]^{2}+\Cc [\mathbf{q}^{(i)}_{\perp}]^{2}}, \quad &(i=a,b) \label{eqn:Q_sup_ph_q}\\
  Q_{\pT}^{(i)}&=\dfrac{-\Ss q^{(i)}_{\mT}+\sqrt{[q^{(i)}_{\mT}]^{2}+\Cc [\mathbf{q}^{(i)}_{\perp}]^{2}}}{\Cc}, \quad &(i=a,b) \label{eqn:Q_sub_ph_q}
\end{align}
and
\begin{align}
  q_{\mT}^{(a)}=-q_{\mT}^{(b)}=p_{1\mT}-p_{3\mT},\label{eqn:q_mh_a_b}\\
  \mathbf{q}_{\perp}^{(a)}=-\mathbf{q}_{\perp}^{(b)}=p_{1\perp}-p_{3\perp}.\label{eqn:q_pp_a_b}
\end{align}
The total energy-momentum conservation in Eq.~(\ref{eqn:M_after_x_T_integration}) as well as the momentum conservation at each vertex were used here. $Q_{\pT}^{(i)}$ and $Q^{\pT(i)}$ satisfy the on-mass-shell condition of the propagating photon with momentum ($q_{\mT}^{(i)}, \mathbf{q}_{\perp}^{(i)}$).
To see this, one can use Eq.~(\ref{eqn:on_shell_4_momentum_useful_relation}) for a massless photon to derive the $Q^{\pT}$ in terms of $q_{\mT}$ and $\mathbf{q}_{\perp}$ noting that $Q^{\pT}$ is positive definite for an on-mass-shell particle due to Eq.~(\ref{eqn:P_interpolation_1}).
The formula for $Q_{\pT}$ can also be obtained in terms of $Q_{\mT}$ and $Q^{\pT}$ by using Eq.~(\ref{eqn:relation_between_covariant_and_contravariant_components_with_any_interpolation}).
Eqs.~(\ref{eqn:Sigma_propagator_a}) and (\ref{eqn:Sigma_propagator_b}) can then be written in a unified form\footnote{Note that the subscripts `$_{\pT}$' and `$_{\mT}$' denote the energy and the longitudinal momentum for a given interpolating momentum, respectively.}:
\begin{align}\label{eqn:unified_time_ordered_photon_propagator}
  \Pi^{(i)}_{\muT\nuT}=\dfrac{1}{2Q^{\pT (i)}}
\dfrac{i\mathscr{T}_{\muT\nuT}{\hat\Theta(q_{\mT}^{(i)})}}{P_{ini\pT}-P_{inter\pT}}, \quad (i=a,b)
\end{align}
where $P_{ini\pT}$ [$P_{inter\pT}$] is the sum of the energy of the initial [intermediate] particles.
All the initial and intermediate particles are now on their mass shells.
This agrees with the familiar time-ordered perturbation theory in the IFD, although we now see it as a generalization to any interpolation angle.
We note here that the propagating photon as the dynamical degree of freedom has only the transverse polarization.
This is consistent with the interpretation that the intermediate particles are now ``on-mass-shell'' or ``physical''.

Besides the two propagating terms, there is a third term in Eq.~(\ref{eqn:M_propagator_after_x_T_integration}) that  represents the instantaneous contribution. For the sake of consistency in our notation, we denote this part of $\Pi_{\muT\nuT}$ as
\begin{align}
  \Pi_{\muT\nuT}^{(c)}=\dfrac{i n_{\muT}n_{\nuT}}{\mathbf{q}_{\perp}^{2}\Cc+q_{\mT}^{2}},\label{eqn:Sigma_propagator_c}
\end{align}
where $q_{\mT}$ and $\mathbf{q}_{\perp}$ are given by Eqs.~(\ref{eqn:q_mh_a_b}) and (\ref{eqn:q_pp_a_b}), respectively.
As noted earlier, the instantaneous contribution stems from the longitudinal polarization of the virtual photon.

We thus have three time-ordered diagrams as shown in FIG.~\ref{fig:time_ordered_photon_propagator}, the first two of which represent time-ordered exchanges of propagating photon and the third of which represents the instantaneous interaction.
The invariant amplitude is then the sum of all three time-ordered amplitudes:
\begin{align}
  i\mathcal{M}=&\sum_{j=a,b,c}{i\mathcal{M}^{(j)}}\nonumber\\
   =&(-ie)^{2}\sum_{j=a,b,c}{(p_{4}^{\muT}+p_{2}^{\muT})\Pi^{(j)}_{\muT\nuT}(p_{3}^{\nuT}+p_{1}^{\nuT})},\label{eqn:M_Ma_Mb_Mc}
\end{align}
where the total energy-momentum conservation factor $(2\pi)^{4}\delta^{4}(p_{4}+p_{3}-p_{2}-p_{1})$ is implied.


\subsection{LFD vs. the Limit to LFD}
\label{sub:discussion}
Since the time-ordered gauge propagators have the interpolating step function given by Eq.~(\ref{eqn:Interpolating_Theta_Function}),
we take a close look at the limiting cases of $\Cc \rightarrow 0$ and compare the results with the exact $\Cc=0$ (LFD) results in this subsection.
For convenience and simplicity of our discussion, we drop the $(i)$ superscript in Eq.~(\ref{eqn:Q_sup_ph_q}) and have
\begin{equation}
   Q^{\pT(a)}=Q^{\pT(b)}=\sqrt{(p_{1\mT}-p_{3\mT})^{2}+(\mathbf{p}_{1\perp}-\mathbf{p}_{3\perp})^{2}\Cc}\equiv Q^{\pT}.\label{eqn:Q_sup_ph_q_final}
\end{equation}
Because the terms proportional to $q_{\muT}$ or $q_{\nuT}$ in $\mathscr{T}_{\muT\nuT}$ can be discarded
due to the current conservation in the physical process, we may also replace $\mathscr{T}_{\muT\nuT}$ by
\begin{align}
 \mathscr{\tilde T}_{\muT\nuT}
  \equiv -g_{\muT\nuT} -\dfrac{q^{2}n_{\muT}n_{\nuT}}{\mathbf{q}_{\perp}^{2}\Cc+q_{\mT}^{2}}.
\label{eqn:Lambda_munu}
\end{align}
Then, Eq.~(\ref{eqn:Sigma_propagator_a}), Eq.~(\ref{eqn:Sigma_propagator_b}) and Eq.~(\ref{eqn:Sigma_propagator_c}) can be rewritten as
\begin{subequations}
  \label{eqn:Sigma_propagator_final}
  \begin{alignat}{3}
      &\Pi_{\muT\nuT}^{(a)}
      =
      &&\dfrac{1}{2 Q^{\pT}}
      \dfrac{i \mathscr{\tilde T}_{\muT\nuT}{\hat\Theta(p_{1\mT}-p_{3\mT})}}{p_{1\pT}-p_{3\pT}-Q_{\pT}^{(a)}}, \label{eqn:Sigma_propagator_a_final}\\
      &\Pi_{\muT\nuT}^{(b)}
      =
      - &&\dfrac{1}{2 Q^{\pT}}
	\dfrac{i \mathscr{\tilde T}_{\muT\nuT}{\hat\Theta(p_{3\mT}-p_{1\mT})}}{p_{1\pT}-p_{3\pT}+Q_{\pT}^{(b)}},  \label{eqn:Sigma_propagator_b_final}\\
      &\Pi_{\muT\nuT}^{(c)}
      =
      &&\dfrac{i n_{\muT}n_{\nuT}}{(Q^{\pT})^{2}}, \label{eqn:Sigma_propagator_c_final}
  \end{alignat}
\end{subequations}
where $Q_{\pT}^{(a)}$ and $Q_{\pT}^{(b)}$ are given by Eq.~(\ref{eqn:Q_sub_ph_q}).

For $\Cc\neq0$, assigning the four-momentum transfer $q$ as
\begin{subequations}
  \label{eqn:transfered_four_momentum_q}
  \begin{align}
    q_{\mT}&\equiv  p_{1\mT}-p_{3\mT}= q_{\mT}^{(a)}=-q_{\mT}^{(b)},\\
    \mathbf{q}_{\perp}&\equiv  p_{1\perp}-p_{3\perp}=\mathbf{q}_{\perp}^{(a)}=-\mathbf{q}_{\perp}^{(b)},\\
    q_{\pT}&\equiv p_{1\pT}-p_{3\pT}=-(p_{2\pT}-p_{4\pT}),
  \end{align}
\end{subequations}
we have
\begin{align}
   Q_{\pT}^{(a)}&=\dfrac{-\Ss q_{\mT}+\sqrt{q_{\mT}^{2}+\Cc \mathbf{q}_{\perp}^{2}}}{\Cc}=\dfrac{-\Ss q_{\mT}+Q^{\pT}}{\Cc}, \label{eqn:Q_sub_ph_q_final(a)}\\
   Q_{\pT}^{(b)}&=\dfrac{\Ss q_{\mT}+\sqrt{q_{\mT}^{2}+\Cc \mathbf{q}_{\perp}^{2}}}{\Cc}=\dfrac{\Ss q_{\mT}+Q^{\pT}}{\Cc}.
   \label{eqn:Q_sub_ph_q_final(b)}
\end{align}
Since both ${\hat\Theta(q_{\mT})}$ and ${\hat\Theta(-q_{\mT})}$ are unity for $\Cc\neq0$, $\Pi_{\muT\nuT}^{(a)}$ and $\Pi_{\muT\nuT}^{(b)}$
can be rewritten as
\begin{subequations}
  \label{eqn:Sigma_propagator_final}
  \begin{alignat}{3}
      &\Pi_{\muT\nuT}^{(a)}
      =
      &&\dfrac{1}{2 Q^{\pT}}
      \dfrac{i \mathscr{\tilde T}_{\muT\nuT}\Cc}{\Cc q_{\pT}+\Ss q_{\mT}-Q^{\pT}}
      =
      &&\dfrac{1}{2 Q^{\pT}}
      \dfrac{i \mathscr{\tilde T}_{\muT\nuT}\Cc}{q^{\pT}-Q^{\pT}}, \label{eqn:Sigma_propagator_a_final_Cc}\\
      &\Pi_{\muT\nuT}^{(b)}
      =
      - &&\dfrac{1}{2 Q^{\pT}}
	\dfrac{i \mathscr{\tilde T}_{\muT\nuT}\Cc}{\Cc q_{\pT}+\Ss q_{\mT}+Q^{\pT}}
      =
      - &&\dfrac{1}{2 Q^{\pT}}
	\dfrac{i \mathscr{\tilde T}_{\muT\nuT}\Cc}{q^{\pT}+Q^{\pT}}. \label{eqn:Sigma_propagator_b_final_Cc}
  \end{alignat}
\end{subequations}
Summing all contributions, we use Eqs.~(\ref{eqn:on_shell_4_momentum_useful_relation}), (\ref{eqn:Q_sup_ph_q_final}) and (\ref{eqn:Lambda_munu}) to verify
\begin{align}\label{eqn:Sigma_a_b_c_sum}
  \Pi_{\muT\nuT}^{(a)}+\Pi_{\muT\nuT}^{(b)}+\Pi_{\muT\nuT}^{(c)}
  =&
  \dfrac{i \mathscr{\tilde T}_{\muT\nuT}\Cc}{(q^{\pT})^{2}-(Q^{\pT})^{2}}+\dfrac{i n_{\muT}n_{\nuT}}{\mathbf{q}_{\perp}^{2}\Cc+q_{\mT}^{2}}\nonumber\\
  =&\dfrac{i \mathscr{\tilde T}_{\muT\nuT}}{q^{2}}+\dfrac{i n_{\muT}n_{\nuT}}{\mathbf{q}_{\perp}^{2}\Cc+q_{\mT}^{2}}\nonumber\\
  =&\dfrac{-ig_{\muT\nuT}}{q^{2}}.
\end{align}
It is clear from this derivation that the contribution from the instantaneous interaction is cancelled by the corresponding
$n_{\muT}n_{\nuT}$ term in the transverse photon propagator and the sum of all the contributions is totally Lorentz invariant.

For $\Cc=0$, ${\hat\Theta(q_{\mT})}$ and ${\hat\Theta(-q_{\mT})}$ are neither unity nor same to each other. Thus,
we need to look into the details of each contribution in Eqs.~(\ref{eqn:Sigma_propagator_a_final}), (\ref{eqn:Sigma_propagator_b_final})
and (\ref{eqn:Sigma_propagator_c_final}) very carefully to understand the Lorentz invariance of the total amplitude.
Taking $\Cc=0$ in Eqs.~(\ref{eqn:Sigma_propagator_a_final}), (\ref{eqn:Sigma_propagator_b_final}) and (\ref{eqn:Sigma_propagator_c_final}),  we get :
\begin{subequations}
  \label{eqn:Sigma_propagator_LFD}
  \begin{alignat}{3}
      &\Pi_{\mu\nu}^{(a)}
      =
      &&\dfrac{1}{2 q^{+}}
      \dfrac{i (-g_{\mu\nu} -\dfrac{q^{2}n_{\mu}n_{\nu}}{(q^+)^{2}}){\Theta(q^+)}}{p_1^- - p_3^- -\frac{ \mathbf{q}_{\perp}^{2}}{2q^{+}}}, \label{eqn:Sigma_propagator_a_final_LF}\\
      &\Pi_{\mu\nu}^{(b)}
      =
      - &&\dfrac{1}{2 |q^{+}|}
	\dfrac{i (-g_{\mu\nu} -\dfrac{q^{2}n_{\mu}n_{\nu}}{(q^+)^{2}}){\Theta(-q^+)}}{p_1^- - p_3^- +\frac{ \mathbf{q}_{\perp}^{2}}{2|q^{+}|}},  \label{eqn:Sigma_propagator_b_final_LF}\\
      &\Pi_{\mu\nu}^{(c)}
      =
      &&\dfrac{i n_{\mu}n_{\nu}}{{q^+}^{2}}, \label{eqn:Sigma_propagator_c_final_LF}
  \end{alignat}
\end{subequations}
where $q^+=p_1^+ - p_3^+$ and of course the conservation of total LF energy in the initial and final states is  imposed such as $p_1^- - p_3^- = p_4^- - p_2^-$. In this case, the contribution from the LF time-ordered process (a) or (b) in Fig.~\ref{fig:time_ordered_photon_propagator}
depends on the values of external momenta $p_1^+$ and $p_3^+$, i.e. whether $p_1^+ > p_3^+$ or $p_1^+ < p_3^+$.
As indicated by $\Theta(q^+)$ and $\Theta(-q^+)$ in Eqs.~(\ref{eqn:Sigma_propagator_a_final_LF}) and (\ref{eqn:Sigma_propagator_b_final_LF}), respectively, only one of the two LF time-ordering processes, (a) or (b), not both contributes
once the external kinematic situation is given. For example, if  $p_1^+ > p_3^+$, then only (a) contributes. Similarly, if $p_1^+ < p_3^+$,
then only (b) contributes. This is dramatically different from the $\Cc\neq0$ case where both (a) and (b) contributes regardless of
$p_{1\mT} > p_{3\mT}$ or $p_{1\mT} < p_{3\mT}$. The limiting case $\Cc\rightarrow 0$ will be separately discussed later after this discussion of
LFD, i.e. $\Cc=0$.
Although either (a) or (b) (not both) contributes to the total amplitude, the Lorentz
invariance of the total amplitude is assured in LFD. For example, if $p_1^+>p_3^+$, then the sum of all contributions is given by
\begin{align}\label{eqn:Sigma_a_c_sum_LF}
  \Pi_{\mu\nu}^{(a)}+\Pi_{\mu\nu}^{(c)}
  =&
 \dfrac{1}{2 q^{+}}
      \dfrac{i (-g_{\mu\nu} -\dfrac{q^{2}n_{\mu}n_{\nu}}{(q^+)^{2}})}{p_1^- - p_3^- -\frac{ \mathbf{q}_{\perp}^{2}}{2q^{+}}}
      +\dfrac{i n_{\mu}n_{\nu}}{{q^+}^{2}}
      \nonumber\\
  =&  \dfrac{i (-g_{\mu\nu} -\dfrac{q^{2}n_{\mu}n_{\nu}}{(q^+)^{2}})}{2 q^+ q^- - \mathbf{q}_{\perp}^{2}}
      +\dfrac{i n_{\mu}n_{\nu}}{{q^+}^{2}}\nonumber\\
  =&\dfrac{-ig_{\mu\nu}}{q^{2}},
\end{align}
where $p_1^- - p_3^- = q^-$ since the four-momentum transfer $q$ is assigned as $p_1- p_3$.
As in the $\Cc\neq0$ case, the contribution from the instantaneous interaction is cancelled by the corresponding
$n_{\mu}n_{\nu}$ term in the transverse photon propagator and the sum of all the contributions is totally Lorentz invariant.
Similarly, if $p_1^+ < p_3^+$, then $\Pi_{\mu\nu}^{(a)}$ is replaced by $\Pi_{\mu\nu}^{(b)}$ with $|q^+|=-q^+$ and the sum
$\Pi_{\mu\nu}^{(b)}+\Pi_{\mu\nu}^{(c)} = \dfrac{-ig_{\mu\nu}}{q^{2}}$ is totally Lorentz invariant.

Now, let's consider the limit $\Cc\rightarrow0$ (i.e., $\delta\rightarrow\pi/4$ or $\Ss\rightarrow1$) from the $\Cc\neq0$ case given by Eqs.~(\ref{eqn:Sigma_propagator_a_final_Cc}) and (\ref{eqn:Sigma_propagator_b_final_Cc}), where we take
$q_{\pT}\rightarrow q_{+}=q^{-}, q_{\mT}\rightarrow q_{-}=q^{+}, Q^{\pT}\rightarrow Q^{+}=q^{+}$.
To obtain the correct limits, we need to make a careful expansion. For $q^+ = p_1^+ - p_3^+ > 0$, we get
\begin{align}
  \Cc q_{\pT}+\Ss q_{\mT}-Q^{\pT}
  &=\Cc q_{\pT}+\Ss q_{\mT}-\sqrt{q_{\mT}^{2}+\Cc \mathbf{q}_{\perp}^{2}} \nonumber\\
  &\rightarrow\Cc q^{-}+ q^{+}-\sqrt{q^{+2}+\Cc \mathbf{q}_{\perp}^{2}} \nonumber\\
  &\rightarrow\Cc q^{-}-\Cc \frac{ \mathbf{q}_{\perp}^{2}}{2q^{+}}+\mathscr{O}(\Cc^{2}), \\
  \Cc q_{\pT}+\Ss q_{\mT}+Q^{\pT}
  &\rightarrow 2q^{+}+\Cc q_{+}+\Cc \frac{ \mathbf{q}_{\perp}^{2}}{2q^{+}}+\mathscr{O}(\Cc^{2}), \\
  \mathscr{\tilde T}_{\muT\nuT}&\rightarrow -g_{\mu\nu}-2\dfrac{q^{-}-\frac{\mathbf{q}_{\perp}^{2}}{2q^{+}}}{q^{+}}n_{\mu}n_{\nu}.
\end{align}
Thus, in the limit $\Cc\rightarrow 0$ for $p_1^+ > p_3^+$, Eq.~(\ref{eqn:Sigma_propagator_a_final_Cc}), Eq.~(\ref{eqn:Sigma_propagator_b_final_Cc}) as well as Eq.~(\ref{eqn:Sigma_propagator_c_final}) become
\begin{subequations}
  \label{eqn:Sigma_propagator_abc_LF}
  \begin{align}
    \Pi_{\mu\nu}^{(a)}=&\dfrac{i\mathscr{\tilde T}_{\muT\nuT}}{2q^{+}}\dfrac{1}{q^{-}-\frac{ \mathbf{q}_{\perp}^{2}}{2q^{+}}}=-i\dfrac{g_{\mu\nu}}{2q^{+}}\dfrac{1}{q^{-}-\frac{ \mathbf{q}_{\perp}^{2}}{2q^{+}}}-i\dfrac{n_{\mu}n_{\nu}}{(q^{+})^{2}},\label{eqn:Sigma_propagator_a_LFlimit}\\
    \Pi_{\mu\nu}^{(b)}=&\lim_{\Cc\rightarrow 0}\dfrac{-i\mathscr{\tilde T}_{\muT\nuT}}{2q^{+}}\dfrac{\Cc}{2q^{+}+\Cc q_{+}+\Cc \frac{ \mathbf{q}_{\perp}^{2}}{q^{+}}} \rightarrow 0,\label{eqn:Sigma_propagator_b_LFlimit}\\
    \Pi_{\mu\nu}^{(c)}=&\dfrac{i n_{\mu}n_{\nu}}{(q^{+})^{2}}. \label{eqn:Sigma_propagator_c_LFlimit}
  \end{align}
\end{subequations}
This shows the agreement between the $\Cc = 0$ result and the $\Cc \rightarrow 0$ result for the kinematic situation $p_1^+ > p_3^+$.
Note here that $\Pi_{\mu\nu}^{(b)} = 0$ was given in LFD ($\Cc=0$) via the factor $\Theta(-q^+)$ in
Eq.~(\ref{eqn:Sigma_propagator_b_final_LF})
while $\Pi_{\mu\nu}^{(b)} \rightarrow 0$ as $\Cc \rightarrow 0$ in Eq.~(\ref{eqn:Sigma_propagator_b_LFlimit}) is obtained without
the factor $\Theta(-q^+)$.
Similarly, the agreement between the $\Cc = 0$ result and the $\Cc \rightarrow 0$ result is also found for the kinematic situation
$p_1^+ < p_3^+$ following the procedure described above.

Having shown the agreement between the $\Cc = 0$ result and the $\Cc \rightarrow 0$ result for the kinematic situation $p_1^+ \neq p_3^+$,
let's now consider the special kinematic situation $p_1^+ = p_3^+$. If this kinematic situation is provided with the non-zero values
of $p_1^+$ and $p_3^+$, i.e. $p_1^+ = p_3^+ \neq 0$, then the LF time-ordered processes (a) and (b) are indistinguishable and
the results of $\Pi_{\mu\nu}^{(a)}$ and $\Pi_{\mu\nu}^{(b)}$ given by Eqs.~(\ref{eqn:Sigma_propagator_a_final_LF}) and (\ref{eqn:Sigma_propagator_b_final_LF}), respectively, are identical with the factor $\Theta(q^+=-q^+=0)=1/2$.
We also find that the limit $\Cc\rightarrow 0$ of Eqs.~(\ref{eqn:Sigma_propagator_a_final_Cc}) and (\ref{eqn:Sigma_propagator_b_final_Cc})
agree with these results, precisely yielding the $1/2$ factor both for (a) and (b) contributions as in the $\Cc=0$ result. Thus, both in $\Cc=0$ and the limit $\Cc\rightarrow 0$, the total
amplitude is given by
\begin{equation}\label{eqn:Sigma_a_b_c_LF}
  \Pi_{\mu\nu}^{(a)} + \Pi_{\mu\nu}^{(b)} + \Pi_{\mu\nu}^{(c)}
  = \dfrac{-ig_{\mu\nu}}{q^{2}},
    \end{equation}
where $q^2=-\mathbf{q}_{\perp}^{2}$ and the divergent instantaneous interaction is cancelled by the corresponding divergent $n_\mu n_\nu$ term in the transverse photon propagator.

Although the LFD results at exact $\Cc=0$ are attainable from the $\Cc\rightarrow 0$ results for all the kinematic regions that we
discussed in this subsection so far such as $p_1^+ > p_3^+$, $p_1^+ < p_3^+$ and $p_1^+ = p_3^+ \neq 0$,
the agreement between the two, i.e. LFD vs. the limit to LFD, should be looked into more carefully for the case $p_1^+ = p_3^+ = 0$
which should be distinguished from the case $p_1^+ = p_2^+ \neq 0$ that we have discussed in this subsection.
This special kinematic situation $p_1^+ = p_3^+ = 0$ involves the discussion of the infinite momentum frame (IMF) with $P^z \rightarrow -\infty$, since all the plus momenta go to zero in this frame. Presenting the numerical results with the frame dependence of the time-ordered amplitudes in the next section, we will discuss a particular case of correlating the two limits, $\Cc \rightarrow 0$ and $q^+ \rightarrow 0$, in conjunction with the J-shaped correlation coined in our previous work~\cite{Ji2012}. As we show in the next section, the results in the limit $\Cc \rightarrow 0$ following the J-curve are different from those in the LFD or at the exact $\Cc = 0$.

\section{Frame Dependence And Interpolation Angle Dependence of Time-ordered Amplitudes}
\label{sec:the_frame_dependence_and_interpolation_angle_dependence_of_time_ordered_amplitudes}

In this section, we numerically compute the scattering amplitudes shown in Fig. \ref{fig:time_ordered_photon_propagator} and
discuss both the frame dependence and the interpolation angle dependence of each and every $x^{\pT}$-ordered amplitudes.
Putting Eq.~(\ref{eqn:Sigma_propagator_a_final}), Eq.~(\ref{eqn:Sigma_propagator_b_final}) and Eq.~(\ref{eqn:Sigma_propagator_c_final}) back into Eq.~(\ref{eqn:M_Ma_Mb_Mc}), we have the time-ordered amplitudes in the following form:
\begin{subequations}
  \label{eqn:M_final}
 \begin{align}
    \mathcal{M}^{(a)}=&-(-ie)^{2}\dfrac{[p_{24}\cdot p_{13}+p_{24}^{\pT}p_{13}^{\pT}q^{2}/(Q^{\pT})^{2}]\Cc{\hat\Theta(p_{1\mT}-p_{3\mT})}}{2Q^{\pT}(q^{\pT}-Q^{\pT})}\label{eqn:M_a_final},\\
    \mathcal{M}^{(b)}=&(-ie)^{2}\dfrac{[p_{24}\cdot p_{13}+p_{24}^{\pT}p_{13}^{\pT}q^{2}/Q^{\pT})^{2}]\Cc{\hat\Theta(p_{3\mT}-p_{1\mT})}}{2Q^{\pT}(q^{\pT}+Q^{\pT})}\label{eqn:M_b_final},\\
    \mathcal{M}^{(c)}=&(-ie)^{2}\dfrac{p_{24}^{\pT}p_{13}^{\pT}}{(Q^{\pT})^{2}}\label{eqn:M_c_final},
 \end{align}
\end{subequations}where
\begin{subequations}
  \label{eqn:p13p24}
  \begin{align}
    p_{24}=p_{2}+p_{4},\label{eqn:p24}\\
    p_{13}=p_{1}+p_{3}.\label{eqn:p13}
  \end{align}
\end{subequations}
As we have shown in the last section, the sum of these three $x^{\pT}$-ordered amplitudes agree with the manifestly invariant total amplitude
regardless of whether $\Cc\neq 0$ or $\Cc=0$:
\begin{align}
  \mathcal{M}=\Sigma_{j=a,b,c}\mathcal{M}^{(j)}=-(-ie)^{2}\dfrac{p_{24}p_{13}}{q^{2}}.\label{eqn:M_tot}
\end{align}

To investigate the frame dependence of these amplitudes we first look at how they change under different transformations.
From Eq.~(\ref{eqn:P_sup_ph_under_T_any_interpolation_angle}), one can see that when $\beta_{3}=0$, $T_{12}^{\dagger}P^{\pT}T_{12}=P^{\pT}$.
So under the kinematic transformation $T_{12}$, all the ``$\pT$'' components, namely $q^{\pT}$, $Q^{\pT}$, $p_{13}^{\pT}$ and $p_{24}^{\pT}$ remain the same. We also note that the factors of ${\hat\Theta(p_{1\mT}-p_{3\mT})}$ and ${\hat\Theta(p_{3\mT}-p_{1\mT})}$ are invariant under
$T_{12}$ because these factors are unity for $\Cc\neq 0$ and become $\Theta(p_1^+ - p_3^+)$ and $\Theta(p_3^+ - p_1^+)$, respectively,
for $\Cc = 0$. Thus, all three $x^{\pT}$-ordered amplitudes are invariant under $T_{12}$ regardless of $\Cc$ values as they should be.

Now applying the longitudinal boost $T_{3}$ to the three time-ordered amplitudes, we note that the operator $K^{3}=M_{\pT\mT}$ changes its characteristic from dynamic for $\Cc\neq 0$ to kinematic in the light-front ($\Cc=0$) as shown in Table~\ref{tab:Kinematic_and_dynamic_generators_for_different_interoplation_angles} which summarized
the set of kinematic and dynamic generators depending on the interpolation angle.
We mentioned this point in Section \ref{sec:interpolation_between_instant_form_and_light_front_form} and provided its elaborate discussion
in Refs.~\cite{Ji2001, Ji2012}. In applying $T_{3}=e^{-i K_{3}\beta^{3}}$, we thus distinguish the values of $\Cc$ between $\Cc\neq 0$ and $\Cc = 0$. We first consider the $\Cc \neq 0$ case and later compare the results in this case with the LFD ($\Cc = 0$) results.

For the kinematics of our two-body scattering process analogous to $e\mu \rightarrow e\mu$, we choose the scattering plane as the $x-z$ plane and the directions of the initial two particles as the parallel/antiparallel to the $z$-axis such that $\mathbf{p}_{1\perp}=\mathbf{p}_{2\perp}=0$ as shown
in FIG.~\ref{fig:time_ordered_photon_propagator}.
In the center of momentum frame (CMF), we have $p_{1}^{z}=-p_{2}^{z}\equiv p$, $\epsilon_{1}\equiv p_{1}^{0}=p_{3}^{0}=\sqrt{p^{2}+m_{1}^{2}}$, $\epsilon_{2}\equiv p_{2}^{0}=p_{4}^{0}=\sqrt{p^{2}+m_{2}^{2}}$ and $M\equiv \epsilon_{1}+\epsilon_{2}$, where we denote the total energy in CMF as ``$M$" for the convenience in the later discussion. We should note that the invariant Mandelstam variable $s = (p_1+p_2)^2$ is given by
$M^2$. The final four-momenta $p_3$ and $p_4$ of the particles that are going back to back at angle $\theta$ in CMF are depicted in FIG.~\ref{fig:time_ordered_photon_propagator}.
Correspondingly, the four-momenta of the initial and final particles in CMF are given by
\begin{subequations}
  \label{eqn:p_i_CM}
  \begin{align}
    p_{1}&=(\epsilon_{1},0,0,p),\\
    p_{2}&=(\epsilon_{2},0,0,-p),\\
    p_{3}&=(\epsilon_{1},p \sin\theta,0,p\cos\theta),\\
    p_{4}&=(\epsilon_{2},-p \sin\theta,0,-p\cos\theta).
  \end{align}
\end{subequations}

To discuss the longitudinal boost ($T_{3}$) effect on time-ordered scattering amplitudes, let's now boost the whole system to the total momentum $P^{z}$. From the Lorentz transformation for a composite free particle system, we know that the total energy in the new frame is $E=\sqrt{(P^{z})^{2}+M^{2}}$.
The Lorentz transformation for each $p_{i}$ $(i=1,2,3,4)$ can then be described in terms of $\gamma=E/M$ and $\gamma\beta=P^{z}/M$:
\begin{subequations}
  \label{eqn:p_i_boosted}
  \begin{align}
    {p'}_{i}^{0}&=\gamma p_{i}^{0}+\gamma\beta p_{i}^{z}\\
    {p'}_{i}^{z}&=\gamma p_{i}^{z}+\gamma\beta p_{i}^{0}\\
    {\mathbf{p'}}_{i}^{\perp}&=\mathbf{p}_{i}^{\perp}
  \end{align}
\end{subequations}
The boosted four momentum transfer can therefore be written as 
\begin{figure}[H]
\centering
\subfloat{
\includegraphics[width=0.82\columnwidth]{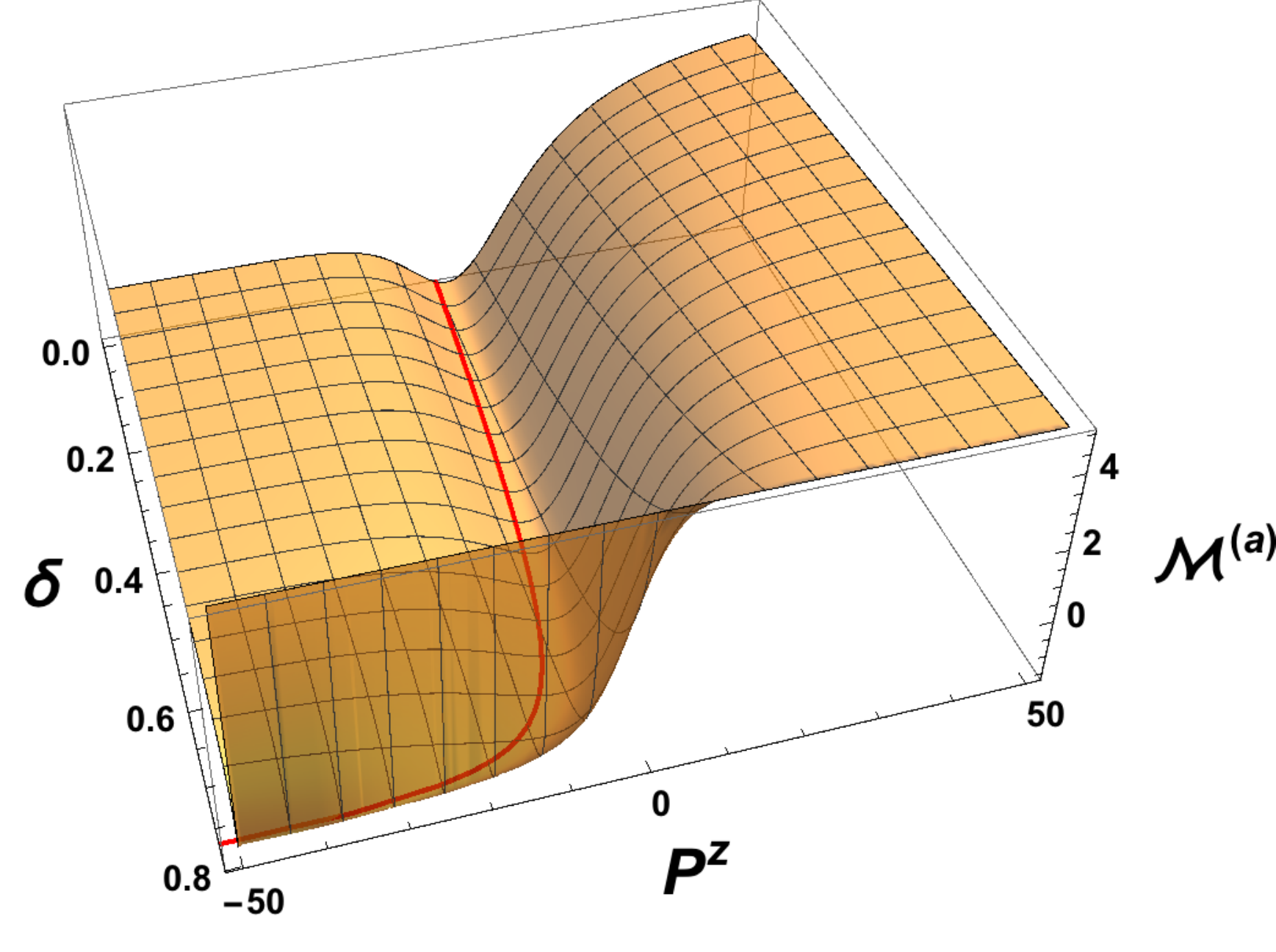}
}
\vspace{0pt}
\centering
\subfloat{
\includegraphics[width=0.82\columnwidth]{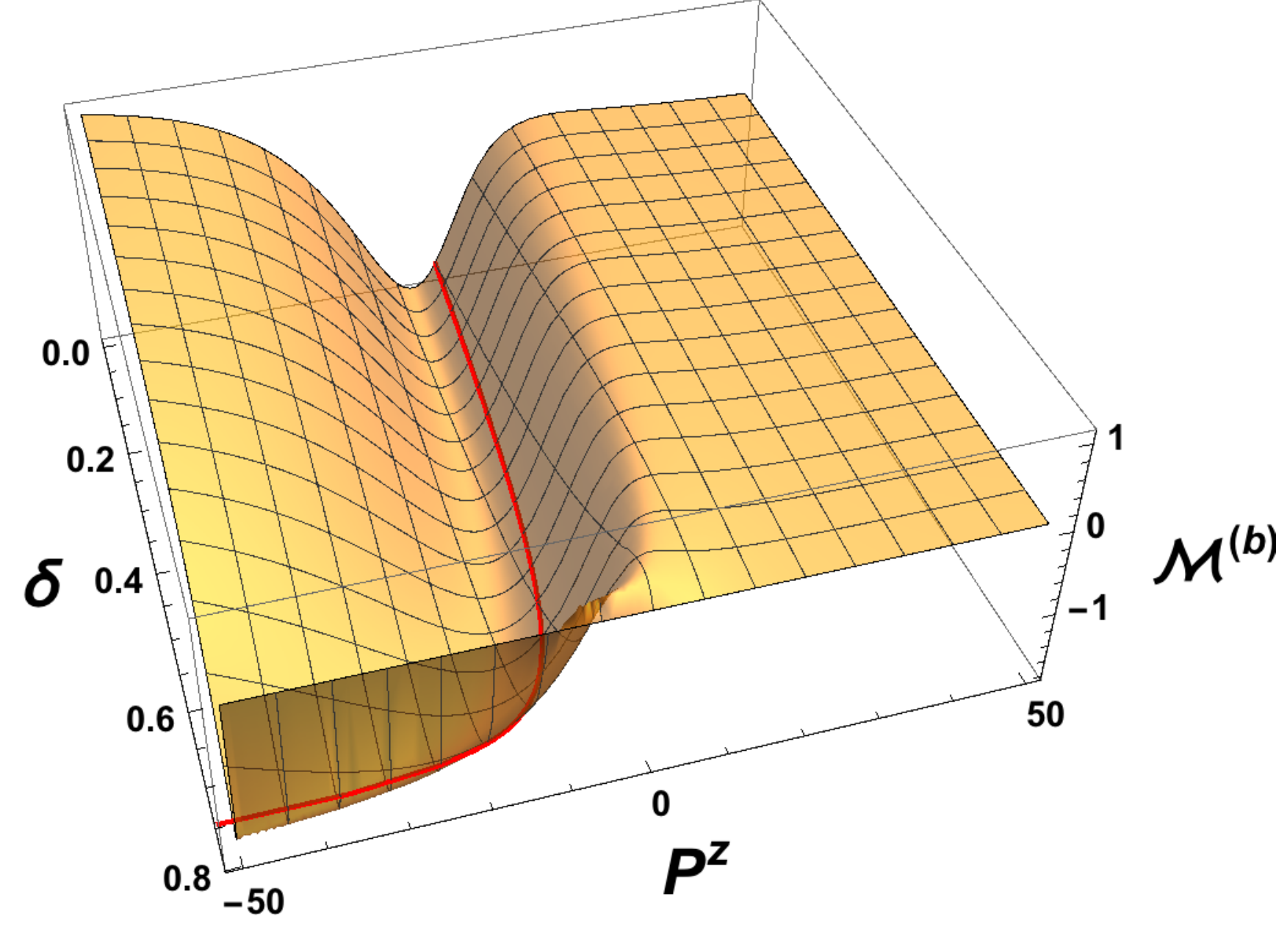}
}
\vspace{0pt}
\centering
\subfloat{
\includegraphics[width=0.82\columnwidth]{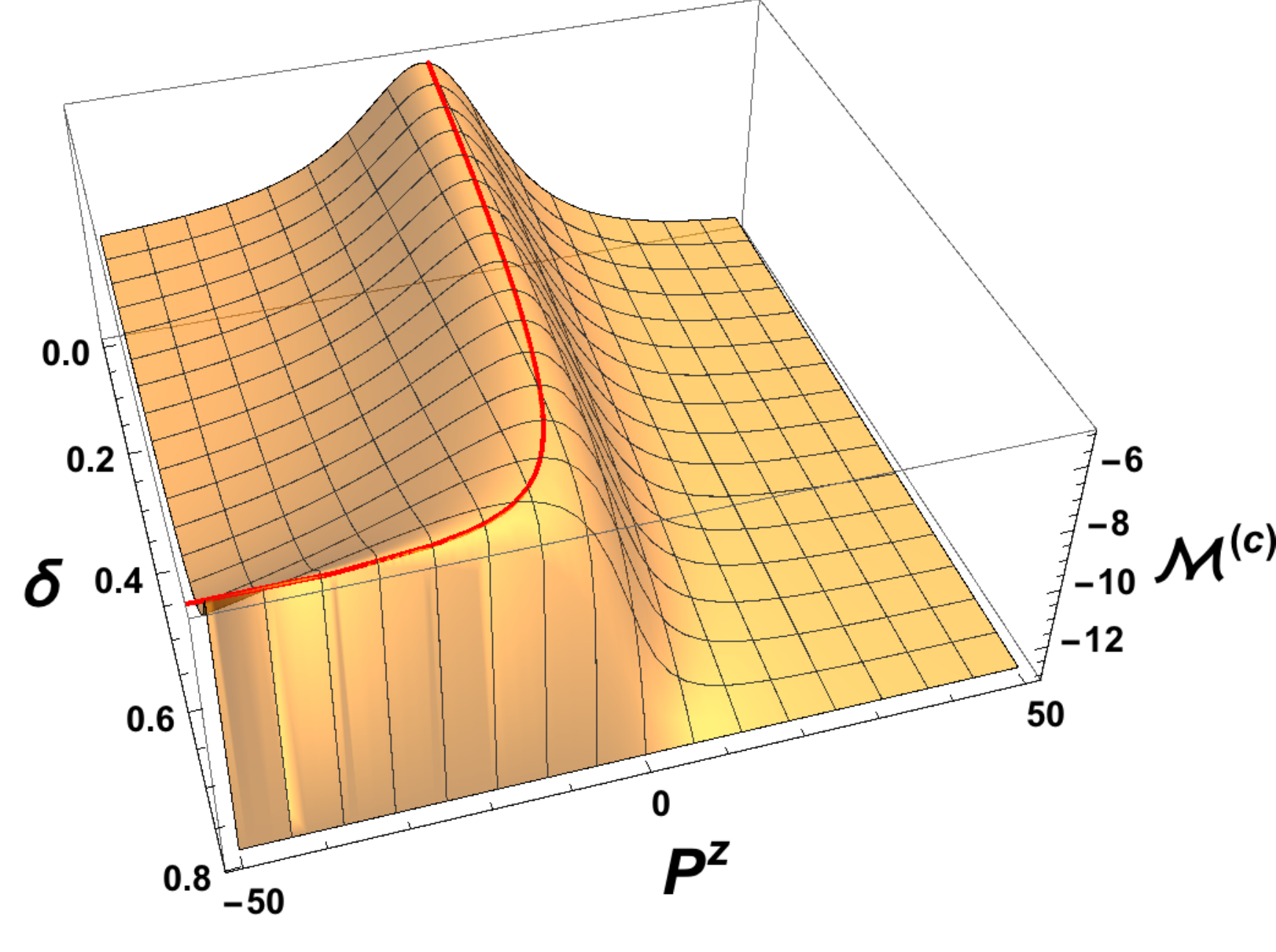}
}
\vspace{0pt}
\centering
\subfloat{
\includegraphics[width=0.82\columnwidth]{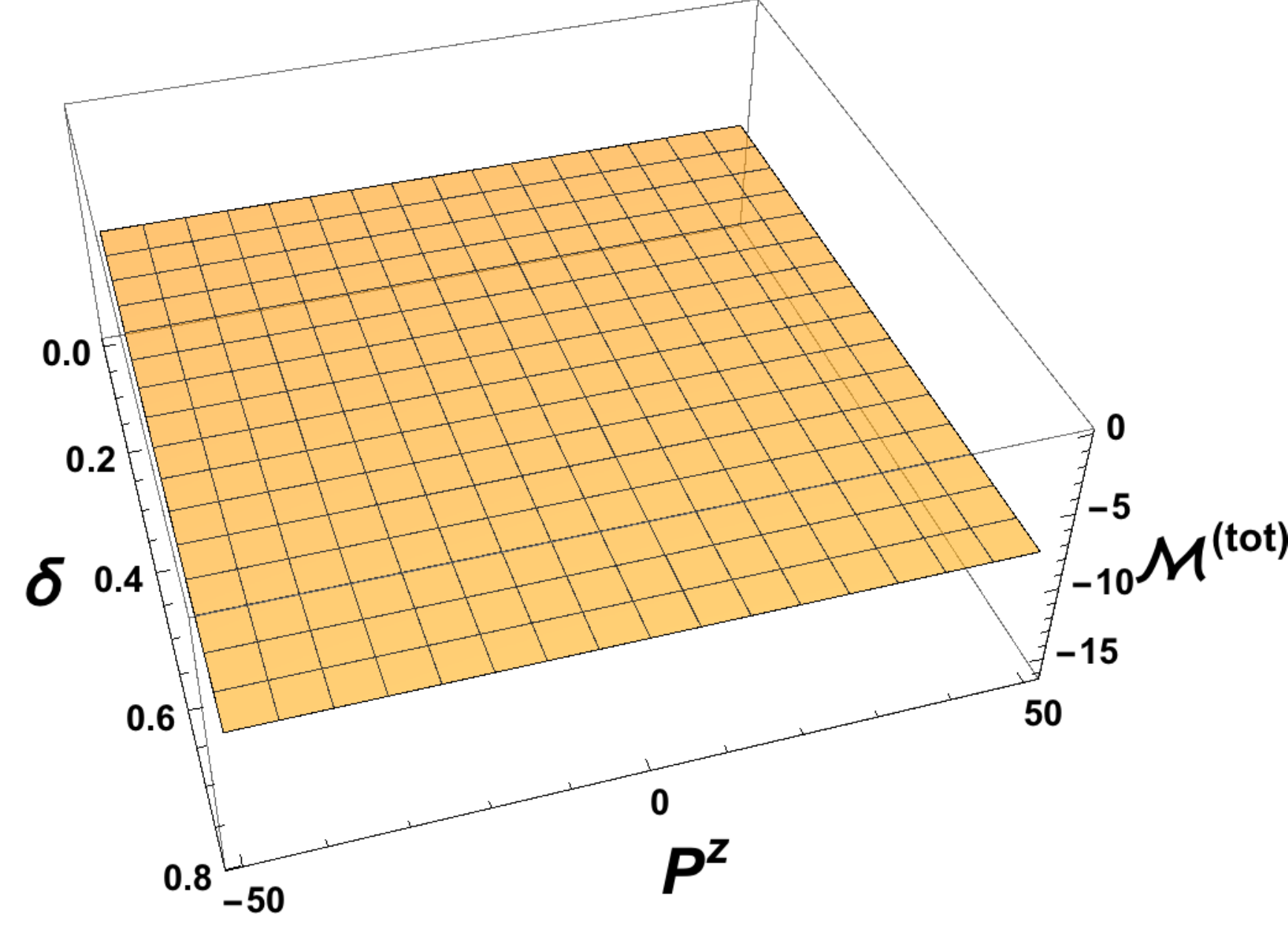}
}
\caption{\label{fig:amp_abc}(color online) Time-ordered amplitudes $\mathcal{M}^{(a)}$, $\mathcal{M}^{(b)}$, $\mathcal{M}^{(c)}$ for $m_{1}=1, m_{2}=2, p=3, \theta=\pi/3$, and their sum $\mathcal{M}^{(tot)}$.}
\end{figure}

\begin{subequations}
  \label{eqn:boosted_p_mu}
  \begin{align}
    {q'}^{x}&={p'}_{1}^{x}-{p'}_{3}^{x}=-p\sin\theta,\\
    {q'}^{y}&={p'}_{1}^{y}-{p'}_{3}^{y}=0,\\
    {q'}^{z}&={p'}_{1}^{z}-{p'}_{3}^{z}=\dfrac{E}{M}p(1-\cos\theta),\\
    {q'}^{0}&={p'}_{1}^{0}-{p'}_{3}^{0}=\dfrac{P^{z}}{M}p(1-\cos\theta),
  \end{align}
\end{subequations}
where $M$ and $E$ are functions of $m_{1}$, $m_{2}$, $p$ and $P^z$.
We can use these four-momentum components given by Eq.~(\ref{eqn:boosted_p_mu}) and apply Eq.~(\ref{eqn:P_interpolation}) to form ${q'}^{\muT}$ and ${q'}_{\muT}$ for any interpolation angle.
Eq.~(\ref{eqn:p_i_boosted}) can also be used together with Eq.~(\ref{eqn:P_interpolation}) and Eq.~(\ref{eqn:p13p24}) to express all the components of the boosted four-momenta. Computing ${p'}_{24}\cdot {p'}_{13}$ and plugging it with other factors such as ${p'}_{24}^{\pT}$ ${p'}_{13}^{\pT}$ and ${q'}_{\muT}$ into Eq.~(\ref{eqn:Q_sup_ph_q_final}) and subsequently Eq.~(\ref{eqn:M_final}), we get the interpolating time-ordered amplitudes $\mathcal{M'}^{(j)}$ in a boosted frame. In CMF, all of these amplitudes are then given by functions of each particle's initial momentum $p$($-p$), individual particle's rest mass $m_{1}$ and $m_{2}$, scattering angle $\theta$, the total momentum $P^{z}$ and the interpolation angle $\delta$. For given values of $m_1, m_2, p$ and $\theta$, the $x^{\pT}$-ordered amplitudes are dependent on the frame ($P^z$) and the interpolation angle ($\delta$ or $\Cc$).

To exhibit this feature quantitatively, we choose $m_{1}=1$, $m_{2}=2$, $p=3$, in the same energy unit (i.e., $m_2$ and $p$ scaled by $m_1$), and $\theta=\pi/3$.
For simplicity, we also take the charge $e$ to be 1 unit in our calculation. In FIG.~\ref{fig:amp_abc}, we plot
$\mathcal{M}^{(a)}$, $\mathcal{M}^{(b)}$, $\mathcal{M}^{(c)}$ as well as the sum of all three amplitudes $\mathcal{M}^{(tot)}$ as functions of both the interpolation angle $\delta$ and the total momentum $P^{z}$ to reflect not only the interpolation angle dependence but also the frame dependence.
The total amplitude $\mathcal{M}^{(tot)}$ shown in FIG.~\ref{fig:amp_abc} is both frame independent and interpolation angle independent as it should be.

The detailed structures of these three time-ordered diagrams are very interesting.
Just like in the $\phi^{3}$ toy model theory studied in \cite{Ji2012}, the $P^{z}>0$ region is smooth for all three amplitudes, while a J-shaped correlation curve exists in the $P^{z}<0$ region, which we plotted as the red solid line in FIG.~\ref{fig:amp_abc} (color online).
This is the curve that starts out in the center of mass frame ($P^{z}=0$) in the $\delta=0$ limit, but maintains the same value of amplitude throughout the whole range of interpolation angle.
The values maintained by these curves can be found for arbitrary $m_{1}$, $m_{2}$, $p$ and $\theta$, and given by
\begin{subequations}
  \label{eqn:M_abc_CM_IF}
  \begin{align}
    \mathcal{M}^{(a)}=\mathcal{M}^{(b)}=&
    -\dfrac{\cot^{2} \frac{\theta}{2}}{2},\label{eqn:Mab_CM_IF}\\
    \mathcal{M}^{(c)}=&
    -\dfrac{\epsilon_{1}\epsilon_{2}}{p^{2}\sin^{2}\frac{\theta}{2}}.\label{eqn:Mc_CM_IF}
  \end{align}
\end{subequations}
Again, $\epsilon_{i}=\sqrt{q^{2}+m_{i}^{2}}$, $(i=1,2)$ and the charge ``$e$'' in Eq.~(\ref{eqn:M_final}) is taken as 1 unit.

\begin{figure}[t]
  \includegraphics[width=\columnwidth]{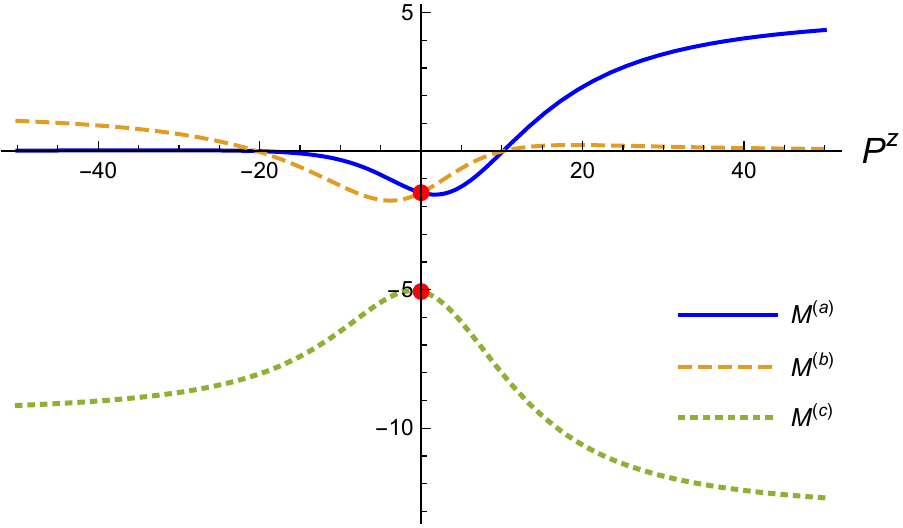}
  \caption{\label{fig:M_abc_IF}(color online) Time-ordered amplitudes $\mathcal{M}^{(a)}$, $\mathcal{M}^{(b)}$ and $\mathcal{M}^{(c)}$ as a function of the total momentum $P^{z}$ at the instant form limit ($\delta=0$).
	The red dots on the $P^{z}=0$ axis denotes the starting position of the J-curves on each surface.}
\end{figure}

In FIG.~\ref{fig:M_abc_IF}, we plot the profiles of $\mathcal{M}^{(a)}$, $\mathcal{M}^{(b)}$ and $\mathcal{M}^{(c)}$ in IFD ($\Cc=1$ or $\delta=0$)
for the given values of $m_{1}, m_{2}, p$ and $\theta$ and depict the starting point of the J-curve corresponding to the values of $\mathcal{M}^{(a)}$, $\mathcal{M}^{(b)}$ and $\mathcal{M}^{(c)}$ at $P^{z}=0$ given by Eqs.~(\ref{eqn:M_abc_CM_IF}). The maxima for $\mathcal{M}^{(a)}$ and $\mathcal{M}^{(b)}$ stay in the $P^{z}>0$ and $P^{z}<0$ regions respectively, but as the scattering angle $\theta$ approaches $0$, the maximum for $\mathcal{M}^{(a)}$ and $\mathcal{M}^{(b)}$ will move towards $P^{z}=0$. As the J-curve does not track the maximum or the minimum in $\mathcal{M}^{(a)}$ and $\mathcal{M}^{(b)}$, we note that
the J-curve does not track the maximum of the surface $M^{(c)}$ either.

It is interesting to note that the J-curve itself does not change with the scattering angle $\theta$.
In fact, we find that it follows exactly the same formula as the one we found in the $\phi^{3}$ theory \cite{Ji2012}:
\begin{align}
  \frac{P^{z}}{M} =-\sqrt{\dfrac{(1-\Cc)}{2\Cc}},\label{eqn:J_curve}
\end{align}
where $M^{2}$ is identical to the invariant Mandelstam variable $s = (p_1+p_2)^2$, i.e. $M=\sqrt{s}$.
We note that the J-curve is ``universal'' in the sense that it doesn't depend on the specific kinematics like the scattering angle or particle masses, but scales with $\sqrt{s}$.

If we follow this J-curve plotted as the red solid line in FIG.~\ref{fig:amp_abc}, we see that as $\Cc\rightarrow 0$ or $\delta\rightarrow\pi/4$,  it is pushed to $P^{z}\rightarrow-\infty$ and the constant values given by Eq.~(\ref{eqn:M_abc_CM_IF}) are maintained throughout the curve.
On the other hand, the $x^+$-ordered light-front amplitudes are invariant under the longitudinal boost $T_{3}=e^{-i K_{3}\beta^{3}}$ and thus
each of $\mathcal{M}^{(a)}$, $\mathcal{M}^{(b)}$ and $\mathcal{M}^{(c)}$ show up individually as a constant independent of $P^{z}$ along the
$\delta=\pi/4$ or $\Cc =0 $ line in FIG.~\ref{fig:amp_abc}. Thus, the values of $\mathcal{M}^{(a)}$, $\mathcal{M}^{(b)}$ and $\mathcal{M}^{(c)}$
in the limit $\Cc \rightarrow 0$ following the J-curve are different from those in the LFD, i.e. at the exact $\Cc = 0$.
This brings up our discussion in the last subsection, Sec. \ref{sub:discussion}, about the issue whether the LFD results with the exact $\Cc=0$ can be reproduced by taking the limit $\Cc \rightarrow 0$. Because all the plus components of the particle momenta vanish in the limit
$P^z \rightarrow -\infty$, we now should look closely at the kinematic situation $p_1^+ = p_3^+ =q^+ = 0$ which we have postponed
its discussion in Sec. \ref{sub:discussion}.

Since the details of each contribution depend on the values of $\Cc$ and $P^z$, the results are in general dependent on
the order of taking the limits to $\Cc \rightarrow 0$ and $P^z \rightarrow -\infty$. In particular, if $\Cc=0$ is taken first, then
the results must be independent of the $P^z$ values due to the longitudinal boost invariance in LFD as discussed before
and the result at $P^z = -\infty$ is identical to that for any other $P^z$ value.
However, if we take $P^z \rightarrow -\infty$ first and consider $\Cc \rightarrow 0$, then we should first examine the $\Cc\neq 0$ results
in the $P^z \rightarrow -\infty$ limit with a great care. We begin with the discussion on the $\Cc =1$ (i.e. IFD) results and vary the interpolation angle from $\delta=0$ to $\delta \rightarrow \pi/4$.

As shown in FIGs.~\ref{fig:amp_abc} and \ref{fig:M_abc_IF}, each of the time-ordered amplitudes $\mathcal{M}^{(a)}$, $\mathcal{M}^{(b)}$ and $\mathcal{M}^{(c)}$ at $\Cc = 1 (\delta = 0) $, are not symmetric under the reflection of $P^z \rightarrow -P^z$. Although the IFD results coincide with the LFD results as $P^z \rightarrow \infty$, they do not agree in the limit $P^z \rightarrow -\infty$ as shown in FIG.~\ref{fig:amp_abc}. Thus, the prevailing notion of the equivalence between IFD and LFD in the infinite momentum frame (IMF) should be taken with a great caution since it works for the limit of $P_z \rightarrow \infty$ but not in the limit $P_z \rightarrow -\infty$. We have already discussed the treachery in taking the limit $P_z \rightarrow -\infty$ even for the amplitudes with the reflection symmetry under $P^z \leftrightarrow -P^z$ in IFD such as the scalar annihilation process analogous to QED $e^+ e^- \rightarrow \mu^+ \mu^-$ in our previous work \cite{Ji2012}. The present analysis of the sQED scattering process analogous to QED $e \mu \rightarrow e \mu$ provides a clear example of breaking the reflection symmetry under $P^z \leftrightarrow -P^z$ in IFD and fortify the clarification of the confusion in the folklore of the equivalence between the IMF and the LFD.
As we vary the interpolation angle from $\delta = 0$ and approach to $\delta = \pi/4$, this broken reflection symmetry under
$P^z \leftrightarrow -P^z$ persists while the LFD results are symmetric under $P^z \leftrightarrow -P^z$ due to the longitudinal boost invariance
discussed previously. The J-curve correlation between $P^z$ and $\Cc$ given by Eq.~(\ref{eqn:J_curve}) provides a simultaneous limit
of  $P^z \sim  - 1/ \Cc^{1/2} \rightarrow -\infty$ as $\Cc \rightarrow 0$ and the corresponding results of $\mathcal{M}^{(a)}$, $\mathcal{M}^{(b)}$ and $\mathcal{M}^{(c)}$ are uniquely given by Eqs.~(\ref{eqn:Mab_CM_IF}) and (\ref{eqn:Mc_CM_IF}).
As mentioned earlier, the numerical values along the J-curve plotted as the red solid line in FIG.~\ref{fig:amp_abc} (color online)
are identical all the way to the $P^z \rightarrow -\infty$ limit and thus the clear distinction is manifest between the results in the
$\Cc \rightarrow 0$ limit with the J-curve correlation and the LFD results with the exact $\Cc=0$.
Since the J-curve exists only for $P^z < 0$, it is also self-evident that the broken reflection symmetry under $P^z \leftrightarrow -P^z$
still persists for the correlated limit of $P^z \sim  - 1/ \Cc^{1/2} \rightarrow -\infty$ and the results obtained in this limit must be different
from the LFD results. Therefore, the limit $P^z \rightarrow -\infty$ is treacherous and
requires a great caution in taking the light-front limit from $\Cc \neq 0$.
As it must be, however, the sum of all time-ordered amplitudes $\mathcal{M}^{(tot)} = \mathcal{M}^{(a)} + \mathcal{M}^{(b)} +\mathcal{M}^{(c)}$
is completely independent of $\Cc$ and $P^z$. The total amplitude $\mathcal{M}^{(tot)}$ is of course identical regardless of whether $\Cc$ is exactly zero or not, however the limit
$P^z \rightarrow -\infty$ is taken.

Finally, we also calculated the time-ordered scattering amplitudes discussed in this section for the scalar $\phi^{3}$ theory that was used
in our previous work~\cite{Ji2012} and summarized the results in
Appendix~\ref{Append:time_ordered_scattering_amplitudes_for_pure_scalar_theory} for a comparison with the sQED results.
For completeness in comparing the results between the $\phi^{3}$ theory and sQED, the time-ordered annihilation amplitudes for sQED are summarized also in
Appendix~\ref{Append:time_ordered_annihilation_amplitudes_for_photon_exchange_process}.


\section{Summary and Conclusion}
\label{sec:conclusions}

In this work, we extended the interpolating scattering amplitudes introduced for the scalar field theory~\cite{Ji2012} to the sQED theory
to discuss the electromagnetic gauge degree of freedom interpolated between the IFD and the LFD.
We developed the electromagnetic gauge field propagator interpolated between the IFD and the LFD and found that the light-front gauge
$A^{+}=0$ in the LFD is naturally linked to the Coulomb gauge $\boldsymbol\nabla \cdot \mathbf{A}=0$ in IFD.
We identified the dynamical degrees of freedom for the electromagnetic gauge fields as the transverse photon fields and clarified
the equivalence between the contribution of the instantaneous interaction and the contribution from the longitudinal polarization of the virtual photon.

Our results for the gauge propagator and time-ordered diagrams clarified whether one should choose the two-term form~\cite{Mustaki1991}
or the three-term form~\cite{Leibbrandt1984, Srivastava2001, Suzuki2003, *Suzuki2004a, *Suzuki2004b, *Misra2005}
for the gauge propagator in LFD. Our transverse photon propagator in LFD assumes the three-term form, but the third term cancels the instantaneous interaction contribution. Thus, one can use the two-term form of the gauge propagator for effective calculation of amplitudes if one also omits the instantaneous interaction from the Hamiltonian.
But if one wants to show equivalence to the covariant theory, all three terms should be kept because the instantaneous interaction is a natural result of the decomposition of Feynman diagrams, and the third term in the propagator is necessary for the total amplitudes to be covariant.
We also see that the photon propagator was derived according to the generalized gauge that links the Coulomb gauge to light-front gauge
and thus the three-term form appears appropriate in order to be consistent with the appropriate gauge.

Using the interpolating photon propagator, we computed the lowest-order scattering process such as
an analogue of the well-known QED process $e \mu \rightarrow e \mu$ in sQED and analyzed the
three corresponding $x^{\pT}$-ordered diagrams,  two of which are associated with the transverse propagating photon and one of which
is associated with the instantaneous interaction.
We analyzed both the frame and interpolation angle dependence of each $x^{\pT}$-ordered diagram including the instantaneous interaction, varying the total momentum $P^z$ of the system and the interpolation angle parameter $\Cc = {\rm cos} 2\delta$.
Our analysis provided a clear example of breaking the reflection symmetry under $P^z \leftrightarrow -P^z$ in IFD and clarified the confusion in the folklore of the equivalence between the IMF and the LFD. As we vary the interpolation angle from $\delta = 0$ ($\Cc = 1$) and approach to $\delta = \pi/4$ ($\Cc = 0$), this broken reflection symmetry under
$P^z \leftrightarrow -P^z$ persists while the LFD results are symmetric under $P^z \leftrightarrow -P^z$ due to the longitudinal boost invariance.
Moreover, the universal correlation between $P^z$ and $\Cc$ given by Eq.~(\ref{eqn:J_curve}) which was coined as the J-curve
in our previous work~\cite{Ji2012} is intact in each of these $x^{\pT}$-ordered diagrams as plotted by the red solid line
in FIG.~\ref{fig:amp_abc}.  The J-curve starts out from the center of mass frame in the instant form and goes to $P^{z}=-\infty$ as it approaches
to the light-front limit. This correlation is independent of the specific kinematics of the scattering process
and manifests the difference in the results between the light-front limit and the exact light-front (LFD).
Since the J-curve exists only for $P^z < 0$, it is also self-evident that the broken reflection symmetry under $P^z \leftrightarrow -P^z$ still persists for the correlated limit of $P^z \sim  - 1/ \Cc^{1/2} \rightarrow -\infty$ and the results obtained in this limit must be different from the LFD results.
The J-curve not only provides a representative characterization of how the $x^{\pT}$-ordered amplitudes change from IFD to LFD, but also gives rise to the issue of zero-modes at $P^{z}=-\infty$ since all the plus momenta of particles vanish in this limit.
The limit $P^z \rightarrow -\infty$ is thus treacherous and
requires a great caution in taking the light-front limit from $\Cc \neq 0$.
The sum of all $x^{\pT}$-ordered amplitudes is however
completely independent of $\Cc$ and $P^z$ and its full
manifest Poincar\'e invariance provides a useful guidance in handling the treacherous zero-mode issue.

\acknowledgments
This work was supported by the US Department of Energy (Grant No. DE-FG02-03ER41260) and the FAPESP-NC grant (NCSU RADAR 2014-0717).

\appendix
\begin{widetext}

\section{Derivation of photon polarization vectors}
\label{sec:derivation_of_photon_polarization_vectors}


We use the following explicit four-vector representation of $\mathbf{K}$ and $\mathbf{J}$ operators given by
\begin{alignat}{3}
  K_{1}=
  \begin{pmatrix}
  0 & i & 0 & 0 \\
  i & 0 & 0 & 0 \\
  0 & 0 & 0 & 0\\
  0 & 0 & 0 & 0
  \end{pmatrix},
  K_{2}=
  \begin{pmatrix}
  0 & 0 & i & 0 \\
  0 & 0 & 0 & 0 \\
  i & 0 & 0 & 0 \\
  0 & 0 & 0 & 0
  \end{pmatrix},
  K_{3}=
  \begin{pmatrix}
  0 & 0 & 0 & i \\
  0 & 0 & 0 & 0 \\
  0 & 0 & 0 & 0 \\
  i & 0 & 0 & 0
  \end{pmatrix},
  J_{1}=
  \begin{pmatrix}
  0 & 0 & 0 & 0 \\
  0 & 0 & 0 & 0 \\
  0 & 0 & 0 & -i\\
  0 & 0 & i & 0
  \end{pmatrix},
  J_{2}=
  \begin{pmatrix}
  0 & 0 & 0 & 0 \\
  0 & 0 & 0 & i \\
  0 & 0 & 0 & 0 \\
  0 & -i & 0 & 0
  \end{pmatrix},
  J_{3}=
  \begin{pmatrix}
  0 & 0 & 0 & 0 \\
  0 & 0 & -i & 0 \\
  0 & i & 0 & 0 \\
  0 & 0 & 0 & 0
  \end{pmatrix}.
  \label{eqn:J_K}
\end{alignat}


Upon the application of $T$ transformation given by Eq.~(\ref{eqn:T_transformation_for_any_interpolation_angle}) with the above four-vector representation for operators $\mathbf{K}$ and $\mathbf{J}$, we find the polarization vectors:
\begin{gather}\label{eqn:Polarization_vectors_in_beta_alpha}
  \epsilon_{\muT}(P,+)= -
  \begin{bmatrix}
    \dfrac{\sin\delta}{\sqrt{2}}\left( \dfrac{\beta_{1}\sin\alpha}{\alpha}-\dfrac{i\beta_{2}\sin\alpha}{\alpha} \right)\\[12pt]
    \dfrac{\Cc}{\sqrt{2}}\left( \dfrac{\beta_{2}^{2}+\beta_{1}^{2}\cos\alpha}{\alpha^{2}} + \dfrac{i\beta_{1}\beta_{2}(-1+\cos\alpha)}{\alpha^{2}} \right)\\[12pt]
    \dfrac{\Cc}{\sqrt{2}}\left( \dfrac{\beta_{1}\beta_{2}(-1+\cos\alpha)}{\alpha^{2}}+\dfrac{i(\beta_{1}^{2}+\beta_{2}^{2}\cos\alpha)}{\alpha^{2}} \right)\\[12pt]
    -\dfrac{\cos\delta}{\sqrt{2}}\left( \dfrac{\beta_{1}\sin\alpha}{\alpha}+\dfrac{i\beta_{2}\sin\alpha}{\alpha} \right)
  \end{bmatrix},
  \epsilon_{\muT}(P,-)=
  \begin{bmatrix}
    \dfrac{\sin\delta}{\sqrt{2}}\left( \dfrac{\beta_{1}\sin\alpha}{\alpha}+\dfrac{i\beta_{2}\sin\alpha}{\alpha} \right)\\[12pt]
    \dfrac{\Cc}{\sqrt{2}}\left( \dfrac{\beta_{2}^{2}+\beta_{1}^{2}\cos\alpha}{\alpha^{2}} - \dfrac{i\beta_{1}\beta_{2}(-1+\cos\alpha)}{\alpha^{2}} \right)\\[12pt]
    \dfrac{\Cc}{\sqrt{2}}\left( \dfrac{\beta_{1}\beta_{2}(-1+\cos\alpha)}{\alpha^{2}} - \dfrac{i(\beta_{1}^{2}+\beta_{2}^{2}\cos\alpha)}{\alpha^{2}} \right)\\[12pt]
    -\dfrac{\cos\delta}{\sqrt{2}}\left( \dfrac{\beta_{1}\sin\alpha}{\alpha}-\dfrac{i\beta_{2}\sin\alpha}{\alpha} \right)
  \end{bmatrix},\\[20pt]
  \epsilon_{\muT}(P,0)=
  \begin{bmatrix}
    \dfrac{\cos\delta(\sin\delta\cosh\beta_{3}+\cos\delta\sinh\beta_{3})-\sin\delta\cos\alpha(\cos\delta\cosh\beta_{3}+\sin\delta\sinh\beta_{3})}{\Cc}\\[12pt]
    \dfrac{\beta_{1} \sin\alpha (\sin\delta\cosh\beta_{3}+\cos\delta\sinh\beta_{3})}{\alpha }\\[12pt]
   \dfrac{\beta_{2} \sin \alpha (\sin\delta\cosh\beta_{3}+\cos\delta\sinh\beta_{3})}{\alpha }\\[12pt]
   \dfrac{\cos\delta\cos\alpha(\cos\delta\cosh\beta_{3}+\sin\delta\sinh\beta_{3})-\sin\delta(\sin\delta\cosh\beta_{3}+\cos\delta\sinh\beta_{3})}{\Cc}
  \end{bmatrix}.
\end{gather}
Using the relations listed in Eqs.~(\ref{eqn:beta123_relation_with_P}) and (\ref{eqn:useful_beta_P_relation}), and multiplying the transverse polarization vectors $\epsilon_{\muT}(P,\pm)$ with the phase factor $\frac{P^{1}\mp P^{2}}{|\mathbf{P}_{\perp}|}$ to simplify, we arrive at the following polarization vectors for the four-momentum $P^{\mu}$:
\begin{align}
  \epsilon_{\muT}(P,+)&=
  - \dfrac{1}{\sqrt{2}\Pp}
  \left(
   \sin\delta|{\bf P}_{\perp}|,
   \dfrac{P^{1}P_{\mT}-iP^{2}\Pp}{|{\bf P}_{\perp}|},
   \dfrac{P^{2}P_{\mT}+i P^{1}\Pp}{|{\bf P}_{\perp}|},
   -\cos\delta|{\bf P}_{\perp}|
  \right)   \label{eqn:polarization_vector_+_in_P_IF}, \\
  \epsilon_{\muT}(P,-)&=
  \dfrac{1}{\sqrt{2}\Pp}
  \left(
  \sin\delta|{\bf P}_{\perp}|,
  \dfrac{P^{1}P_{\mT}+iP^{2}\Pp}{|{\bf P}_{\perp}|},
  \dfrac{P^{2}P_{\mT}-i P^{1}\Pp}{|{\bf P}_{\perp}|},
  -\cos\delta|{\bf P}_{\perp}|
  \right)   \label{eqn:polarization_vector_-_in_P_IF}, \\
\epsilon_{\muT}(P,0)&=
\dfrac{1}{M \Pp}
\left(
\frac{\Pp^{2}\cos\delta-P_{\mT}P^{\pT}\sin\delta}{\Cc},
P^{1}P^{\pT},
P^{2}P^{\pT},
\frac{P^{\pT}P_{\mT}\cos\delta - \Pp^{2}\sin\delta}{\Cc}
\right) \label{eqn:polarization_vector_0_in_P_IF}.
\end{align}
The polarization vectors listed above are written in the form $\epsilon_{\muT}(P,\lambda)=(\epsilon^{0},\epsilon^{1},\epsilon^{2},\epsilon^{3})$.
We then change the basis to $(\epsilon_{\pT},\epsilon_{1},\epsilon_{2},\epsilon_{\mT})$ using the relations listed in Eq.~(\ref{eqn:relation_between_covariant_and_contravariant_components_with_any_interpolation}).
Finally, we obtain the polarization vectors given by Eq.~(\ref{eqn:polarization_vector_in_P_any_interpolation}):
\begin{subequations}
		\label{eqn:polarization_vector_appendix}
		\begin{align}
				\epsilon_{\muT}(P,+)&=
				- \dfrac{1}{\sqrt{2}\Pp}
				\left( \Ss |{\bf P}_{\perp}|,  \dfrac{P_{1}P_{\mT}-iP_{2}\Pp}{|{\bf P}_{\perp}|}, \dfrac{P_{2}P_{\mT}+i P_{1}\Pp}{|{\bf P}_{\perp}|},
						-\Cc |{\bf P}_{\perp}| \right) \label{Aeqn:polarization_vector_+_in_P_any_interpolation}, \\
				\epsilon_{\muT}(P,-)&=
				\dfrac{1}{\sqrt{2}\Pp}
				\left(
						\Ss|{\bf P}_{\perp}|,
						\dfrac{P_{1}P_{\mT}+i P_{2}\Pp}{|{\bf P}_{\perp}|},
						\dfrac{P_{2}P_{\mT}-i P_{1}\Pp}{|{\bf P}_{\perp}|},
						-\Cc|{\bf P}_{\perp}|
				\right) \label{Aeqn:polarization_vector_-_in_P_any_interpolation}, \\
				\epsilon_{\muT}(P,0)&=
				\dfrac{P^{\pT}}{M \Pp}
				\left(
						P_{\pT}-\dfrac{M^{2}}{P^{\pT}},
						P_{1},
						P_{2},
						P_{\mT}
				\right) \label{Aeqn:polarization_vector_0_in_P_any_interpolation},
		\end{align}
\end{subequations}
where the relation $\Pp^{2}=(P^{\pT})^{2}-M^{2}\Cc$ is used to derive the first component of $\epsilon_{\muT}(P,0)$.
\end{widetext}


\section{\texorpdfstring{$\mathscr{T}_{\muT\nuT}$}{Tmunv} and \texorpdfstring{$\mathscr{L}_{\muT\nuT}$}{Lmunv} derived directly from the interpolating polarization vectors}
\label{sec:photon_propagator_derived_directly_from_polarization_vectors_in_the_interpolating_form}

In this Appendix, we derive the numerator of the photon propagator directly from the polarization vectors listed in Eq.~(\ref{eqn:polarization_vector_in_P_any_interpolation}) and Eq.~(\ref{eqn:polarization_vector_appendix}).

We evaluate the relevant matrix elements. With the four-momentum $P=q$ for virtual photon, these are, for $\lambda=+1$:
\begin{eqnarray}
\epsilon^*_{\pT}(q,\lambda=+)\epsilon_{\pT}(q,\lambda=+) & = & \frac{\mathbb{S}^2|{\bf q}_\perp|^2}{2\Qq^{2}} ,\\
\epsilon^*_{\pT}(q,\lambda=+)\epsilon_1(q,\lambda=+) & = & \frac{\mathbb{S}\left(q_{\mT}q_1-iq_2\Qq\right)}{2\Qq^{2}} ,\\
\epsilon^*_{\pT}(q,\lambda=+)\epsilon_2(q,\lambda=+) & = & \frac{\mathbb{S}\left(q_{\mT}q_2+iq_1\Qq\right)}{2\Qq^{2}} ,\\
\epsilon^*_{\pT}(q,\lambda=+)\epsilon_{\mT}(q,\lambda=+) & = & -\frac{\mathbb{C}\mathbb{S}|{\bf q}_\perp|^2}{2\Qq^{2}} ,
\end{eqnarray}

\begin{eqnarray}
\epsilon^*_{\mT}(q,\lambda=+)\epsilon_{\pT}(q,\lambda=+) & = & -\frac{\mathbb{C}\mathbb{S}|{\bf q}_\perp|^2}{2\Qq^{2}} ,\\
\epsilon^*_{\mT}(q,\lambda=+)\epsilon_1(q,\lambda=+) & = & -\frac{\mathbb{C}\left(q_{\mT}q_1-iq_2\Qq\right)}{2\Qq^{2}} ,\\
\epsilon^*_{\mT}(q,\lambda=+)\epsilon_2(q,\lambda=+) & = & -\frac{\mathbb{C}\left(q_{\mT}q_2+iq_1\Qq\right)}{2\Qq^{2}} ,\\
\epsilon^*_{\mT}(q,\lambda=+)\epsilon_{\mT}(q,\lambda=+) & = & \frac{\mathbb{C}^2|{\bf q}_\perp|^2}{2\Qq^{2}} ,
\end{eqnarray}

\begin{eqnarray}
\epsilon^*_1(q,\lambda=+)\epsilon_{\pT}(q,\lambda=+) & = & \frac{\mathbb{S}\left(q_{\mT}q_1+iq_2\Qq\right)}{2\Qq^{2}} ,\\
\epsilon^*_1(q,\lambda=+)\epsilon_1(q,\lambda=+) & = & \frac{q_{\mT}^2q_1^2+q_2^2\Qq^{2}}{2|{\bf q}_\perp|^2\Qq^{2}} ,\\
\epsilon^*_1(q,\lambda=+)\epsilon_2(q,\lambda=+) & = & \frac{-\mathbb{C}q_1q_2+iq_{\mT}\Qq}{2\Qq^{2}} ,\\
\epsilon^*_1(q,\lambda=+)\epsilon_{\mT}(q,\lambda=+) & = & \frac{-\mathbb{C}\left(q_{\mT}q_1+iq_2\Qq\right)}{2\Qq^{2}},
\end{eqnarray}

\begin{eqnarray}
\epsilon^*_2(q,\lambda=+)\epsilon_{\pT}(q,\lambda=+) & = & \frac{\mathbb{S}\left(q_{\mT}q_2-iq_1\Qq\right)}{2\Qq^{2}} ,\\
\epsilon^*_2(q,\lambda=+)\epsilon_1(q,\lambda=+) & = &\frac{-\mathbb{C}q_1q_2-iq_{\mT}\Qq}{2\Qq^{2}} ,\\
\epsilon^*_2(q,\lambda=+)\epsilon_2(q,\lambda=+) & = & \frac{q_{\mT}^2q_2^2+q_1^2\Qq^{2}}{2|{\bf q}_\perp|^2\Qq^{2}} ,\\
\epsilon^*_2(q,\lambda=+)\epsilon_{\mT}(q,\lambda=+) & = & \frac{-\mathbb{C}\left(q_{\mT}q_2-iq_1\Qq\right)}{2\Qq^{2}},
\end{eqnarray}

and for $\lambda=-1$:
\begin{eqnarray}
\epsilon^*_{\pT}(q,\lambda=-)\epsilon_{\pT}(q,\lambda=-) & = & \frac{\mathbb{S}^2|{\bf q}_\perp|^2}{2\Qq^{2}} ,\\
\epsilon^*_{\pT}(q,\lambda=-)\epsilon_1(q,\lambda=-) & = & \frac{\mathbb{S}\left(q_{\mT}q_1+iq_2\Qq\right)}{2\Qq^{2}} ,\\
\epsilon^*_{\pT}(q,\lambda=-)\epsilon_2(q,\lambda=-) & = & \frac{\mathbb{S}\left(q_{\mT}q_2-iq_1\Qq\right)}{2\Qq^{2}} ,\\
\epsilon^*_{\pT}(q,\lambda=-)\epsilon_{\mT}(q,\lambda=-) & = & -\frac{\mathbb{C}\mathbb{S}|{\bf q}_\perp|^2}{2\Qq^{2}} ,
\end{eqnarray}

\begin{eqnarray}
\epsilon^*_{\mT}(q,\lambda=-)\epsilon_{\pT}(q,\lambda=-) & = & -\frac{\mathbb{C}\mathbb{S}|{\bf q}_\perp|^2}{2\Qq^{2}} ,\\
\epsilon^*_{\mT}(q,\lambda=-)\epsilon_1(q,\lambda=-) & = & -\frac{\mathbb{C}\left(q_{\mT}q_1+iq_2\Qq\right)}{2\Qq^{2}} ,\\
\epsilon^*_{\mT}(q,\lambda=-)\epsilon_2(q,\lambda=-) & = & -\frac{\mathbb{C}\left(q_{\mT}q_2-iq_1\Qq\right)}{2\Qq^{2}} ,\\
\epsilon^*_{\mT}(q,\lambda=-)\epsilon_{\mT}(q,\lambda=-) & = & \frac{\mathbb{C}^2|{\bf q}_\perp|^2}{2\Qq^{2}} ,
\end{eqnarray}

\begin{eqnarray}
\epsilon^*_1(q,\lambda=-)\epsilon_{\pT}(q,\lambda=-) & = & \frac{\mathbb{S}\left(q_{\mT}q_1-iq_2\Qq\right)}{2\Qq^{2}} ,\\
\epsilon^*_1(q,\lambda=-)\epsilon_1(q,\lambda=-) & = & \frac{q_{\mT}^2q_1^2+q_2^2\Qq^{2}}{2|{\bf q}_\perp|^2\Qq^{2}} ,\\
\epsilon^*_1(q,\lambda=-)\epsilon_2(q,\lambda=-) & = & \frac{-\mathbb{C}q_1q_2-iq_{\mT}\Qq}{2\Qq^{2}} ,\\
\epsilon^*_1(q,\lambda=-)\epsilon_{\mT}(q,\lambda=-) & = & \frac{-\mathbb{C}\left(q_{\mT}q_1-iq_2\Qq\right)}{2\Qq^{2}},
\end{eqnarray}

\begin{eqnarray}
\epsilon^*_2(q,\lambda=-)\epsilon_{\pT}(q,\lambda=-) & = & \frac{\mathbb{S}\left(q_{\mT}q_2+iq_1\Qq\right)}{2\Qq^{2}} ,\\
\epsilon^*_2(q,\lambda=-)\epsilon_1(q,\lambda=-) & = &\frac{-\mathbb{C}q_1q_2+iq_{\mT}\Qq}{2\Qq^{2}} ,\\
\epsilon^*_2(q,\lambda=-)\epsilon_2(q,\lambda=-) & = & \frac{q_{\mT}^2q_2^2+q_1^2\Qq^{2}}{2|{\bf q}_\perp|^2\Qq^{2}} ,\\
\epsilon^*_2(q,\lambda=-)\epsilon_{\mT}(q,\lambda=-) & = & \frac{-\mathbb{C}\left(q_{\mT}q_2+iq_1\Qq\right)}{2\Qq^{2}},
\end{eqnarray}
where $\Qq\equiv\sqrt{q_{\mT}^{2}+\mathbf{q}_{\perp}^{2}\Cc}$.

By defining $\mathscr{T}_{\muT\nuT} = \sum_{\lambda=+,-}\epsilon^*_{\muT}(q,\lambda)\epsilon_{\nuT}(q,\lambda)$, we obtain the following matrix form:
\begin{alignat}{3}
  \mathscr{T}_{\muT\nuT} & = \frac{1}{\Qq^{2}}
  \begin{pmatrix}
  \mathbb{S}^2|{\bf q}_\perp|^2 & \mathbb{S}q_{\mT}q_1 & \mathbb{S}q_{\mT}q_2 & -\mathbb{C}\mathbb{S}|{\bf q}_\perp|^2 \\
  \mathbb{S}q_{\mT}q_1 & q_{\mT}^2+\mathbb{C}q_2^2 & -\mathbb{C}q_1q_2 & -\mathbb{C}q_{\mT}q_1\\
  \mathbb{S}q_{\mT}q_2 & -\mathbb{C}q_1q_2 & q_{\mT}^2+\mathbb{C}q_1^2 & -\mathbb{C}q_{\mT}q_2 \\
-\mathbb{C}\mathbb{S}|{\bf q}_\perp|^2 & -\mathbb{C}q_{\mT}q_1 & -\mathbb{C}q_{\mT}q_2 & \mathbb{C}^2|{\bf q}_\perp|^2
  \end{pmatrix}.\label{eqn:T_matrixform}
\end{alignat}
This can be written in general as
\begin{align}
  \mathscr{T}_{\muT\nuT}=-g_{\muT\nuT}+X(q_{\muT}n_{\nuT}+n_{\muT}q_{\nuT})+Yn_{\muT}n_{\nuT}+Z q_{\muT}q_{\nuT},\label{eqn:T_X_Y_Z}
\end{align}
where the coefficients $X$, $Y$, $Z$ can be fixed by comparing this equation with the explicit matrix form given by Eq.~(\ref{eqn:T_matrixform}).
We find
\begin{subequations}
		\begin{align}
				X&=\frac{q^{\pT}}{\Qq^{2}}=\frac{(q\cdot n)}{\Qq^{2}},\\
				Y&=-\frac{q^{2}}{\Qq^{2}},\\
				Z&=-\frac{\Cc}{\Qq^{2}}.
		\end{align}\label{eqn:X_Y_Z}
\end{subequations}
Thus, our result agrees with Eq.~(\ref{eqn:transverse_photon_propagator_numerator_in_radiation_gauge}) derived in the main text.

Similar manipulation can be done for the longitudinal part, $\mathscr{L}_{\muT\nuT}=\epsilon^*_{\muT}(q,0)\epsilon_{\nuT}(q,0)$, yielding
\begin{alignat}{3}
  \mathscr{L}_{\muT\nuT} & = \frac{(q^{\pT})^{2}}{(q)^{2}\Qq^{2}}
	\begin{pmatrix}
  L^2 & Lq_1 & Lq_2 & Lq_{\mT} \\
  Lq_1 & q_1^2& q_1q_2 & q_{\mT}q_1\\
  Lq_2 & q_1q_2 & q_2^2 & q_{\mT}q_2 \\
 Lq_{\mT} & q_{\mT}q_1 & q_{\mT}q_2 & q_{\mT}^2
  \end{pmatrix},
\end{alignat}
where we introduced for convenience, $$L = q_{\pT}-\frac{(q)^{2}}{q^{\pT}}.$$
Notice that because we are dealing with virtual photons, we have replaced $M^2$ by $P^{\muT}P_{\muT}=q^2$ in the longitudinal polarization vector given by Eq.~(\ref{eqn:polarization_vector_0_in_P_any_interpolation}).
Written in a general form, this gives Eq.~(\ref{eqn:Longitudinal_photon_propagator_numerator}).


\section{Photon propagator decomposition on the light front}
\label{sec:photon_propagator_decomposition_on_the_light_front}
Because $\Cc=0$, there's only one pole at $\mathbf{q}_{\perp}^{2}/2q^{+}-i\epsilon/2q^{+}$ for the first term in Eq.~(\ref{eqn:photon_propagator_gauge_decomposition_2}).
When $q^{+}>0$, the pole is in the lower half plane, and when $q^{+}<0$, the pole is in the upper half plane.
So, for the first term in Eq.~(\ref{eqn:photon_propagator_gauge_decomposition_2}), we separate the $q^{+}$ integral into two regions:
\begin{align}
  \int\frac{d^{2}\mathbf{q}_{\perp}}{(2\pi)^{4}}
	\left[
	\int_{0}^{\infty}dq^{+}e^{-i(q^{+}x^{-}-\mathbf{q}^{\perp}\mathbf{x}^{\perp})}I\right. \nonumber\\
	\left.+\int_{-\infty}^{0}dq^{+}e^{-i(q^{+}x^{-}-\mathbf{q}^{\perp}\mathbf{x}^{\perp})}I
	\right],
	\label{eqn:Photon_propagator_LF_decomposition_term1}
\end{align}
where
\begin{align}
  I=\int_{-\infty}^{\infty}dq^{-} e^{-iq^{-}x^{+}}\frac{\mathscr{T}_{\mu\nu}}{2q^{-}q^{+}-\mathbf{q}_{\perp}^{2}+i\epsilon}.\label{eqn:Inner_most_integral_in_photon_propagator}
\end{align}
To evaluate $I$, we close the contour in the plane where the arc contribution is zero.
This means we close the contour in the upper (lower) half plane when $x^{+}<0$ ($x^{+}>0$).
For the factor $I$ in the first term of Eq.~(\ref{eqn:Photon_propagator_LF_decomposition_term1}), the only none-zero contribution comes from the pole that's in the lower half plane, which is picked up when we close the contour from below ($x^{+}>0$).
Similarly, for the factor $I$ in the second term of Eq.~(\ref{eqn:Photon_propagator_LF_decomposition_term1}), the only none-zero contribution comes from the pole that's in the upper half plane, which is picked up when we close the contour from above ($x^{+}<0$).
Thus, Eq.~(\ref{eqn:Photon_propagator_LF_decomposition_term1}) becomes
\begin{align}
  \int\frac{d^{2}\mathbf{q}_{\perp}}{(2\pi)^{4}}\mathscr{T}_{\mu\nu}
	\left[
	\int_{0}^{\infty}dq^{+}e^{-i(q^{+}x^{-}-\mathbf{q}^{\perp}\mathbf{x}^{\perp})}(-2\pi i)\Theta(x^{+})\frac{e^{-iHx^{+}}}{2q^{+}}\right. \nonumber\\
	\left.+\int_{-\infty}^{0}dq^{+}e^{-i(q^{+}x^{-}-\mathbf{q}^{\perp}\mathbf{x}^{\perp})}(2\pi i)\Theta(-x^{+})\frac{e^{iHx^{+}}}{2q^{+}}
	\right],
	\label{eqn:Photon_propagator_LF_decomposition_term1_step2}
\end{align}
where $H=|\mathbf{q}_{\perp}^{2}/2q^{+}|$.

We now make changes of the variables $q^{+}\rightarrow -q^{+}$ and $\mathbf{q}_{\perp}\rightarrow - \mathbf{q}_{\perp}$ in the second term that's proportional to $\Theta(-x^{+})$ in order to have the limits on the $q^{+}$ integral become the same as the first term.
We can then combine these two terms.
This gives us
\begin{align}
  i\int\frac{d^{2}\mathbf{q}_{\perp}}{(2\pi)^{3}}
	\int_{0}^{\infty}\frac{dq^{+}}{2q^{+}}\mathscr{T}_{\mu\nu}
	\left[\Theta(x^{+})e^{-iHx^{+}}e^{-i(q^{+}x^{-}-\mathbf{q}^{\perp}\mathbf{x}^{\perp})}\right. \nonumber\\
	\left.+\Theta(-x^{+})e^{iHx^{+}}e^{i(q^{+}x^{-}-\mathbf{q}^{\perp}\mathbf{x}^{\perp})}
	\right] \nonumber\\
	=i\int\frac{d^{2}\mathbf{q}_{\perp}}{(2\pi)^{3}}
	\int_{0}^{\infty}\frac{dq^{+}}{2q^{+}}\mathscr{T}_{\mu\nu}
	\left[\Theta(x^{+})e^{-iq_{\mu}x^{\mu}}+\Theta(-x^{+})e^{iq_{\mu}x^{\mu}}\right].
	\label{eqn:Photon_propagator_LF_decomposition_term1_step3}
\end{align}
Since we've made the change of variables and the $q^{+}$ can not be negative anymore, we now have $q^{-}=H=\mathbf{q}_{\perp}^{2}/2q^{+}$ in $q_{\mu}$.

The second term in Eq.~(\ref{eqn:photon_propagator_gauge_decomposition_2}) gives straightforwardly the same result as the last term in Eq.~(\ref{eqn:photon_time_ordered_propagator}), where $\Cc=0$ (LFD).
Putting everything together, we can write Eq.~(\ref{eqn:photon_propagator_gauge_decomposition_2}) in LFD as
\begin{multline}
  i\int\frac{d^{2}\mathbf{q}_{\perp}}{(2\pi)^{3}}
	\int_{0}^{\infty}\frac{dq^{+}}{2q^{+}}\mathscr{T}_{\mu\nu}
	\left[\Theta(x^{+})e^{-iq_{\mu}x^{\mu}}+\Theta(-x^{+})e^{iq_{\mu}x^{\mu}}\right]\\
		+i\delta(x^{+})\int \dfrac{d^{2}\mathbf{q}_{\perp} }{(2\pi)^{3}}\int_{-\infty}^{\infty}d q^{+} \dfrac{n_{\mu}n_{\nu}}{{q^{+}}^{2}} e^{-i(q^{+}x^{-}+\mathbf{q}_{\perp}\mathbf{x}^{\perp})}.
	\label{eqn:append_LF_propogator_decomposition}
\end{multline}


\section{Time-ordered scattering amplitudes for $\phi^{3}$ Theory}
\label{Append:time_ordered_scattering_amplitudes_for_pure_scalar_theory}
\begin{figure}[!hb]
\centering
\subfloat{\includegraphics[width=\columnwidth]{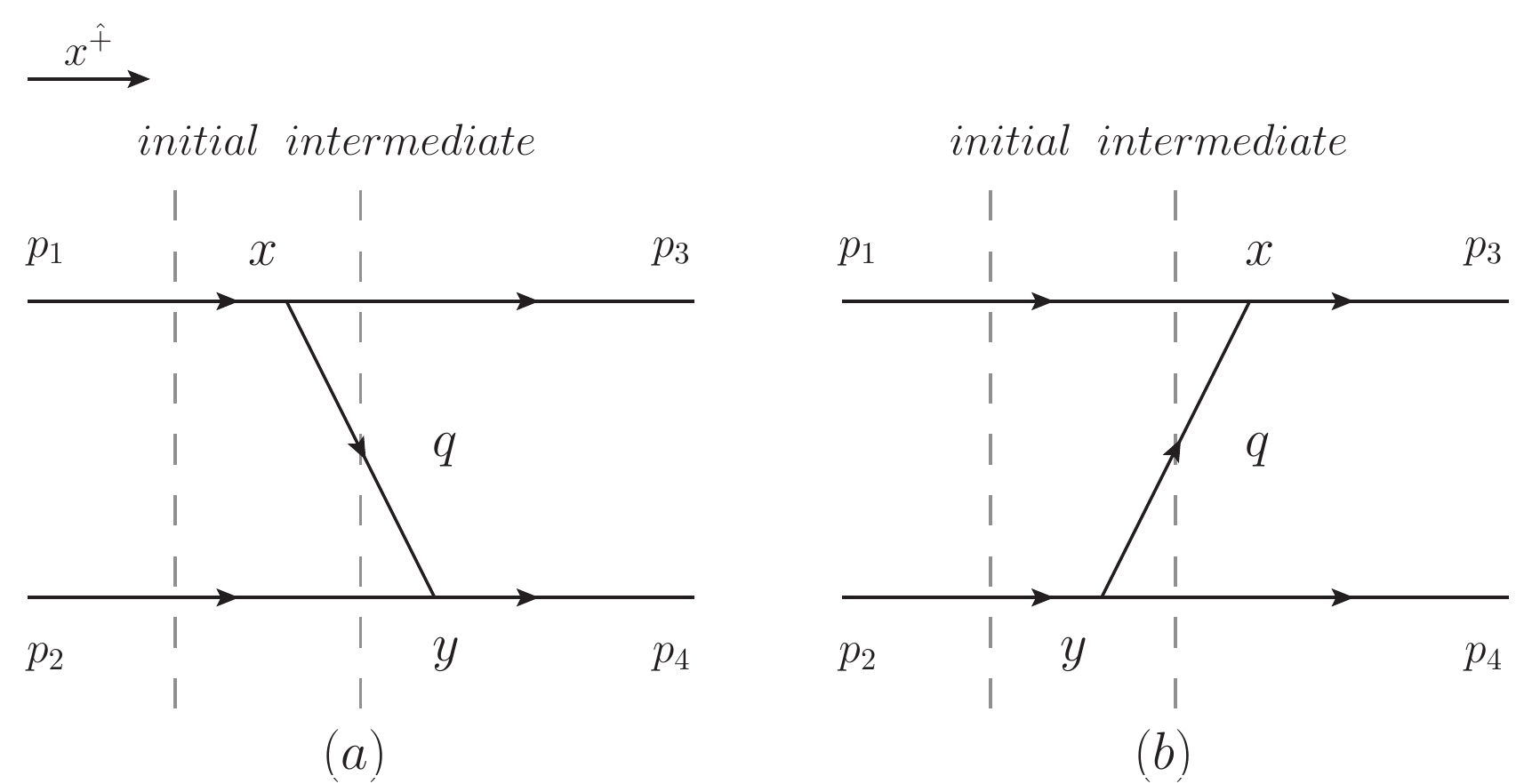}}
\vspace{0pt}
\centering
\subfloat{
\includegraphics[width=0.82\columnwidth]{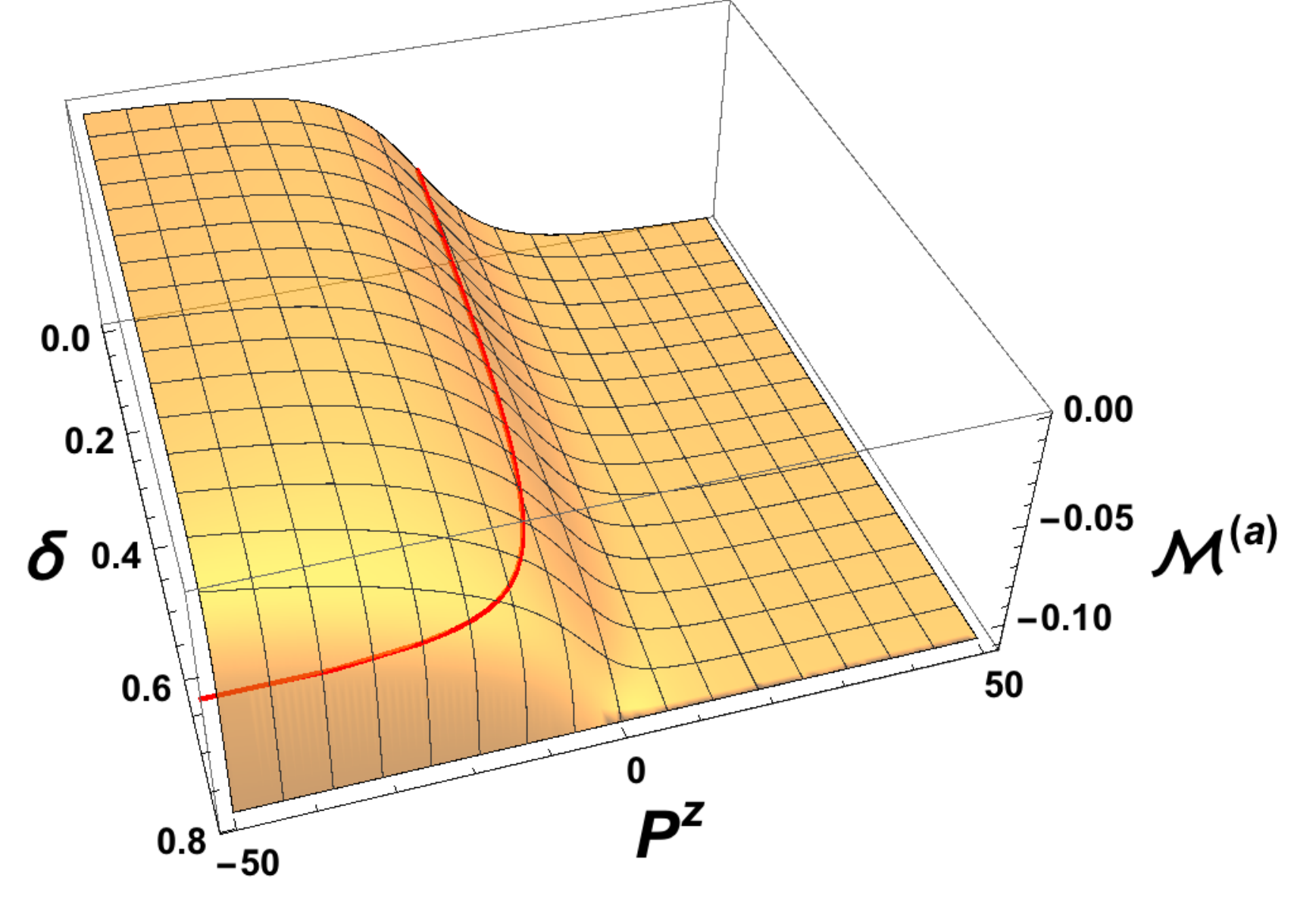}
}
\vspace{0pt}
\centering
\subfloat{
\includegraphics[width=0.82\columnwidth]{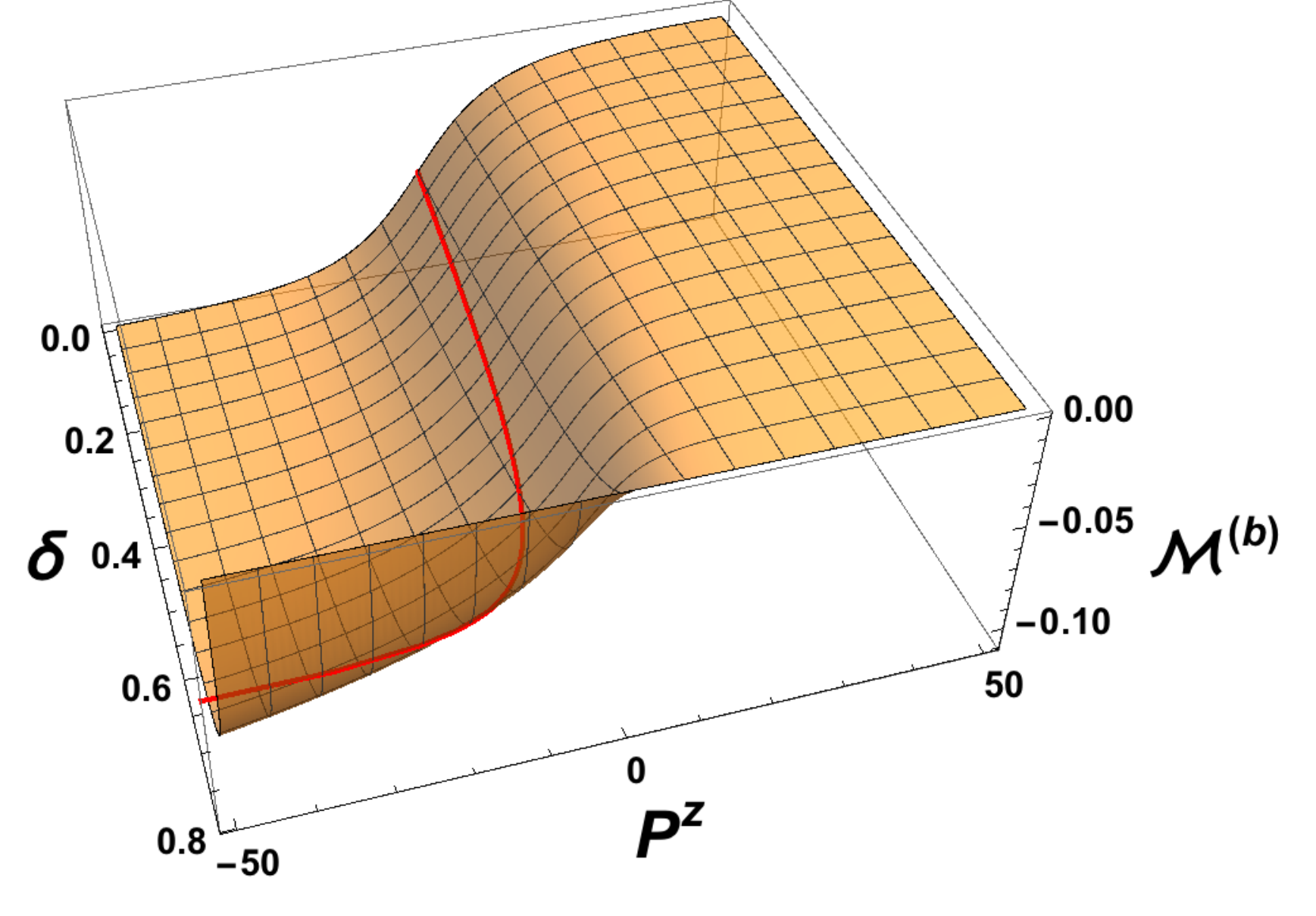}
}
\caption{\label{fig:scattering_phi3}(Color Online) Time-ordered scattering diagrams and corresponding amplitudes for the $\phi^{3}$ theory.}
\end{figure}
In $\phi^{3}$ theory, the scattering shown in FIG.~\ref{fig:time_ordered_photon_propagator} involves the exchange of scalar particle.
 Ignoring the inessential factors including the square of the coupling constant, we can write the time-ordered scattering amplitudes as those used in Ref.~\cite{Ji2012}:
\begin{subequations}
  \label{eqn:Mab_scalar_scattering}
  \begin{alignat}{2}
     &\mathcal{M}^{(a)}
     =
     &&\dfrac{1}{2 Q^{\pT}}
     \dfrac{1}{q^{\pT}-Q^{\pT}}, \label{eqn:Ma_scalar_scattering}\\
     &\mathcal{M}^{(b)}
     =
     - &&\dfrac{1}{2 Q^{\pT}}
    \dfrac{1}{q^{\pT}+Q^{\pT}},  \label{eqn:Mb_scalar_scattering}
  \end{alignat}
\end{subequations}
where $Q^{\pT}$ is defined in Eq.~(\ref{eqn:Q_sup_ph_q_final}).

We use the same kinematics used for the sQED case, i.e. $\mathbf{q}_{\perp}=0$.
Because the kinematics is the same, the boosted $q'$ satisfies the same equations as listed in Eq.~(\ref{eqn:boosted_p_mu}).
Using Eq.~(\ref{eqn:P_interpolation}), we can also get the boosted ${q'}_{\muT}$ and ${q'}^{\muT}$, which can then be used to calculate the scattering amplitudes in the boosted frame.

The amplitudes are given by the functions of $m_{1}$, $m_{2}$, $p$ and $\theta$.
For comparison, we use the same parameter values that we used for plotting FIG.~\ref{fig:amp_abc} in sQED, i.e. $m_{1}=1$, $m_{2}=2$, $p=3$, and $\theta=\pi/3$.
$\mathcal{M}^{(a)}$ and $\mathcal{M}^{(b)}$ with the J-curve depicted by the solid red line are plotted in FIG.~\ref{fig:scattering_phi3}.
Comparing it to FIG.~\ref{fig:amp_abc}, we see that the photon propagator modifies the profiles of the time-ordered amplitudes.
Nevertheless, the J-curve remains the same and is still given by Eq.~(\ref{eqn:J_curve}).



\section{Time-ordered annihilation amplitudes in sQED}
\label{Append:time_ordered_annihilation_amplitudes_for_photon_exchange_process}
\begin{figure}[h!]
\centering
\vspace{10pt}
\includegraphics[width=\columnwidth]{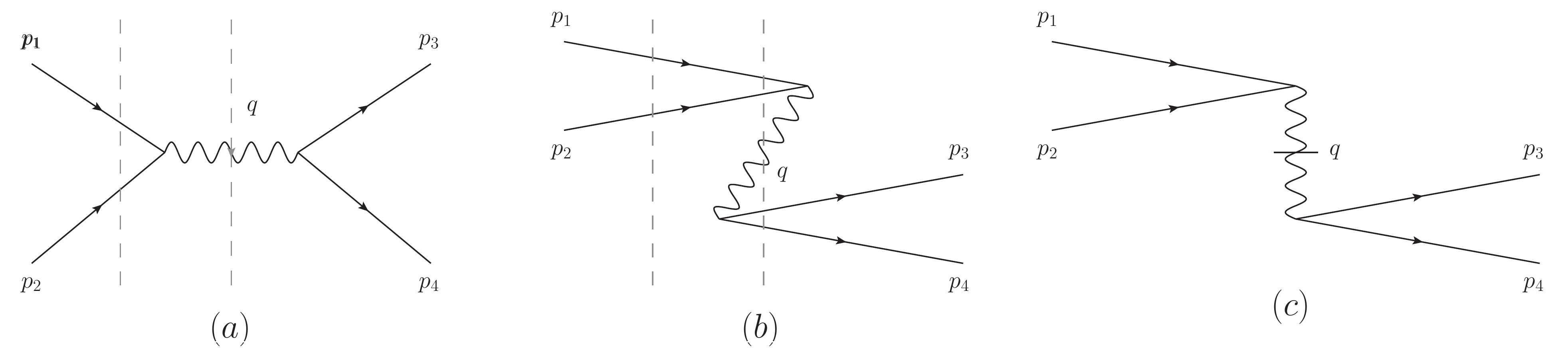}\\[10pt]
\vspace{0pt}
\centering
\subfloat{
\includegraphics[width=0.75\columnwidth]{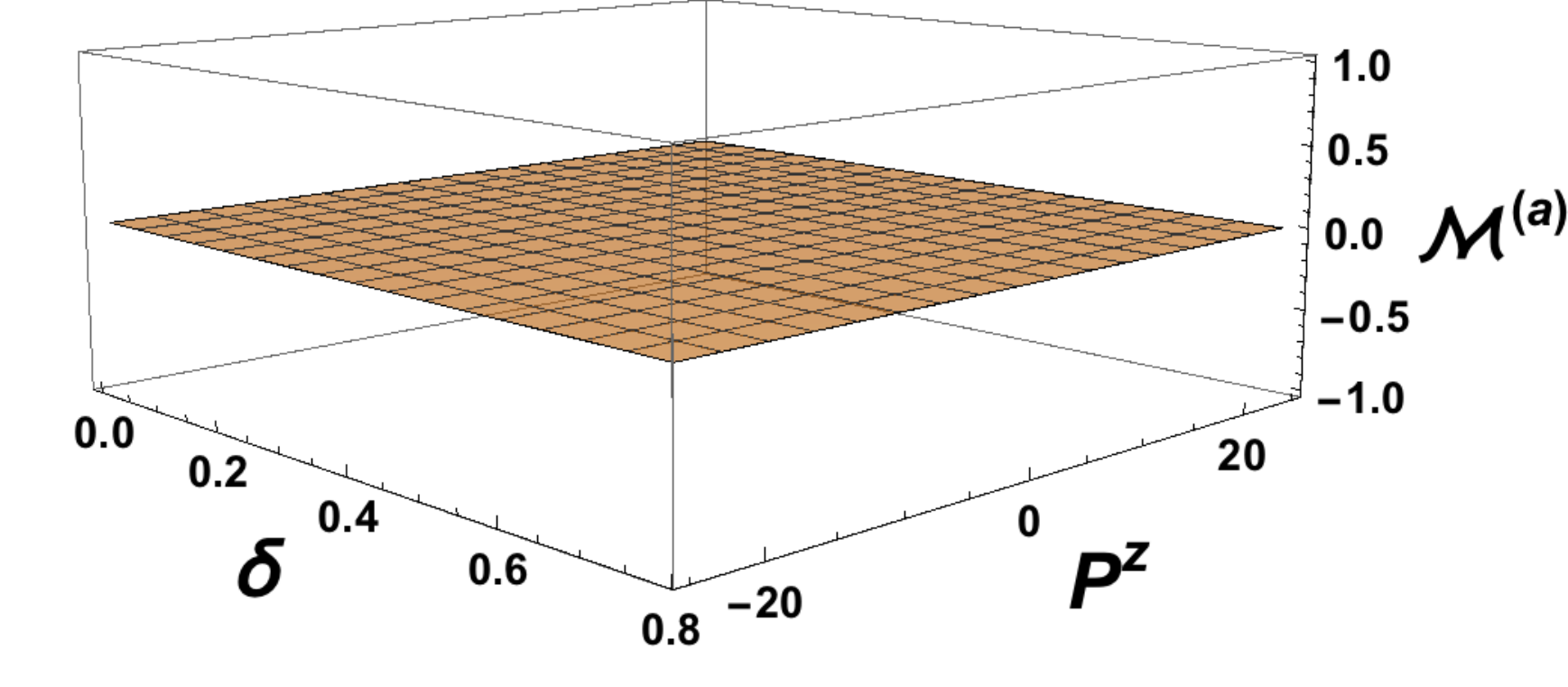}
}
\vspace{0pt}
\centering
\subfloat{
\includegraphics[width=0.75\columnwidth]{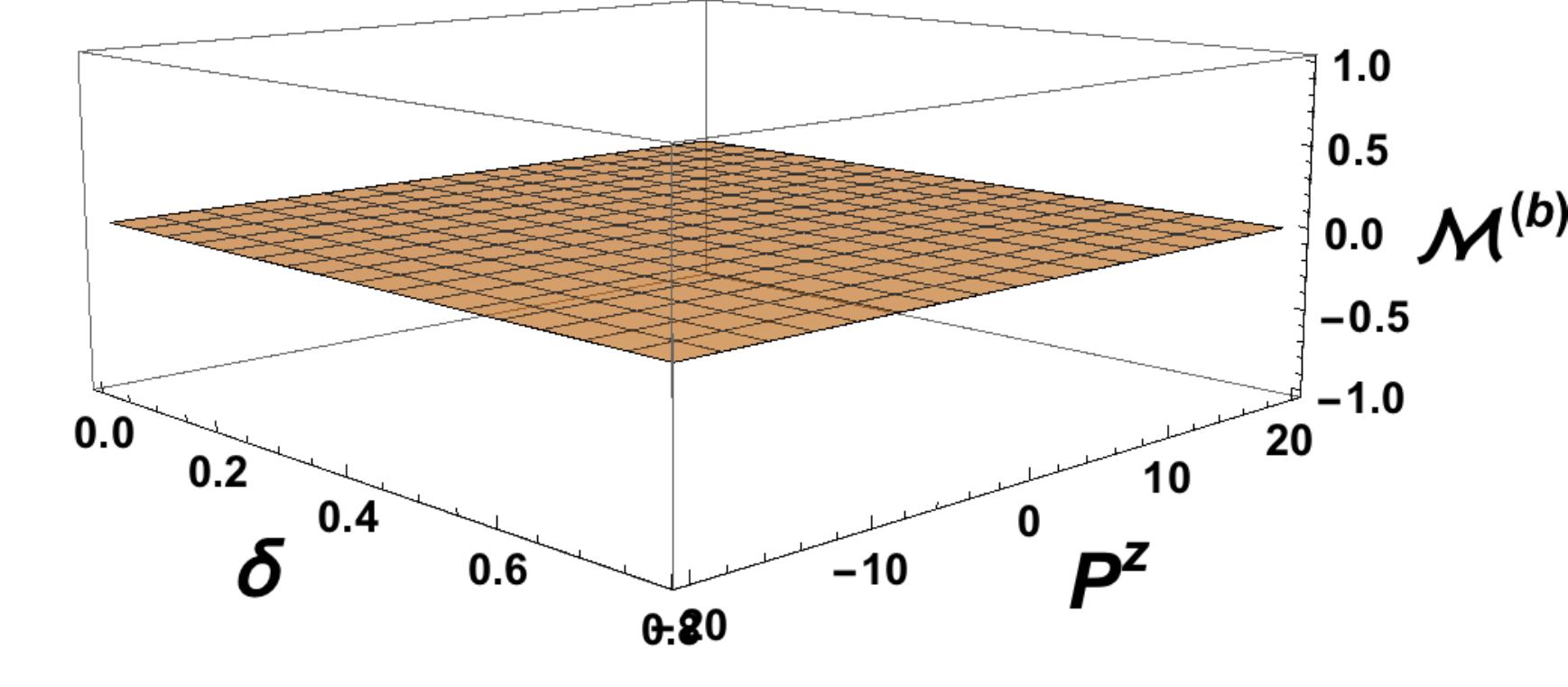}
}
\vspace{0pt}
\centering
\subfloat{
\includegraphics[width=0.75\columnwidth]{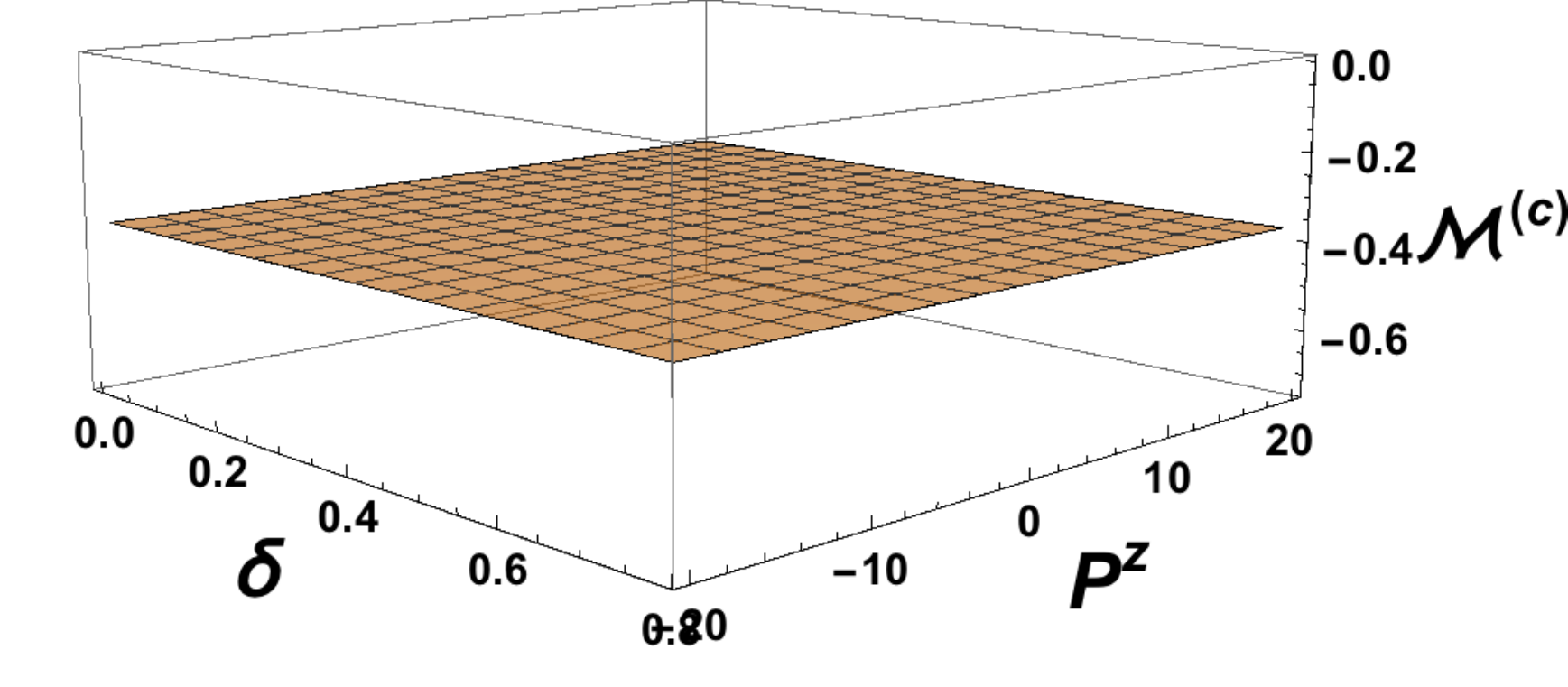}
}
\caption{Time-ordered annihilation diagrams and their amplitudes for sQED\label{fig:annihilation_sQED}}
\end{figure}

For the sQED annihilation process analogous to $e^+ e^- \rightarrow \mu^+ \mu^-$ in QED, we can still use FIG.~\ref{fig:time_ordered_photon_propagator} for the kinematics.
Three time-ordered diagrams corresponding to the lowest order amplitudes are shown in FIG.~\ref{fig:annihilation_sQED}.
Here, the amplitude of Eq.~(\ref{eqn:M_Ma_Mb_Mc}) changes to
\begin{align}
  i\mathcal{M}=&\sum_{j=a,b,c}{i\mathcal{M}^{(j)}}\nonumber\\
   =&(-ie)^{2}\sum_{j=a,b,c}{(p_{3}^{\muT}-p_{4}^{\muT})\Sigma^{(j)}_{\muT\nuT}(p_{1}^{\nuT}-p_{2}^{\nuT})}.\label{eqn:M_annihilation}
\end{align}
The time-ordered amplitudes are given by
\begin{subequations}
  \label{eqn:M_final_anni}
 \begin{align}
    \mathcal{M}^{(a)}=&-(-ie)^{2}\dfrac{[p_{34}\cdot p_{12}+p_{34}^{\pT}p_{12}^{\pT}q^{2}/(Q^{\pT})^{2}]\Cc}{2Q^{\pT}(q^{\pT}-Q^{\pT})}\label{eqn:M_a_final_anni},\\
    \mathcal{M}^{(b)}=&(-ie)^{2}\dfrac{[p_{34}\cdot p_{12}+p_{34}^{\pT}p_{12}^{\pT}q^{2}/Q^{\pT})^{2}]\Cc}{2Q^{\pT}(q^{\pT}+Q^{\pT})}\label{eqn:M_b_final_anni},\\
    \mathcal{M}^{(c)}=&(-ie)^{2}\dfrac{p_{34}^{\pT}p_{12}^{\pT}}{(Q^{\pT})^{2}}\label{eqn:M_c_final_anni},
 \end{align}
\end{subequations}
where
\begin{subequations}
  \begin{align}
    p_{34}=p_{3}-p_{4},\label{eqn:p34m}\\
    p_{12}=p_{1}-p_{2},\label{eqn:p12m}
  \end{align}
\end{subequations}
and $q=p_{1}+p_{2}$.
We use the same kinematics (CMF) as before so that $p_{i}$'s are given by
\begin{subequations}
  \label{eqn:p_i_CM_anni}
  \begin{align}
    p_{1}&=(\epsilon,0,0,p),\\
    p_{2}&=(\epsilon,0,0,-p),\\
    p_{3}&=(\epsilon,p_{f} \sin\theta,0,p_{f}\cos\theta),\\
    p_{4}&=(\epsilon,-p_{f} \sin\theta,0,-p_{f}\cos\theta).
  \end{align}
\end{subequations}
Since we are considering the annihilation process, the initial particles 1 and 2 should have the same mass, and the final particles 3 and 4 should have the same mass.
So, we set $m_{1}=m_{2}=m$ and $m_{3}=m_{4}=m_{f}$.
Due to the energy conservation, all four particles should have the same energy in CMF, and $p$ and $p_{f}$ are therefore related by $p_{f}=\sqrt{q^{2}+m^{2}-m_{f}^{2}}$.
The boosted ${p'}_{i}$'s still follow Eq.~(\ref{eqn:p_i_boosted}).
Thus, $q'$ in the annihilation process is rather simple:
\begin{subequations}\label{eqn:q_boosted for annihilation}
  \begin{align}
    {q'}^{x}&={p'}_{1}^{x}+{p'}_{2}^{x}=0,\\
    {q'}^{y}&={p'}_{1}^{y}+{p'}_{2}^{y}=0,\\
    {q'}^{z}&={p'}_{1}^{z}+{p'}_{2}^{z}=P^{z},\\
    {q'}^{0}&={p'}_{1}^{0}+{p'}_{2}^{0}=E.
  \end{align}
\end{subequations}

Applying Eq.~(\ref{eqn:p_i_boosted}) to Eqs.~(\ref{eqn:p34m}) and (\ref{eqn:p12m}) with $p_{i}$ given by Eq.~(\ref{eqn:p_i_CM_anni}), we find that
\begin{subequations}
\begin{align}
  {p'}_{34}\cdot{p'}_{12}=-4p \sqrt{m^{2}-m_{f}^{2}+p^{2}}\cos\theta, \label{eqn:p34p12_annihilation}\\
  {q'}^{2}\dfrac{{p'}_{34}^{\pT}{p'}_{12}^{\pT}}{({Q'}^{\pT})^{2}}=4p \sqrt{m^{2}-m_{f}^{2}+p^{2}}\cos\theta. \label{eqn:p34pp12p_annihilation}
\end{align}
\end{subequations}
So the numerators in Eqs.~(\ref{eqn:M_a_final_anni}) and (\ref{eqn:M_b_final_anni}) are 0, regardless of the frame or the interpolation angle $\delta$.
Therefore, $\mathcal{M}^{(a)}=\mathcal{M}^{(b)}=0$, and the only non-zero contribution comes from the instantaneous interaction.
One can verify that indeed the covariant total amplitude is the same as $\mathcal{M}^{(c)}$:
\begin{align}
  \mathcal{M}^{(c)}=\mathcal{M}=-\dfrac{p_{34}\cdot p_{12}}{q^{2}}=-\dfrac{p \sqrt{m^{2}-m_{f}^{2}+p^{2}}\cos\theta}{m^{2}+p^{2}},\label{eqn:Mc_anni}
\end{align}
where the coupling constant $e$ is taken as 1.
We can see that this result is also independent of the frame and interpolation angle, just as expected.

In FIG.~\ref{fig:annihilation_sQED}, we again use the same parameter values as before: $m_{1}=1$, $m_{2}=2$, $p=3$, and $\theta=\pi/3$.
The corresponding time-ordered amplitudes shown here are all flat planes.

\bibliography{Interpolation_photon}
\bibliographystyle{apsrev4-1}

\end{document}